\documentclass[aps,prd,twocolumn,showpacs,notitlepage,eqsecnum,
superscriptaddress,nofootinbib]{revtex4-1}

\usepackage[centertags]{amsmath} \usepackage{amssymb} \usepackage{latexsym}
\usepackage{enumerate} \usepackage{graphicx} \usepackage{mathrsfs}
\usepackage[colorlinks]{hyperref} \usepackage{stmaryrd} \usepackage{import}
\usepackage{tensor} \usepackage[usenames,dvipsnames]{xcolor} \usepackage{bm}
\usepackage{multirow} \usepackage{breakurl} \usepackage{float}
\usepackage{verbatim} \usepackage{mathrsfs}

\allowdisplaybreaks[1]

\definecolor{CiteColor}{rgb}{0,0.5,0} \hypersetup{citecolor=CiteColor}
\definecolor{RefColor}{rgb}{0.55,0,0} \hypersetup{linkcolor=RefColor}
\definecolor{darkgreen}{rgb}{0.2,0.7,0.2}

\newcommand{\bi}{\begin{itemize}} \newcommand{\ei}{\end{itemize}}
\newcommand{\be}{\begin{equation}} \newcommand{\ee}{\end{equation}}
\newcommand{\nn}{\nonumber} 

\renewcommand{\l}{\left(} \renewcommand{\r}{\right)} \renewcommand{\a}{\alpha}
\renewcommand{\b}{\beta} \newcommand{\g}{\gamma} 
\renewcommand{\d}{\delta} \newcommand{\D}{\Delta} \newcommand{\e}{\epsilon}
 
\newcommand{\la}{\lambda} \renewcommand{\O}{\Omega} \renewcommand{\o}{\omega}
\renewcommand{\th}{\theta}  \newcommand{\q}{\quad}
 \newcommand{\s}{\sigma} \newcommand{\vp}{\varphi}

\newcommand{\pa}{\partial}

\begin{document}

\title{Evolution of small-mass-ratio binaries with a spinning secondary}

\author{Niels Warburton} \affiliation{School of Mathematics and Statistics and
UCD Institute for Discovery, University College Dublin, Belfield, Dublin 4,
Ireland} \author{Thomas Osburn} \affiliation{Department of Physics and
Astronomy, Oxford College of Emory University, Oxford, Georgia 30054, USA}
\author{Charles R.~Evans} \affiliation{Department of Physics and Astronomy,
University of North Carolina, Chapel Hill, North Carolina 27599, USA}

\begin{abstract} We calculate the evolution and gravitational-wave emission of
a spinning compact object inspiraling into a substantially more massive
(non-rotating) black hole. We extend our previous model for a non-spinning
binary [Phys.~Rev.~D 93, 064024] to include the Mathisson-Papapetrou-Dixon
spin-curvature force.  For spin-aligned binaries we calculate the dephasing 
of the inspiral and associated waveforms relative to models that do not 
include spin-curvature effects. We find this dephasing can be either positive or negative depending on the initial separation of the binary. For binaries in which the spin and orbital 
angular momentum are not parallel, the orbital plane precesses and we use 
a more general osculating element prescription to compute inspirals. 
\end{abstract}

\pacs{04.25.dg, 04.30.-w, 04.25.Nx, 04.30.Db}

\maketitle

\section{Introduction}\label{sec:intro}

The era of gravitational wave astronomy has recently dawned
\cite{LIGO16a,LIGO16b,LIGO17a}, being ushered in by tremendous advances in
experimental physics, data analysis, and theoretical source modeling.  Source 
modeling is necessary, at minimum, for estimating the physical parameters of 
relativistic compact binary inspirals.  It can also be required to even detect 
systems with comparable masses \cite{LIGO16b} and it is expected to be 
crucial for detection of extreme-mass-ratio binaries.  For both detection and 
parameter estimation, searches are run over a large parameter space 
(using e.g., Monte-Carlo-based methods), at each step convolving the data
with theoretical waveform templates. The signal-to-noise ratio (SNR) will only
coherently grow with the number of oscillations in the signal if the phase 
of the template and the signal are closely matched, typically to less than 
one radian over the observed time span.  Waveform templates that do
not meet this requirement will result in a substantial loss of accuracy for
parameter estimation, or even worse provide no detection at all.  Tracking 
the signal phase to within one radian is a particularly stringent requirement 
for small mass-ratio systems which accumulate hundreds to many hundreds of 
thousands of radians of phase whilst in the detector band.

The gravitational waves emitted from the inspiral of a stellar mass black hole
or other comparable mass compact object into a massive black hole with mass
$M\sim10^4$--$10^7M_\odot$ are in the frequency band of the Laser
Interferometer Space Antenna, LISA, which has recently been selected as the 
European Space Agency's L3 mission \cite{LISA}.  These extreme mass-ratio 
inspirals (EMRIs)
are key sources for LISA \cite{Babak:2017tow}. The information carried in the
waveforms from EMRIs will allow us to precisely measure properties of massive
black holes and their stellar environment as well as providing unprecedented
tests of general relativity in the strong-field regime
\cite{VigeHugh10,BaraCutl07,BrowETC07,Gair:2012nm,Barausse:2014tra}. 

Another class of interesting systems involve intermediate mass black holes
(IMBHs) with $M\sim10^2$--$10^3 M_\odot$.  Such black holes can form 
intermediate mass-ratio inspirals (IMRIs) that fall into two categories 
depending on whether there is a stellar mass compact object inspiraling into 
the IMBH or the IMBH is inspiraling into a massive black hole.  The former 
will merge in the band of
ground-based detectors if the chirp mass of the binary is $\lesssim 350M_\odot$
\cite{Brown:2006pj} (this is slightly increased for ground-based observatories
beyond the advanced detector era \cite{Huerta:2010un,Huerta:2010tp}). The
inspiral phase of such binaries should also be detectable in LISA many weeks
before their merger in the LIGO band \cite{Miller:2002vg,Sesana:2016ljz}. The
second type of IMRI would be a very loud source in the LISA band (possibly not
even requiring matched filtering to find \cite{Miller:2004va}). As the
population of IMBHs is poorly understood the event rates for both types of
IMRIs are not well constrained \cite{Mandel:2007hi}. For this reason we will
concentrate our discussion around EMRIs whilst bearing in mind that IMRIs are
an exciting potential class of sources that can be modeled with a similar
setup.

The leading-order (in the mass-ratio) dissipative dynamics of EMRIs is now
understood when the primary is rotating \cite{DrasHugh05,FujiHikiTago09} but
has yet to be compiled into complete inspiral models (many so-called `kludge'
models have been developed \cite{Babak:2006uv} which use some of this
information but to date the development of these models has been driven by the
need to rapidly compute waveform templates to train data analysis algorithms
rather than a need for high precision).  Producing waveform models that track
the phase evolution of the binary whilst it is emitting gravitational waves in
LISA's band to within one radian requires including subleading-order
corrections to the orbital dynamics.

These subleading-order corrections include the following effects: (1) orbital 
resonances, (2) first-order conservative terms, (3) oscillatory first-order 
dissipative terms, (4) second-order in the mass-ratio, orbit-averaged 
dissipative contribution, and (5) spin-orbit coupling effects 
\cite{HindFlan08}.  There has been great progress calculating some of these 
effects within black hole perturbation theory and,
in particular, within the self-force program \cite{PoisPounVega11}. For
geodesic equatorial motion we have calculations at first order in the mass
ratio of conservative dynamics about non-rotating \cite{Detw08, BaraSago09,
BaraSago11, DolaETC14a, DolaETC14b, Nolan:2015vpa,Akcay:2015pza,ShahPoun15} and
rotating black holes \cite{ShahFrieWhit14, Isoyama:2014mja, VandShah15,
vandeMeent:2016hel,Kavanagh:2015lva}. These results have been successfully
compared and synergized with results from other approaches to the two-body
problem \cite{BlanETC09, LetiETC11, Akcay:2012ea, Bini:2013rfa, Bini:2014ica,
Bini:2014zxa, Isoyama:2014mja, ShahFrieWhit14,  Akcay:2015pza, VandShah15,
Zimmerman:2016ajr,Akcay:2016dku}. Orbital resonances are known to be important
\cite{FlanHind10, Brink:2013nna, Ruangsri:2013hra} but a precise understanding
of them awaits calculations for generic orbits about rotating black holes.
Steady progress is also being made on second order in the mass-ratio
calculations \cite{Poun12a,Gral12,Detw12, Pound:2014koa, WarbWard14,
PounMill14, WardWarb15, Pound:2015wva, Miller:2016hjv, Pound:2017psq}.
Including these various results into self-force inspiral models has also been
an active field of research \cite{WarbETC12, OsbuWarbEvan16, DienETC12}.

The primary focus of this work is to incorporate effects associated with the
spin of the secondary into inspiral models for small mass-ratio systems. This
has been explored previously using models incorporating post-Newtonian results
\cite{Huerta:2011kt,Huerta:2011zi,Huerta:2011zi} and in models using
strong-field self-force results restricted to quasi-circular inspirals
\cite{Burko:2015sqa}. In this work we compute the effect of spin-curvature
coupling on generic inspirals into a non-rotating black hole. 

The next section details the different drivers of an inspiral and highlights
which ones we are including in our current model. The rest of the paper is
organized as followed. In Sec.~\ref{sec:overview} we give an overview of our
approach to modeling inspirals. In Sec.~\ref{sec:osculate} we describe the
osculating element prescription we use, extending previous formulations to
generic motion (not confined to the equatorial plane). In
Sec.~\ref{sec:forcing_terms} we describe the calculation of the self-force and
spin-curvature forcing terms. In Sec.~\ref{sec:waveforms} we describe how, once
we have the inspiral trajectory, we compute the associated waveform by moving
through a sequence of so-called snapshot waveforms. In Sec.~\ref{sec:aligned}
we give results for spin-aligned binaries where we observe the difference in the accumulated inspiral phase with respect to a non-spinning binary can be either positive or negative depending on the initial separation of the binary. In \ref{sec:inclined} we give results for spin-unaligned binaries which precess out of the equatorial plane. Finally, we give some concluding remarks in Sec.~\ref{sec:conclusions}.

Throughout this work we will use geometrized units such that the speed of 
light and the gravitational constant are equal to unity ($c=G=1$).  We will 
denote the mass of the primary by $M$ and the mass of the secondary by $\mu$, 
and will adopt standard Schwarzschild coordinates 
$x^{\a} = (t, r, \theta, \varphi )$. 

\section{Physical drivers of an inspiral} 
\label{sec:effects_of_SF}

The physical mechanism of an inspiral can be viewed as a force that drives the
secondary's orbit away from geodesic motion in the (background) spacetime of 
the primary.  This force has a non-local contribution arising from the body's 
interaction with its own metric perturbation, commonly called the self-force.  
If the secondary is spinning there is an additional non-local contribution 
resulting from perturbing the orbit and the stress-energy tensor of the body 
as well as a local contribution arising from the coupling between the spin of 
the body and the tidal field (background curvature) of the primary.  The 
latter of these is the well-known Mathisson-Papapetrou-Dixon (MPD) 
spin-curvature force \cite{Papa51,Dixo70}.  By expanding the Einstein field 
equations perturbatively
in powers of the mass ratio, $\epsilon = \mu/M \ll 1$, the equations of motion
for the inspiraling body can be written as 
\begin{align}
\label{eqn:forced} 
\mu u^\beta \nabla_\beta u^\alpha =& \mu^2\left(F^{(1)\alpha}_\text{mono} + \mu
F^{(2)\alpha}_\text{mono}\right) 
\\ 
& + S\left(F^\alpha_\text{spin-curvature}
\nonumber + \mu F^{(1)\alpha}_\text{dipole}\right) \equiv F^\alpha ,
\end{align} 
where $u^\a$ is the body's four-velocity, $\nabla$ denotes the
covariant derivative with respect to the background metric of the primary, $S$
is the spin magnitude, and $F^\alpha$ is the net force. By $F^{(n)\alpha}$ we
denote the $n$th-order self-force, i.e., the part proportional to the $n+1$
power of the mass ratio. The `mono/dipole' subscripts denote whether the force
arises from the mass or spin of the secondary respectively. The subscript
`spin-curvature' denotes the MPD force. As we discuss below, in
Eq.~\eqref{eqn:forced} we have truncated the expansion in the mass ratio at
high enough order to include all important effects.

When the secondary moves along a geodesic of the background spacetime the
influence on the inspiral of the forces on the righthand side of
Eq.~\eqref{eqn:forced} can be split into conservative (time-symmetric) and
dissipative (time-antisymmetric) pieces such that 
\begin{align} 
F^\a = F^\a_\text{diss} + F^\a_\text{cons} .  
\end{align} 
Conservative forces act to perturb the orbital parameters, but do not cause 
a secular decay of the orbit.  The self-force has a conservative component 
and the leading-order MPD force $F^\a_\text{spin-curvature}$ is also 
conservative in nature.  Dissipative forces are responsible for radiation 
reaction effects that lead to, e.g., the decay of orbital energy and angular 
momentum.  This secular decay can be calculated by averaging 
$F^\a_\text{diss}$, which motivates a decomposition of the net force
into an adiabatic part, $F^\a_\text{ad}$, and an oscillatory part,
$F^\a_\text{osc}$ 
\begin{align} 
F^\a 			&= F^\a_\text{ad} +
F^\a_\text{osc}, \notag \\ F^\a_\text{ad}  &\equiv \langle F^\a_\text{diss}
\rangle ,\\ F^\a_\text{osc} &\equiv F^\a - \langle F^\a_\text{diss} \rangle .
\end{align} 
The adiabatic part varies slowly over an inspiral on the radiation reaction 
timescale and represents some average over the orbital timescale (see
\cite{OsbuWarbEvan16} for details about an appropriate averaging procedure).
The oscillatory part varies more rapidly on the orbital time scale. 

A number of authors have considered how these different forces influence the
phase of an inspiral \cite{HughETC05,DrasFlanHugh05,Tanaka:2006} with one of
the most rigorous discussions given by Hinderer and Flanagan \cite{HindFlan08}.
We now briefly review several key results and highlight where previous work has
employed the various components of the self-force in computing inspirals.

EMRIs accumulate tens to hundreds of thousands of radians of orbital phase
whilst the binary is in the LISA passband.  The leading-order contribution to
the inspiral phase enters at $\mathcal{O}(\epsilon^{-1})$ and is driven by the
adiabatic, first-order-in-the-mass-ratio, self-force 
$F^{(1)\alpha}_{\text{ad}}$.  A number of authors have used
$F^{(1)\alpha}_{\text{ad}}$ to calculate the leading-order phase evolution of
generic inspirals into Kerr black holes
\cite{DrasHugh06,FujiHikiTago09,Babak:2006uv}.  At subleading order (ignoring
resonances that only affect generic inspirals into rotating black holes
\cite{FlanHind10}) the next contributions to the orbital phase enter at
$\mathcal{O}(\epsilon^0)$.  These include the oscillatory part of the
first-order force, $F^{(1)\alpha}_{\text{osc}}$, and the adiabatic part of the
second-order force, $F^{(2)\alpha}_{\text{ad}}$. 

In order to classify the effects of the secondary's spin in this hierarchy we
must relate the spin magnitude to the mass ratio.  If the secondary is a
rotating black hole we can write 
\begin{align}
\label{eq:S_small} 
S &\equiv |s| \mu^2 , \quad\text{where}\quad|s| \le 1 .  
\end{align} 
all other reasonable stellar objects have an even smaller spin--see e.g., 
Sec. II.B.1 of \cite{Hartl:2002ig}.  With this definition we can, by comparing 
the two terms in Eq.~\eqref{eqn:forced}, conclude that the (conservative) MPD 
and the adiabatic part of the $F^{(1)\alpha}_\text{dipole}$ will contribute 
to the orbital phase at subleading-order.

To summarize, the influence of each component of the force on the phase of the
waveform in the inspiral is 
\begin{align} 
\label{eq:phase_scaling} 
\Phi = &\underbrace{\kappa_{0} \; \epsilon^{-1}}_\text{adiabatic:
$F^{(1)\alpha}_{\text{ad}}$} +\underbrace{\kappa_{1/2} \;
\epsilon^{-1/2}}_\text{resonances (Kerr only)} 
\\ 
\nn 
&\q\q\q\q\q\q\q\q +
\underbrace{\kappa_{1} \; \epsilon^0}_{\substack{\text{post-1-adiabatic:} \\
\text{$F^{(1)\alpha}_{\text{osc}} + F^{(2)\alpha}_{\text{ad}}$} }} + \cdots ,
\q\q 
\end{align} 
where the $\kappa$ coefficients are dimensionless, of order
unity, and depend on the ingress and egress (or merger) frequencies in a
particular detector, but not on the mass ratio $\epsilon$. 

\section{Overview of our approach}
\label{sec:overview}

The MPD force is calculated by evaluating spin and curvature quantities at the
instantaneous position of the smaller body. In contrast, the self-force is a
functional of the smaller body's past worldline. To compute an inspiral in a
self-consistent manner one solves for the worldline using
Eq.~\eqref{eqn:forced} while simultaneously calculating the perturbation in the
gravitational field to generate the local self-force \cite{DienETC12}. In this
work at each instance along the worldline we approximate the true (inspiraling) 
past worldline of the small body with the geodesic that is tangent at that 
instance. The tangent geodesic is periodic allowing the self-force to be
computed efficiently in the frequency-domain. We use make of this approach to
compute the self-force in the Lorenz gauge
\cite{WarbETC12,AkcaWarbBara13,OsbuETC14,OsbuWarbEvan16}. This \emph{geodesic
self-force approximation} introduces a discrepancy with the true inspiral at
post-1-adiabatic order \cite{Pound:2015tma}, but preliminary evidence suggests
that the coefficient of this error term is small \cite{Warb13,Warb14a,Dien15}. 

Approximating the true self-force by the self-force calculated for motion along
geodesics tangent to the worldline naturally suggests evolving the inspirals
using a relativistic osculating elements prescription \cite{PounPois08b,
GairETC11}. These prescriptions recast of the equation of motion
Eq.~\eqref{eqn:forced} (making no small force assumption) and describe the
inspiral in terms of geometric quantities. The derivation in Schwarzschild
spacetime by Pound and Poisson \cite{PounPois08b} is restricted to motion in
the equatorial plane so in Sec.~\ref{sec:osculate} we extend it to generic
motion required when the spin of the secondary and the orbital angular momentum
are not aligned.

Second-order self-force results have not yet been computed.  From Eq.
\eqref{eq:phase_scaling} we see that neglecting $F^{(2)\a}_{\text{ad}}$
introduces error at post-1-adiabatic order. Once second-order results are known
they will be straightforward to incorporate into our scheme, but in this
paper only first-order self-force effects are included.  Similarly, there
are dissipative effects from the spin of the secondary that enter at
post-1-adiabatic order.  These have been calculated for circular orbits in
Schwarzschild \cite{Harms:2016ctx} and Kerr \cite{Harms:2015ixa,Han:2010tp}
spacetime, but to the best of our knowledge have not yet been calculated for
eccentric orbits. Again, once $F^{(1)\alpha}_\text{dipole}$ is known it can be
straightforwardly incorporated into our long term evolution scheme.  As we do 
not yet have calculations for these pieces of the force we will hereafter 
often adopt the notation 
\begin{align} F^\a_\text{self} &\equiv \mu^2
F^{(1)\alpha}_\text{mono},	
\\ 
F^\a_\text{spin} &\equiv \mu^2 s F^\a_\text{spin-curvature}.  
\end{align}

Finally, from Eq.~\eqref{eq:phase_scaling}, we note that the adiabatic
self-force enters at lower order than the other components of the force, and
accordingly must be computed with greater accuracy in order to affect the 
phase error at the same level.  To ensure the self-force is sufficiently 
accurate we use a hybrid scheme that computes the adiabatic component of the 
self-force with a highly accurate Regge-Wheeler code and computes the other 
components with a Lorenz-gauge code.  This scheme is detailed in previous 
papers \cite{OsbuETC14,OsbuWarbEvan16}.

\section{Osculating element description of motion} \label{sec:osculate}

In this section we recast the equations of motion \eqref{eqn:forced} into ones
for the evolution of the osculating elements of the inspiral. This procedure is
analogous to Lagrange's equations of planetary motion in Newtonian celestial
mechanics. Relativistic osculating element prescriptions of motion were first
given for Schwarzschild spacetime by Pound and Poisson \cite{PounPois08b}. In
that work they specialized to motion in the equatorial plane and here we
generalize their result to generic motion to allow for a description of the
inspiral when the spin of the secondary is not aligned or anti-aligned with the
orbital angular momentum. In deriving the osculating equations of motion we do
not make any small force approximation or other assumptions about the forcing
terms--those are independent assumptions that we will discuss in the next
section. In describing the motion of the secondary we will treat it as a
point particle, with discussion in the next section on the justification for 
this.  We will parameterize the particle's motion by its proper time $\tau$ and
distinguish the particle's time-dependent coordinate location from general 
spacetime coordinates with a subscript `$p$'.

The central idea to the osculating elements prescription arises from noting 
that at each point along the accelerated worldline, $z^\mu(\tau)$, there is a 
one-to-one relation between the particle's position and velocity and a 
tangent (or osculating) geodesic.  Each tangent geodesic has associated with 
it a set of orbital elements $I^A$ (such as energy, angular momentum, 
azimuthal angle at periastron, etc) that uniquely identifies the geodesic.  
Consequently the
worldline can be described either by a sequence of spacetime coordinates,
$z^\mu(\tau)$, or as a sequence of orbital elements of the osculating
geodesics, $I^A(\tau)$. With the four-velocity of the tangent geodesic given by
$u^\a_G(I^A,\tau) = \pa_\tau z^\a_G(I^A,\tau)$, we can write 
\begin{align}
z^\a(\tau) = z^\a_G(I^A,\tau), \q\q u^\a(\tau) = u^\a_G(I^A,\tau) , 
\end{align}
where hereafter a sub/superscript `$G$' denotes a quantity related to a
geodesic. The equations of motion take the following form in terms of the
osculating elements~\cite{PounPois08b} 
\begin{align} \label{eqn:Ievolve}
\frac{\pa z^\a_G}{\pa I^A} \frac{\pa I^A}{\pa\tau} = 0 , 
\q\q\q 
\mu \frac{\pa u^\a_G}{\pa I^A}\frac{\pa I^A}{\pa\tau} = F^\a .  
\end{align} 
Our explicit choice of osculating elements $I^A$ for bound motion and the 
resulting equations of motion are given in the following subsections.

\subsection{Bound geodesics in Schwarzschild spacetime} 
\label{sec:geo}

During the adiabatic stages of an inspiral (before the particle plunges into
the black hole) the geodesics tangent to the inspiral are bound and generically
eccentric. In this section we describe these tangent geodesics. We begin by
examining the case where the motion is confined to the equatorial plane. Later,
we will generalize to inclined geodesics (though ones that are still confined 
to some plane, as must occur in the Schwarzschild background).  We
denote equatorial geodesics by a set of functions $z_G^{\prime\a}(\tau)
=\left[t'_p(\tau),r'_p(\tau),\pi/2,\varphi'_p(\tau)\right]$, parameterized by
proper time $\tau$.  For later convenience we will omit the `prime' from
coordinates that are invariant under rotations, i.e., $t_p=t'_p$ and
$r_p=r'_p$. The geodesic four-velocity $u^{\prime\a}_G$ is given by
\begin{align} 
\label{eqn:fourvelocity} 
u^{\prime\a}_G = \l
\frac{{\mathcal{E}^G}}{f_p}, u^{r}_G, 0, \frac{{\mathcal{L}^G}}{r_p^2} \r,
\end{align} 
where $f_p\equiv 1-2M/r_p$, and ${\mathcal{E}^G}$ and
${\mathcal{L}^G}$ are the specific orbital energy and angular momentum,
respectively.  The constraint $u^{\a} u_{\a} = -1$ yields an expression for
$u^r_G$: 
\begin{align} 
\label{eqn:ur} 
\l u^{r}_G \r^2 =
(\mathcal{E}^G)^2-f_p\left(1+\frac{(\mathcal{L}^G)^2}{r_p^2}\right) .
\end{align} 
We parameterize the geodesic with the orbital eccentricity, $e$, and 
semi-latus rectum, $p$, which are related to the radial turning points
$r_{\rm min}$ and $r_{\rm max}$ via 
\begin{align} 
\label{eqn:defeandp} 
p = \frac{2 r_{\rm max} r_{\rm min}}{M (r_{\rm max} + r_{\rm min})} , \q \q e =
\frac{r_{\rm max} - r_{\rm min}}{r_{\rm max} + r_{\rm min}}.  
\end{align}
Eq.~\eqref{eqn:defeandp} and the roots of Eq.~\eqref{eqn:ur} give the
relationship between ($p$,$e$) and ($\mathcal{E}$,$\mathcal{L}$): 
\begin{align}
\label{eqn:specEL} 
\mathcal{E}^G &= \sqrt{\frac{(p-2)^2-4 e^2}{p(p-3-e^2)}},
\q\q\;\; 
\mathcal{L}^G = \frac{pM}{\sqrt{p-3-e^2}} .  
\end{align} 
Bound orbits exist with $e<1$ and $p>6+2e$.  For $e<1$ the line $p=6+2e$ is 
a separatrix between bound and plunging orbits \cite{CutlKennPois94}.

In self-force calculations it is convenient to reparameterize the orbital
motion (i.e., all the curve functions) with the relativistic anomaly $\chi$
\cite{Darw59}, defined so that 
\begin{align}
\label{eq:r_chi} 
r_p(\chi) &= \frac{p M}{1+e \cos\left[\chi-\chi_0\right]} .  
\end{align} 
The parameter $\chi_0$ specifies the value of $\chi$ at periastron passage.  

Eq.~\eqref{eq:r_chi} can be used with Eqs.~\eqref{eqn:fourvelocity} and
\eqref{eqn:specEL} to derive the following initial value equations for the
evolution of the orbit 
\begin{align} 
\label{eqn:tau} 
\frac{d\tau_p}{d\chi} &=
\frac{Mp^{3/2}}{(1+e \cos v)^2}\sqrt{\frac{p-3-e^2}{p-6-2 e \cos v}} ,
\\
\label{eqn:t} 
\frac{dt_p}{d\chi} &= \frac{r_p^2}{M(p-2-2 e\cos
v)}\sqrt{\frac{(p-2)^2-4e^2}{p-6-2 e\cos v}} , 
\\ 
\label{eqn:phi}
\frac{d\varphi'_p}{d\chi} &= \sqrt{\frac{p}{p-6-2 e\cos v}} , 
\end{align} 
where $v\equiv \chi-\chi_0$. It is useful to introduce initial values $T$ 
and $\Phi$ to Eqs.~\eqref{eqn:t} and \eqref{eqn:phi} respectively 
\begin{align} 
t_p(\chi) &= T + \int_{\chi_0}^{\chi-\chi_0} \frac{dt_p}{d\chi} dv , 
\\ 
\vp'_p(\chi) &= \Phi + \int_{\chi_0}^{\chi-\chi_0} \frac{d\vp'_p}{d\chi} dv .  
\end{align}
These integrals have analytic solutions in terms of special functions
\cite{FujiHiki09}, and we find the analytic solution for $\vp'_p$ to be useful
in this work 
\begin{align} 
\label{eqn:phiEllip} 
\vp'_p(\chi) &= \Phi +
2\sqrt{\frac{p}{p-6-2e}}\bar{F}\Big(\frac{v}{2}\Big|\frac{-4e}{p-6-2e}\Big),
\end{align} 
where $\bar{F}$ is the incomplete elliptic integral of the first kind 
\begin{align} 
\bar{F}(a|b) &\equiv \int_0^a \left( 1-b \sin^2{x} \right)^{-1/2} dx.  
\end{align} 
We also find use for the incomplete elliptic integral of the second kind 
$\bar{E}$ 
\begin{align} 
\bar{E}(a|b) &\equiv \int_0^a \left( 1-b \sin^2{x} \right)^{1/2} dx.  
\end{align}

In a coordinate system where the geodesic appears inclined, the worldline
is given by 
$z_G^{\a} =\left[t_p(\tau),r_p(\tau),\th_p(\tau),\varphi_p(\tau)\right]$.  
To transform from $z_G^{\prime\a}$ to $z_G^{\a}$ we will introduce a rotation 
matrix with Euler angles $\O$ and $\iota$ 
\begin{align} 
&\left[ \begin{array}{c} x_p \\ y_p
\\ z_p \end{array} \right] = \left[ \begin{array}{ccc} \cos{\O} &
-\sin{\O}\cos{\iota} & \sin{\O}\sin{\iota} \\ \sin{\O} & \cos{\O}\cos{\iota} &
-\cos{\O}\sin{\iota} \\ 0 & \sin{\iota} & \cos{\iota} \end{array} \right]
\left[ \begin{array}{c} x'_p \\ y'_p \\ z'_p \end{array} \right] .  
\end{align}
Here $\O$ is the longitude of the ascending node and $\iota$ is the orbital
inclination. Using $(x'_p,y'_p,z'_p)=(r_p \cos\vp'_p,r_p \sin\vp'_p,0)$ the
following equations are derived 
\begin{align} 
x_p &= r_p(\cos\O\cos\vp'_p-\sin\O\cos\iota\sin\vp'_p) , 
\\ 
y_p &= r_p(\sin\O\cos\vp'_p+\cos\O\cos\iota\sin\vp'_p) , 
\\ 
z_p &= r_p \sin\iota\sin\vp'_p , 
\\ 
\phi &\equiv \int_0^{\vp'_p}(\sec\iota\cos^2u+\cos\iota\sin^2u)^{-1}du , 
\\ 
\label{eqn:phip}
\vp_p &= \phi+\O, 
\\ 
\label{eqn:thetap} 
\th_p &= \cos^{-1}(\sin\iota\sin\vp'_p) .  
\end{align} 
The $\th$ and $\vp$ components of $u^\a_G$ are given by
\begin{align} 
u^\th_G &= -\frac{(1+e\cos{v})^2\sin{\iota}\,\cos{\vp'_p}}{pM 
\sqrt{(p-3-e^2)(1-\sin^2{\iota}\,\sin^2{\vp'_p})}},
\\ 
u^\vp_G &=  \frac{(1+e \cos{v})^2(p-3-e^2)^{-1/2}}{pM(\sec\iota\cos^2\vp'_p+ 
\cos\iota\sin^2\vp'_p)}.
\end{align} 
The $u^t_G$ and $u^r_G$ four-velocity components are unaffected by the rotation.

\subsection{Evolution of the orbital elements}
\label{sec:evolution_of_osc_els}

The complete set of orbital elements we choose to describe the tangent
geodesics are given by 
\begin{align} 
I^A &= \{e,p,\chi_0,\iota,\O,\Phi,T\} .
\end{align} 
Any fixed set of these elements uniquely specifies a bound,
eccentric, possibly inclined, geodesic in the spacetime. To describe the
inspiraling worldline we promote these elements to have time-dependence. The
rate-of-change of $I^A$ is described by Eq.~\eqref{eqn:Ievolve}. In order to
explicitly give the evolution equations, let us define the operator
\begin{align} 
\mathscr{D} \equiv \frac{de}{d\chi} \frac{\pa}{\pa e}  +
\frac{dp}{d\chi} \frac{\pa}{\pa p} + \frac{d\chi_0}{d\chi}
\frac{\pa}{\pa\chi_0}, 
\end{align} 
Parameterizing the elements using $\chi$ we can then write 
\begin{align} &0 = \mathscr{D}[t_p] + \frac{\pa t_p}{\pa T}
\frac{dT}{d\chi} , 
\label{eqn:tOsc} 
\\ 
&0 = \mathscr{D}[r_p], 
\label{eqn:rOsc}
\\ 
&0 = \mathscr{D}[\theta_p] + \frac{\pa \th_p}{\pa\iota} \frac{d\iota}{d\chi}
+ \frac{\pa \th_p}{\pa\Phi} \frac{d\Phi}{d\chi}, 
\label{eqn:thOsc} 
\\ 
&0 = \mathscr{D}[\varphi_p] + \frac{\pa \vp_p}{\pa\iota} \frac{d\iota}{d\chi} +
\frac{\pa \vp_p}{\pa\O} \frac{d\O}{d\chi} + \frac{\pa \vp_p}{\pa\Phi}
\frac{d\Phi}{d\chi}, 
\label{eqn:phOsc} 
\\ 
&\frac{1}{\mu}\frac{d\tau_p}{d\chi}
F^\alpha = \mathscr{D}[u^\alpha_G] \qquad \text{where}\quad \alpha=\{t,r\},
\label{eqn:utOsc} 
\\ 
&\frac{1}{\mu}\frac{d\tau_p}{d\chi} F^\beta =
\mathscr{D}[u^\beta_G] + \frac{\pa u^\beta_G}{\pa\iota} \frac{d\iota}{d\chi} +
\frac{\pa u^\beta_G}{\pa \Phi} \frac{d\Phi}{d\chi} 
\label{eqn:uthOsc}  
\\
&\qquad\qquad\qquad\qquad\qquad\text{where} \quad \beta =\{\th,\vp\},
\nonumber
\end{align} 
Eqs.~\eqref{eqn:rOsc} and \eqref{eqn:utOsc} form a closed system
that describes the evolution of the elements $\alpha=\{e,p,\chi_0\}$
\begin{align}
\label{eq:d_dchi} 
\frac{d\alpha}{d\chi} &= c^{(t)}_\alpha F^t + c^{(r)}_\alpha F^r , 
\end{align} 
The $c$-coefficients depend on $e$, $p$, and
$v$ and, as they are a little unwieldy, are given explicitly in
Appendix~\ref{apdx:osculating}.

To take advantage of our hybrid self-force
technique~\cite{OsbuETC14,OsbuWarbEvan16} we eliminate $F^r$ from the 
$\alpha = \{p,e\}$ versions of Eq.~\eqref{eq:d_dchi} 
using the orthogonality condition $u_\mu F^\mu=0$ 
\begin{align} 
F^r &= \frac{f_p \left(f_p u^t F^t-r_p^2 u^\th
F^\th-r_p^2 \sin^2{\th_p}\,u^\vp F^\vp\right)}{u^r} .  
\end{align} 
This form
appears inconvenient because $u^r$ vanishes at the radial turning points.
However, the presence of a $\sin{v}$ factor in Eqs.~\eqref{eqn:cre} and
\eqref{eqn:crp} allows the resulting expressions to be simplified in such a way
that the denominator does not vanish. Eq.~\eqref{eqn:tOsc} describes the
rate-of-change of $T$. However, instead of calculating $T$ we dynamically
evolve $t_p$ using Eq.~\eqref{eqn:t}. Eq.~\eqref{eqn:thOsc} and
\eqref{eqn:uthOsc} describe the evolution of the elements
$\alpha=\{\iota,\Phi\}$ 
\begin{align}
\label{eq:osc_els_i} 
\frac{d\a}{d\chi} &=
c^{(\th)}_\alpha F^\th + c^{(e)}_\alpha\frac{de}{d\chi} +
c^{(p)}_\alpha\frac{dp}{d\chi} + c^{(\chi_0)}_\alpha\frac{d\chi_0}{d\chi} ,
\end{align} 
Similarly, Eq.~\eqref{eqn:phOsc} describes the evolution of $\O$
\begin{align}
\label{eq:osc_els_O} 
\frac{d\O}{d\chi} &= c^{(e)}_\O
\frac{de}{d\chi} +c^{(p)}_\O \frac{dp}{d\chi} +c^{(\chi_0)}_\O
\frac{d\chi_0}{d\chi} +c^{(\iota)}_\O \frac{d\iota}{d\chi} +c^{(\Phi)}_\O
\frac{d\Phi}{d\chi} .  
\end{align} 
The $c$-coefficients in these two equations depend on $e$, $p$, $v$, $\iota$, 
and $\vp'_p$ and their explicit form can be
found in Appendix~\ref{apdx:osculating}. Note that in the $c^{(\th)}_\Phi$
coefficient given in Eq.~\eqref{eqn:cthPh} there is a $1/\sin{\iota}$ factor
that diverges when $\iota=0$. This is a familiar property of osculating orbits
in Newtonian celestial mechanics. To circumvent this issue one can change to a
different set of orbital elements or, as we do in this work, leverage the
spherical symmetry of the problem. Without loss of generality, we choose the
initial condition $\iota_0=\pi/4$ which avoids the worldline ever crossing
$\iota=\{0,\pi/2\}$ as under our small mass-ratio approximation $\iota$ will
oscillate with a small amplitude.

\section{Forcing terms}
\label{sec:forcing_terms}

The forcing terms that drive the inspiral are on the righthand side of
Eq.~\eqref{eqn:forced}. In this work we do not include the second-order or
spin-dipole forcing terms (as they have yet to be calculated) so the force,
$F^\mu$, has two parts 
\begin{align} F^\alpha =
\mu^2\left(F^{(1)\alpha}_\text{mono} + s F^\alpha_\text{spin-curvature}\right)
= F^\alpha_\text{self} + F^\alpha_\text{spin} .  
\end{align} 
The first term arises from the interaction of the secondary with its own
(first-order in the
mass-ratio) metric perturbation and is known as the (first-order) self-force.
The second term arises from the interaction of the spin of the secondary with
the background spacetime of the primary. This force is known as the
Mathisson-Papapetrou-Dixon (MPD) spin-curvature force. We discuss how these
two forcing terms are calculated in the following subsections.

\subsection{Self-force}

In the small mass ratio limit the spacetime metric of the binary can be written
as $g_{\mu\nu} + h_{\mu\nu}$ where $g_{\mu\nu}$ is the background metric of the
primary and $h_{\mu\nu}$ is a first-order-in-the-mass-ratio perturbation
arising from the secondary (we do not include higher-order in the mass ratio
corrections in this work). Within this description the secondary interacts with
its own metric perturbation. This self-interaction gives rise to a self-force
that acts to drive the inspiral. Being sourced by the metric perturbation
backscattering off the background curvature of the primary, this
self-interaction is non-local so that the self-force at any instance is a
functional of the the entire past (inspiraling) history of the secondary. This
makes the self-force particularly challenging to calculate.

One way to make the calculation more amenable is to restrict the secondary's
motion to a geodesic of the background spacetime. The periodic nature of the
geodesic then allows efficient frequency-domain techniques to be employed
\cite{DetwPois04,BaraLous05,Bern07,BaraSago10,HoppEvan10,Akca11,WarbETC12,
AkcaWarbBara13,HoppEvan13,OsbuETC14,HoppETC15,OsbuWarbEvan16}
and we make use of this approach in this work. Approximating the true
self-force at each instance (a functional of the inspiraling worldline) with
the geodesic self-force introduces a discrepancy with the true inspiral at
post-1-adiabatic order \cite{Pound:2015tma}, but preliminary evidence suggests
that the coefficient of this error term is very small
\cite{Warb13,Warb14a,Dien15}. When the spin of the secondary is not aligned
with the orbital angular momentum the orbit will precess about the initial 
equatorial plane. The effect of this precession on the self-force is captured 
by the $F^{(1)\alpha}_\text{dipole}$ term that we are not currently including 
in our model (but once it has been calculated it is straightforward to
incorporate).

We now briefly review our calculation of the self-force. We model the secondary
as a point particle to give an unambiguous result that does not depend on the
higher multipole structure of the body (though we will add dipole structure to
account for the body's spin in the next subsection). This point particle model
necessitates a regularization procedure to handle the divergence in the metric
perturbation at the particle's location \cite{MinoSasaTana97,QuinWald97,
PoisPounVega11}. Practically, we make use of the mode-sum scheme
\cite{BaraOri00,Bara01,BaraETC02,BaraOri03,DetwMessWhit03,HeffOtteWard12a,
HeffOtteWard12b}.  The regularization procedure was, until recently 
\cite{PounMerlBara13}, only understood in the Lorenz gauge. In this gauge the 
trace-reversed metric perturbation, $\bar{h}_{\mu\nu}\equiv
h_{\mu\nu}-\frac{1}{2}g_{\mu\nu}(h_{\a\b}g^{\a\b})$, is governed by the
linearized Einstein equations subject to the constraint
$\nabla^\mu\bar{h}_{\mu\nu}=0$ 
\begin{align} 
\Box \bar{h}_{\mu\nu} + 2
R^{\a\;\,\b}_{\;\,\mu\;\,\nu} \bar{h}_{\a\b} = -16 \pi T_{\mu\nu} , 
\end{align}
where $\Box\equiv g^{\a\b}\nabla_\a\nabla_\b$, $R^{\a\;\,\b}_{\;\,\mu\;\,\nu}$
is the Riemann tensor of the background spacetime, and $T_{\mu\nu}$ is the
stress energy tensor of a point mass following a geodesic. A number of authors
have considered Lorenz-gauge metric perturbations in Schwarzschild spacetime
\cite{DetwPois04,BaraLous05,Bern07,BaraSago07,BaraSago10,Akca11,WarbETC12,
AkcaWarbBara13,DolaBara13,HoppEvan13,OsbuETC14,HoppETC15} and Kerr spacetime 
\cite{Isoyama:2014mja}. This work utilizes the frequency domain code 
presented in Ref.~\cite{OsbuETC14} with refinements described in 
Refs.~\cite{HoppETC15,OsbuWarbEvan16}.

The adiabatic part of the self-force, which contributes at leading-order in
Eq.~\eqref{eq:phase_scaling}, must be calculated to within an error tolerance
smaller than the mass ratio whereas the terms contributing at post-1-adiabatic
order require only a few digits of accuracy. We use a hybrid scheme
\cite{OsbuETC14,OsbuWarbEvan16} that has been developed to achieve these
tolerances. This hybrid scheme uses a highly accurate Regge-Wheeler-Zerilli
(RWZ) \cite{ReggWhee57,Zeri70} code (based on
Refs.~\cite{HoppEvan10,HoppETC15}) to generate $F^\mu_{\text{ad}}$ while
post-processing the Lorenz gauge self-force (and spin-curvature force) results
to calculate $F^\mu_{\text{osc}}$. This scheme relaxes the requirements on our
Lorenz-gauge self-force code (which takes substantially longer to run than the
RWZ code at high precision) while maintaining target phase accuracies in the
resulting inspiral.

Our hybrid code outputs the Fourier coefficients of the self force for a given
orbital configuration.  We calculate these Fourier coefficients for tens of
thousands of orbital configurations and perform an interpolation over the
relevant parameter space. We minimize interpolation error by performing
multiple local interpolations, modifying our discretization in regions where
the self-force varies more rapidly e.g., near the separatrix. For full details
see Ref.~\cite{OsbuWarbEvan16}. This interpolant can be rapidly evaluated for
any orbit configuration (including those not in the original dataset) and we
couple the interpolant with the osculating element equations of motion to
compute inspirals.

Our hybrid code computes the self-force in a coordinate system where the
instantaneous orbital motion is equatorial (the `prime' coordinate system). In
this coordinate system $F^{\prime\th}_\text{self}$ vanishes. To transform to
the inclined frame we apply a rotation giving 
\begin{align} 
F^t_\text{self} &= F^{\prime t}_\text{self}, \q\q\q\q F^r_\text{self} 
= F^{\prime r}_\text{self},
\\ 
F^\th_\text{self} &= -\frac{4\,F^{\prime \vp}_\text{self}\, \sin{\iota} \, 
\cos{\vp'_p}\,\sqrt{1-\sin^2\iota\,\sin^2\vp'_p}}
{3+\cos{(2\iota)}+2\cos{(2\vp'_p)}\,\sin^2\iota},
\\ 
F^\vp_\text{self} &= \frac{4\,F^{\prime\vp}_\text{self} \, 
\cos{\iota}}{3+\cos{(2\iota)}+2\cos{(2\vp'_p)}\,\sin^2{\iota}}.
\end{align}

\subsection{Spin-curvature force}
\label{sec:spin_curvature}

When multipole-moments beyond the monopole are endowed to the orbiting particle
there is an interaction between them and the Riemann tensor of the background
spacetime. Ignoring the self-force and multipole-moments beyond the spin of
the body (the `pole-dipole' approximation) gives rise to the
Mathisson-Papapetrou-Dixon equations of motion
\cite{Mathisson2010,Papa51,Dixo70} 
\begin{align} 
u^\beta\nabla_\beta p^\alpha 
&= -\frac{1}{2} R^\alpha_{\;\,\nu\la\s}u^\nu S^{\la\s}
\label{eq:MPD-p}		
\\ 
u^\beta\nabla_\beta S^{\mu\nu} 	&= p^\mu u^\nu - p^\nu u^\mu .
\label{eq:MPD-S} 
\end{align} 
where $p^\alpha$ is the particle's four-momentum
and $S^{\a\b}$ is the spin tensor of the orbiting body. Generically, the
four-velocity and the four-momentum are not aligned, i.e., $p^\alpha \neq \mu
u^\alpha$.

The systems of equations \eqref{eq:MPD-p} and \eqref{eq:MPD-S} do not form a
closed system and an additional spin-supplementary condition (SSC) must be
imposed. The choice of SSC is not unique owing to the fact that in relativity
the center of mass of a spinning body is an observer dependent point. The
choice of SSC is fundamentally arbitrary but might be motivated by simplifying
the equations of motion for a particular setup. Careful comparison shows that
each SSC leads to the same physics
\cite{Costa:2014nta,2007MNRAS.382.1922K,Lukes-Gerakopoulos:2014dma}. In this
work we adopt the Pirani condition \cite{Pira56} 
\begin{align} u_\mu S^{\mu\nu} = 0. 
\label{eqn:SSC} 
\end{align} 
It is convenient to introduce the spin vector, $S^\mu$, from which the spin 
tensor can be constructed 
\begin{align}
\label{eqn:spin_vector} 
S^{\mu\nu}=\e^{\mu\nu\a\b}u_\a S_\b , 
\end{align} 
where $\e^{\mu\nu\a\b}$ is the Levi-Civita tensor. The magnitude of the spin 
vector $S^2\equiv S^\alpha S_\alpha$ is a constant \cite{RuanVigeHugh15}.

The set of equations \eqref{eq:MPD-p}, \eqref{eq:MPD-S} and \eqref{eqn:SSC} can
be integrated to compute the wordline of a spinning body. For each Killing
vector $\xi^\a_{(j)}$ of the spacetime the quantity \cite{RuanVigeHugh15}
\begin{align} 
\mathcal{C}^S = p_\a\xi^\a_{(j)}-\frac{1}{2\mu} S^{\a \b}
\nabla_\b \xi^{(j)}_\a .  
\label{eq:cons_spin} 
\end{align} 
is conserved along
the body's worldline. The first term of Eq.~\eqref{eq:cons_spin} is conserved
for a geodesic (e.g. $\mu\, \mathcal{E}^G=p_\a \xi^\a_{(t)} $) while the second
term accounts for the MPD force. Here a superscript $S$ denotes a conserved
quantity related to a spinning body (recall that a superscript $G$ denotes a
quantity conserved for a geodesic). The four Killing vectors of Schwarzschild
spacetime admit four conserved quantities in accordance with
Eq.~\eqref{eq:cons_spin}. We demonstrate conservation of two of these
quantities as a consistency check in Section~\ref{sec:check} by disabling 
(for testing purposes) the self-force in our numerical scheme. 

As we are working in the small mass-ratio limit $\e\ll1$ Eq.~\eqref{eq:S_small}
tells us that the spin magnitude is also small. We thus linearize our
calculation in $S$. This results in substantially simpler equations of motion,
though it is an open question whether the non-linear (in $S$) terms might lead
to interesting resonance phenomenon that the linear approximation does not
capture \cite{RuanVigeHugh15}. To linear order in $S$ the four-velocity and
angular-momentum are parallel 
\begin{align} 
u^\alpha = p^\alpha/\mu + \mathcal{O}(S^2) .
\end{align} 
The equations of motion \eqref{eq:MPD-p}-\eqref{eq:MPD-S} reduce to 
\begin{align} 
u^\beta\nabla_\beta
u^\alpha 	&= -\frac{1}{2} R^\alpha_{\;\,\nu\la\s}u^\nu S^{\la\s}
\label{eq:MPD-p_linear_S}		
\\ 
u^\beta\nabla_\beta S^{\mu\nu}
&= 0 .
\label{eq:MPD-S_linear_S} 
\end{align} 
Comparing Eq.~\eqref{eqn:forced} and
Eq.~\eqref{eq:MPD-p_linear_S} we identify the spin-curvature force as
\begin{align}
\label{eq:Fspin} 
F^\alpha_\text{spin} = -\frac{1}{2} R^\alpha_{\;\,\nu\la\s}u^\nu S^{\la\s} 
\end{align} 
Equation \eqref{eq:MPD-S_linear_S} tells us that, to linear order in $S$, the 
spin vector is parallel transported along the worldline. In Schwarzschild 
spacetime this gives us explicitly 
\begin{align} &\frac{dS^\a}{d\tau} =
-\Gamma^\a_{\mu\nu}S^\mu u^\nu, \\ &\frac{dS^t}{d\chi} =
-\frac{d\tau_p}{d\chi}\frac{M}{f_p r_p^2}(S^t u^r+S^r u^t) , 
\\
&\frac{dS^r}{d\chi} = \frac{d\tau_p}{d\chi}\bigg(\frac{M}{f_p r_p^2}(S^r
u^r-f_p^2 S^t u^t) 
\notag 
\\ 
&\q\q\q\q\q\q\; +f_p r_p(S^\th u^\th+S^\vp u^\vp \sin^2{\th_p}) \bigg), 
\\ 
&\frac{dS^\th}{d\chi} = \frac{d\tau_p}{d\chi}\left(
S^\vp u^\vp \cos{\th_p}\,\sin{\th_p}-\frac{S^\th u^r+S^r u^\th}{r_p} \right) ,
\\ 
&\frac{dS^\vp}{d\chi} = -\frac{d\tau_p}{d\chi}\left(
\frac{\cos{\th_p}}{\sin{\th_p}}(S^\vp u^\th+S^\th u^\vp) + \frac{S^r
u^\vp+S^\vp u^r}{r_p} \right) .  
\end{align} 
To couple these spin evolution
equations to the osculating elements equations we rewrite then as forcing
terms. Eqs.~\eqref{eqn:SSC} and \eqref{eqn:spin_vector} imply that the spin
vector is purely spatial in the rest frame of the small body 
\begin{align}
u_\mu S^\mu = 0.  
\end{align} 
We use this condition to determine $S^t$
\begin{align} S^t &= \frac{S^r u^r+f_p r_p^2(S^\th u^\th+S^\vp u^\vp
\sin^2{\th_p})}{f_p^2 u^t} .  
\end{align} 
This leaves three degrees of freedom for choosing the initial conditions of 
$S^\mu$.  Using Eq.~\eqref{eq:Fspin} and \eqref{eqn:spin_vector} to express 
the components of $F^\mu_\text{spin}$ in terms of the components of the spin 
vector gives 
\begin{align} 
F^t_\text{spin}
&= \frac{3Mu^r\sin{\th_p}(S^\vp u^\th-S^\th u^\vp)}{r_pf_p},
\label{eq:Ft_spin} 
\\ 
F^r_\text{spin} &= \frac{3Mf_p u^t\sin{\th_p}(S^\vp
u^\th-S^\th u^\vp)}{r_p},		
\label{eq:Fr_spin} 
\\ 
F^\th_\text{spin} &= \frac{3M u^\vp\sin{\th_p}(S^t u^r-S^r u^t)}{r_p^3},
\label{eq:Ftheta_spin} 
\\ 
F^\vp_\text{spin} &= -\frac{3M u^\th(S^t u^r-S^r u^t)}{r_p^3\sin{\th_p}}.
\label{eq:Fphi_spin}
\end{align}

\section{Waveforms}
\label{sec:waveforms}

Coupling the force terms in the previous section with the osculating element
equations of motion in Sec.~\ref{sec:osculate} allows for the computation of
inspiral trajectories. With a trajectory in hand there are a number of ways to
compute the waveform. 

The most accurate method is to use the trajectory as a source in a time-domain
code. Via extrapolation or via hyperboloidal compactification, the waveform can
be extracted at null infinity \cite{Zenginoglu:2011zz, DienETC12,
Sundararajan:2008zm}. The high computational cost of these time-domain
simulations means they cannot provide a wide survey of waveforms across the
parameter space but they are invaluable to assess the accuracy of other
waveform generation methods.

One of the most common methods for computing waveforms is to use the weak-field
quadrupole formula (sometimes supplemented by octupolar or fast-motion
corrections). This technique is taken by so-called `kludge' methods
\cite{Barack:2003fp,Babak:2006uv,Chua:2015mua} that combine input from a number
of different techniques (e.g., post-Newtonian and black hole perturbation
theory) to model the trajectory. These techniques allow waveforms to be
computed rapidly but they may not faithfully represent the true waveform in
the very strong-field \cite{Babak:2006uv}.

In this work we take a different approach inspired by our geodesic self-force
interpolation model. Using our frequency-domain codes we compute so-called
`snapshot' waveforms \cite{DrasHugh06,FujiHikiTago09} for a large number of
parameters in the $(p,e)$ space. We then interpolate between these snapshot
waveforms to create a continuously evolving waveform over the entire inspiral.

\subsection{Frequency-domain waveforms}
\label{sec:freq_waveforms}

The radiation from the binary can be extracted from the (complex) Weyl
curvature scalar 
\begin{align} \Psi_4 =
-C_{\alpha\beta\gamma\delta}n^\alpha\bar{m}^\beta n^\gamma \bar{m}^\delta
\end{align} 
where $C_{\a\b\g\d}$ is the Weyl curvature tensor (equal to the
Riemann tensor in vacuum), and $n^\a$ and $\bar{m}^\a$ are components of the
Kinnersly tetrad \cite{Teuk73}. Far from the source $\Psi_4$ can be related to
the gravitational radiation via 
\begin{align} 
\label{eqn:waveDeriv}
\Psi_4(r\rightarrow \infty) &\simeq \frac{1}{2}\left(\ddot{h}_+ - i
\ddot{h}_\times \right) .  
\end{align} 
where $h_+$ and $h_\times$ are the two
independent polarizations of the gravitational waves and an overdot denotes
differentiation with respect to coordinate time.

The scalar $\Psi_4$ can be decomposed into spin-weighted spherical harmonics
$\,_{-2}Y_{lm}$ as 
\begin{align} 
\Psi_4(t,r,\th,\vp) &=
\sum_{l=2}^{l_\text{max}} \sum_{m=-l}^l 
\psi_4^{lm}(t,r)\,_{-2}Y_{lm}(\th,\vp) .  
\end{align} 
The function $\psi_4$ can be obtained directly using the Teukolsky formalism 
\cite{Teuk73}.  As our hybrid self-force scheme employs highly accurate 
results in the Regge-Wheeler gauge we already have precomputed data for the 
asymptotic amplitudes of the RWZ fields.  These amplitudes can be related to 
$\psi_4$ via 
\begin{align}
\psi^{lm}_4(r\rightarrow \infty) &\simeq \sum_{n=n_\text{min}}^{n_\text{max}}
\frac{\o_{mn}^2}{4r} \sqrt{(l+2)(l+1)l(l-1)} 
\notag 
\\
&\q\q\; 
\times \left(i C^{\text{odd}}_{lmn} -C^{\text{even}}_{lmn} \right) 
e^{-i\o_{mn}(t-r)} , 
\end{align} 
where $C^{\text{even}}_{lmn}$ is the coefficient of the
Zerilli-Moncrief master function \cite{Zeri70,Monc74}, and
$C^{\text{odd}}_{lmn}$ is the coefficient of the Cunningham-Price-Moncrief
master function \cite{CunnPricMonc78,CunnPricMonc79} (according to the
conventions of \cite{HoppEvan10}) and $\omega_{mn} = m\Omega_\varphi +
n\Omega_r$ is the mode frequency. The waveform is calculated by integrating
Eq.~\eqref{eqn:waveDeriv} twice with respect to time, giving 
\begin{align} &h_+
- i h_\times = \frac{1}{r}\sum_{l=2}^{l_\text{max}} \sum_{m=-l}^l H_{lm}(t,r)
\,_{-2}Y_{lm}(\th,\vp) , \\ &H_{lm}(t,r) \equiv
\sum_{n=n_\text{min}}^{n_\text{max}} \frac{1}{2}\sqrt{(l+2)(l+1)l(l-1)} 
\notag
\\
&\q\q\q\q\q\q\q\; \times \left(C^{\text{even}}_{lmn}-i C^{\text{odd}}_{lmn}
\right) e^{-i\o_{mn}(t-r)} , 
\label{eqn:hlm} 
\end{align}

In order to evaluate $C^{\text{even/odd}}_{lmn}$ for arbitrary values of
$(p,e)$ we interpolate over the parameter space using the same scheme that we
used for the self-force \cite{OsbuWarbEvan16}. In total we computed $11,234$
orbital configurations in the range $p<57$ and $e<0.82$.  For purposes of 
reconstructing waveforms, the $C$-coefficients are interpolated for 
a range of indices: $n_{\text{min}} = -40$ through $n_{\text{max}} = +40$,
and for every $m$ for $l=2$ and $l=3$. 

Using this prescription the waveform for a given $(p,e)$, corresponds to a
geodesic with periastron passage at $t=\varphi=0$. To compute the waveform
associated with an inspiral we update the snapshot parameters $(p,e,t,\vp)$ at
each periastron passage to maintain consistency between the inspiraling orbital
motion and the waveform's amplitude and phase. The values of $(p,e)$ jump
discontinuously at periastron passages under this method. This jump has a
minimal effect on our inspiral waveform, as the dephasing between the
geodesic waveform and the true inspiral waveform takes places over the
radiation reaction timescale which is much longer than the orbital timescale.
These discontinuous changes will thus be negligible for small mass-ratio 
binaries while the inspiral is evolving adiabatically (as it does away from 
the separatrix). As
the waveform and inspiral parameters are synchronized at each periastron
passage our inspiral waveform should be a good representation of the true
waveform throughout the entire adiabatic inspiral.

\section{Spin-aligned inspirals (planar motion)}
\label{sec:aligned}

For orbits where the spin and orbital angular momentum are aligned the inspiral
motion is confined to a plane. In this scenario the osculating elements
equations simplify greatly ($d\iota/d\chi=d\Omega/d\chi=0$). When $\iota$ and
$\Omega$ are constant the troublesome $1/\sin{\iota}$ terms can be avoided by
calculating $\vp'_p$ dynamically using Eq.~\eqref{eqn:phi} instead of evolving
$\Phi$. Under these simplifications our osculating elements scheme is applied
to equatorial inspirals by enforcing the condition $\iota=0$ or, equivalently,
$\th_p=\pi/2$. In this case the only non-zero component of the spin vector,
$S^\alpha$, is 
\begin{align} 
S^\theta = \frac{s\mu^2}{r_p} .
\label{eq:Stheta_spin_aligned} 
\end{align} 
Consequently, from Eqs.~\eqref{eq:Ft_spin}-\eqref{eq:Fphi_spin}, 
only $F^t_\text{spin}$ and $F^r_\text{spin}$ are non-zero 
\begin{align} 
F^t_\text{spin} &= -\frac{3s\mu^2Mu^r u^\vp}{r_p^2 f_p} 
\\ 
F^r_\text{spin} &= -\frac{3s\mu^2Mf_p u^t u^\vp}{r_p^2} .  
\end{align} 
We now present some sample inspirals and waveforms for the spin-aligned case.

\subsection{Sample results} 
\label{sec:AlignedResults}

\begin{figure} \includegraphics[width=8.4cm]{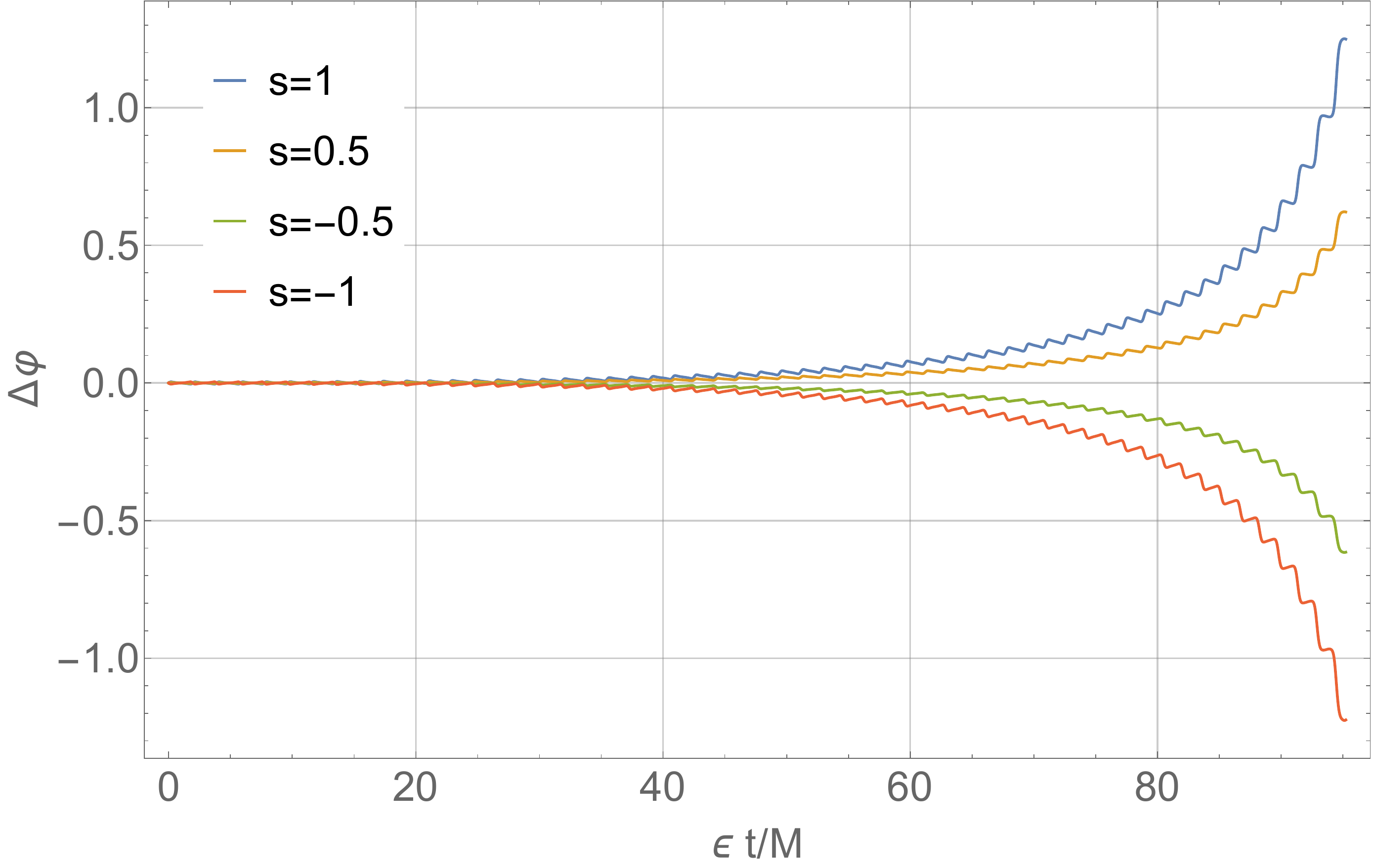}
\caption{Dephasing of inspirals with a spinning secondary with respect to a
non-spinning inspiral. The dephasing is defined as $\Delta\varphi \equiv
\varphi_p^{s\neq0} -\varphi_p^{s=0}$. The initial parameters for the
non-spinning inspiral are $(p_0,e_0)=(10,0.4)$ and the initial frequencies of
the spinning inspirals are matched to the initial frequencies of the
non-spinning case.  For these calculations we choose a mass ratio of 
$\mu/M=5\times10^{-3}$, which is large enough that the eccentric oscillations 
of the inspirals are visible in the figure.  Reading from top to bottom the 
spin values are $s=\{+1,+0.5,-0.5,-1\}$.}
\label{fig:Delta_phi_spin_aligned} 
\end{figure}

\begin{figure}
\includegraphics[width=8.4cm]{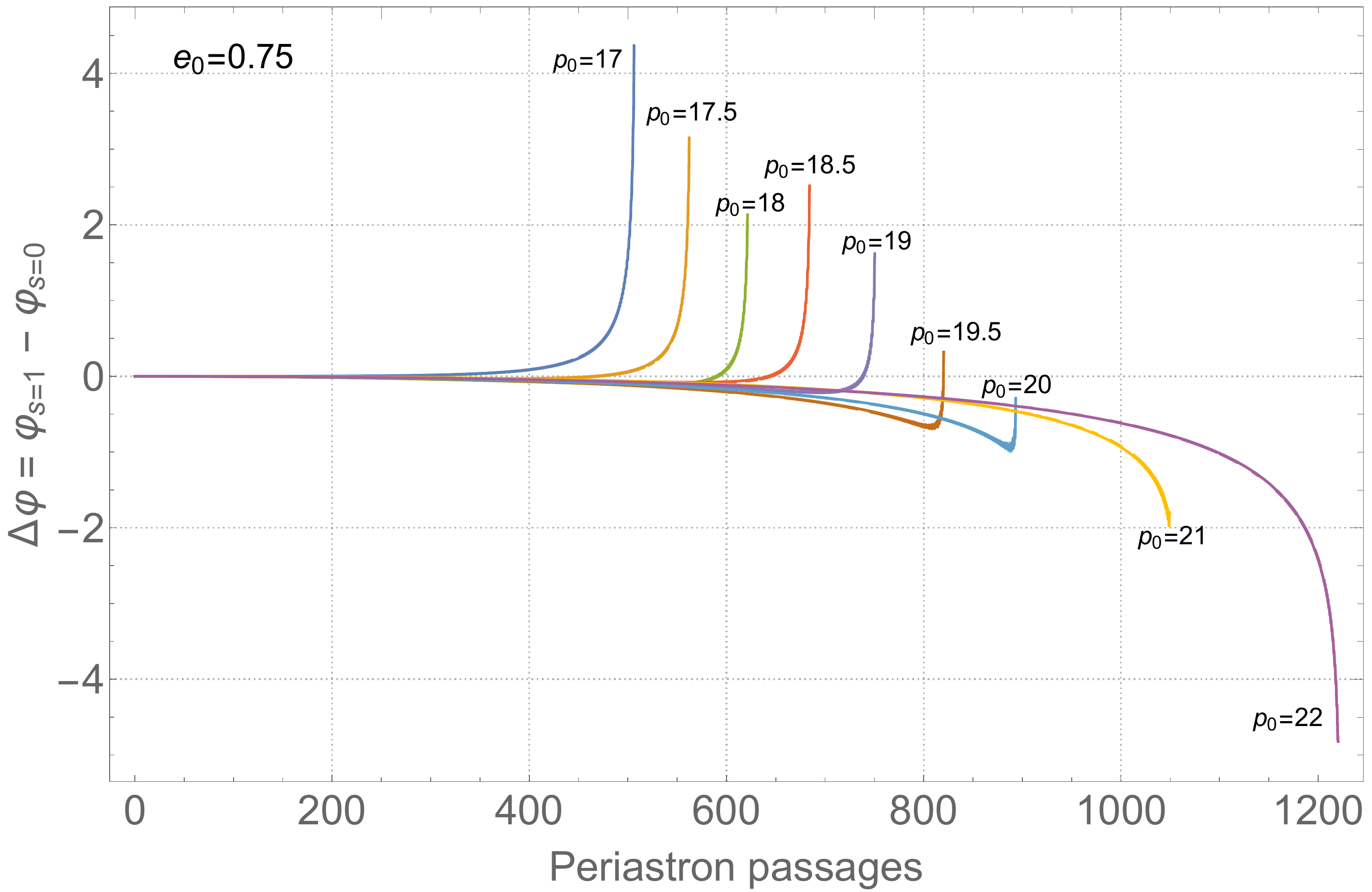}
\caption{Dephasing of a spin-aligned binary with respect to a
(initially frequency matched) non-spinning binary. All inspirals initially have
eccentricity $e_0=0.75$ and we show $\Delta\varphi = \varphi_{s=1} -
\varphi_{s=0}$ for a variety of inspirals with differing initial $p_0$ values.
For inspirals with a strong-field $p_0 \lesssim 19$ the accumulation of
$\Delta\varphi$ is positive and monotonic (modulo oscillations on the orbital
timescale).  For $p_0 \gtrsim 21$ the accumulation of $\Delta\varphi$ is
negative and monotonic. Between these two regimes the accumulation of
$\Delta\varphi$ is not monotonic and the spinning inspiral can initially
lose phase with respect to the non-spinning case before catching up again,
even returning to being in phase ($\Delta\varphi=0$) as plunge is approached.
The opposite behavior is observed for spin anti-aligned ($s=-1$) binaries. The
inspirals in this figure were computed with mass ratio $\epsilon =
5\times10^{-3}$.}
\label{fig:Delta_phi_sign_change}
\end{figure}

\begin{figure} 
\includegraphics[width=8.4cm]{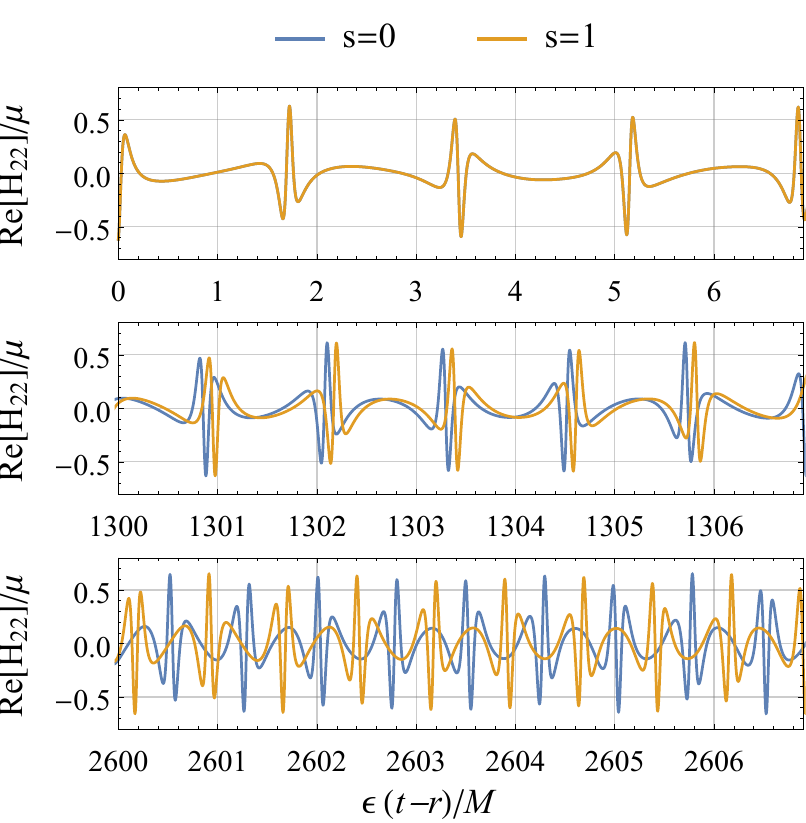}
\caption{Sample waveforms for inspirals with a spinning and non-spinning
secondary with $\e = 10^{-3}$. The initial orbital parameters are
$(p_0,e_0)=(20,0.7)$. We match the initial frequencies of the spinning and
non-spinning configuration so that the two waveforms are initially in phase 
(top panel).  After $\sim 140$ radial oscillations the dephasing of the 
waveforms becomes noticeable (middle panel).  After $\sim 360$ radial 
oscillations the waveforms have dephased by a half cycle (bottom panel).  
Close to the plunge ($\sim 700$ radial oscillations, 
$\epsilon(t-r)/M\simeq 3700$) the waveforms have dephased by 2 complete 
cycles.  
\label{fig:e7_waves}} 
\end{figure}

\begin{figure} 
\includegraphics[width=8.4cm]{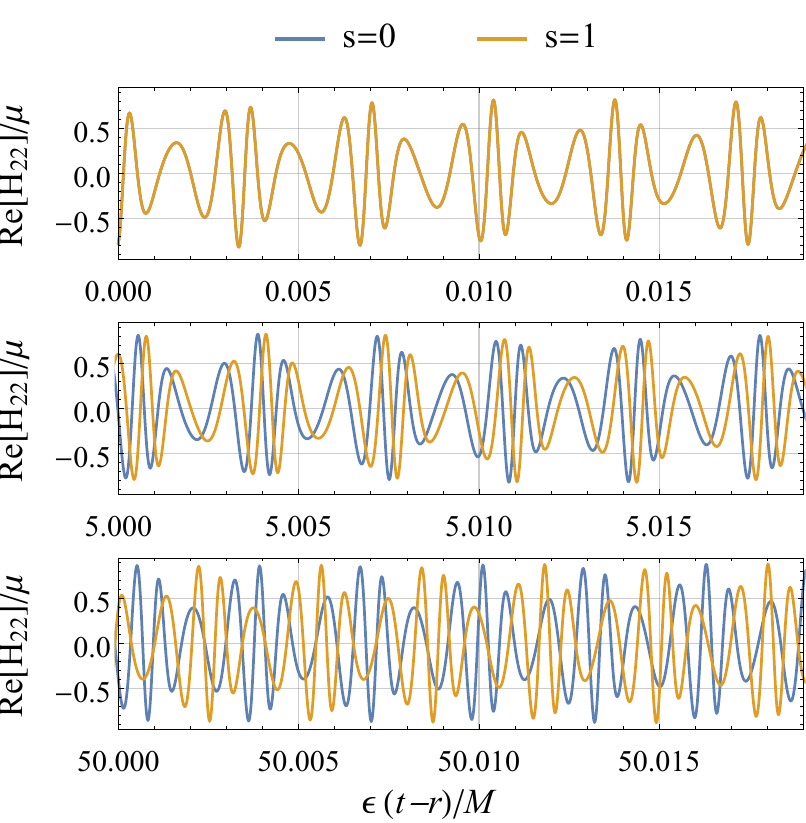}
\caption{Sample waveforms for inspirals with a spinning and non-spinning
secondary with $\e = 10^{-5}$.  The initial orbital parameters are
$(p_0,e_0)=(10,0.3)$. We match the initial frequencies of the spinning and
non-spinning configuration so that the two waveforms are initially in phase 
(top panel).  After $\sim 230$ radial oscillations the dephasing of the 
waveforms becomes noticeable (middle panel).  After $\sim 2400$ radial 
oscillations the waveforms have dephased by a half cycle (bottom panel).  
Close to the plunge ($\sim 4400$ radial oscillations, 
$\epsilon(t-r)/M\simeq 90$) the waveforms have dephased by 15 complete 
cycles.  
\label{fig:e3_waves}} 
\end{figure}

It is key to assess how a spinning secondary influences the phasing of the
inspiral. We compare the inspiral phase for binaries with a spinning secondary
against non-spinning binaries and show the results for the inspiral
trajectories in Figs.~\ref{fig:Delta_phi_spin_aligned} and 
\ref{fig:Delta_phi_sign_change}
and sample waveforms in Figs.~\ref{fig:e7_waves} and \ref{fig:e3_waves}. When
we compare inspirals we match the initial frequencies of the spinning inspiral
with the frequencies of the non-spinning inspiral using the technique described
in our previous paper \cite{OsbuWarbEvan16}. 

For Fig.~\ref{fig:Delta_phi_spin_aligned} the initial parameters for the
non-spinning binary are $(p,e) = (10,0.4)$. Over the course of the inspiral we
find for the maximum spin $|s| = 1$ that the orbit dephases by 
$\Delta\varphi \sim 1.2$ radians.  In this example spin-aligned binaries act 
to increase the phase of the
inspiral with respect to the non-spinning case. Similarly, anti-aligned
binaries act to decrease the inspiral phase.  Burko and Khanna
\cite{Burko:2015sqa} consider a similar setup for the evolution of
quasi-circular inspirals including spin-curvature effects. In their Fig.~6 they
consider the effect of varying the spin on the secondary for inspirals which
start at $r_0=10M$.  In the quasi-circular case, they find $\sim4$ radians of 
phase difference for the $|s| = 1$ inspiral.  Our results and theirs cannot 
be directly compared, however, as they attempt to include second-order 
radiative effects (via a post-Newtonian approximation) in their inspiral 
calculation, which are not a part of our model.

The phase difference observed in the example presented in
Fig.~\ref{fig:Delta_phi_spin_aligned}, whereby the spin-aligned binary takes
longer to merge than the spin anti-aligned binary, is reminiscent of the
`orbital hangup' observed for circular binaries in numerical relativity
simulations \cite{Campanelli:2006uy}. A more exhaustive search through the
$(p_0,e_0)$ parameter space reveals that this behavior is not universal and for
some configurations the opposite is observed, i.e., a spin-aligned binary can
accumulate less phase than the associated non-spinning binary. Figure
\ref{fig:Delta_phi_sign_change} demonstrates this change in behaviour for a
number of inspirals that differ in $p_0$ but each begin with the same high 
initial eccentricity $e_0=0.75$.  In general we find that spin-aligned
inspirals with a $p_0$ in the strong-field ($p\lesssim 20$) accumulate more
phase before the onset of the plunge than the non-spinning case, whereas
spin-aligned inspirals with $p_0$ in the weak field ($p\gtrsim 20$) experience 
the opposite and accumulate less phase before the onset of the plunge than 
the non-spinning case.  We also
observe that the accumulation of the $\Delta\varphi$ is not always monotonic,
e.g., for $p_0\simeq 20$ a spin-aligned binary can initially lose phase
with respect to the non-spinning case but can catch up with it later, even to
the extent that the two inspirals realign in phase ($\Delta\varphi=0$). We
observe this change of sign of $\Delta\varphi$ occurs regardless of the initial
eccentricity, including for quasi-circular inspirals with $e_0 = 0$.\footnote{In the literature it is common to read statements like `circular orbits remain circular as they adiabatically evolve due to radiation reaction'. This statement is true when one is concerned only with the leading-order phase evolution. When modeling post-1-adiabatic corrections, as in this work, it is important to note that inspirals with $e_0=0$ develop eccentricity which oscillates near $e=0$ with an amplitude that scales with the mass ratio.} For these inspirals we observe the change in sign of $\Delta\varphi$ occurs for inspirals with $p_0\simeq30$.

The influence of the spin-curvature force on $\Delta\varphi$ contrasts with that of the conservative self-force, which always acts to
reduce the inspiral phase \cite{BaraSago11,OsbuWarbEvan16}. In regions of the
parameter space where the effect of the spinning secondary acts to
monotonically increase the inspiral phase this raises the possibility that
waveforms associated with an
inspiral computed using the radiative approximation \cite{DrasHugh06,
FujiHikiTago09} plus a spinning secondary might closely mimic a waveform
computed using the full dissipative and conservative self-force with a
non-spinning binary. Exploring the possible degeneracies between these two
models we leave for future work.

In Figs.~\ref{fig:e7_waves} and \ref{fig:e3_waves} we show gravitational
waveforms from a high and a medium eccentricity inspiral, respectively.  Each 
figure contains three panels, displaying three different epochs in the 
inspiral.  The figures show the waveforms at an early time, when the 
dephasing first becomes noticeable, and later when the waveforms first 
dephase by a half cycle.  To display the dephasing, each panel shows the
waveforms for both the spinning and non-spinning cases.  Initially, the
frequencies of the two waveforms are matched but over time the waveforms 
dephase by a number of cycles.

\section{Arbitrary spin inspirals with orbital plane precession} 
\label{sec:inclined}

In cases where the secondary's spin and the orbital angular momentum are not 
initially aligned, the orbital plane and spin will subsequently precess 
during the inspiral.  To capture this motion we evolve the inspiral using
the osculating element equations we derived in Sec.~\ref{sec:osculate}, 
though with the initial condition $\iota_0 = \pi/4$ to avoid coordinate 
divergences in the osculating element equations (recall the discussion at 
the end of Sec.~\ref{sec:evolution_of_osc_els}). In the previous section 
the influence of
a spin-aligned secondary on the inspiral phase was assessed by comparing to a
non-spinning inspiral with the same two initial orbital frequencies. In the
case of generic spin orientation there are three orbital frequencies, the
additional one associated with the precession of the orbital plane. It may be
possible to make a matched-frequency comparison between a spin-unaligned
inspiral in Schwarzschild spacetime and an inspiral with a non-spinning
secondary in Kerr spacetime.  At the present time fully relativistic, generic 
orbit Kerr inspirals are not available for comparison (though kludge models
do exist for this case \cite{Babak:2006uv}). Our waveform generation scheme,
presented in Sec.~\ref{sec:freq_waveforms}, is only applicable to inspirals in
the equatorial plane. Consequently, in this section we showcase generic spin
inspirals without assessing the effect of the spin on the inspiral phase or
attempting to compute the associated waveforms (these are left for future
work). 

\subsection{Consistency checks}

\label{sec:check}

\begin{figure} \includegraphics[width=8.5cm]{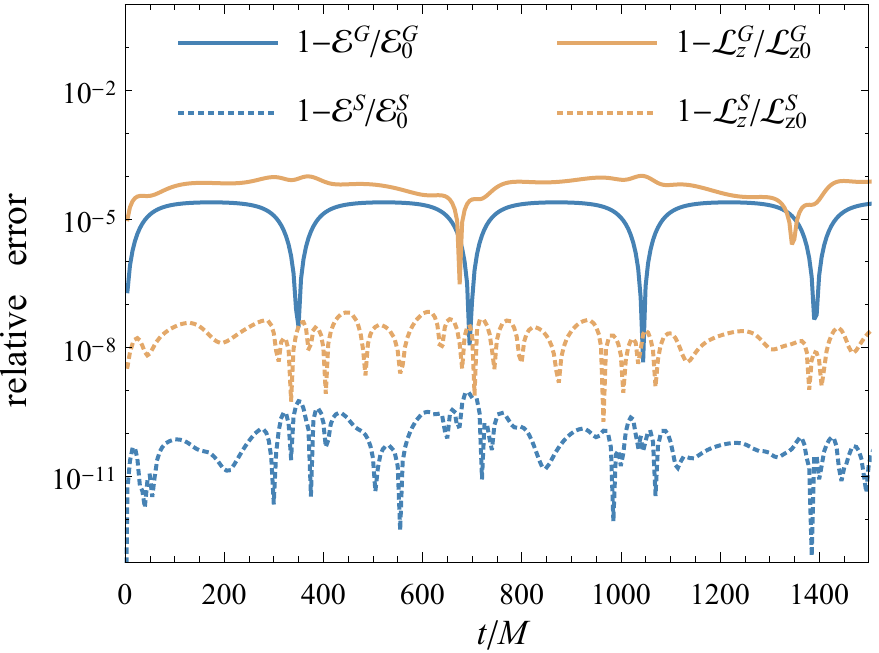}
\caption{Energy ($\mathcal{E^S}$) and angular momentum ($z$-component
$\mathcal{L}_z^S$) conservation in the presence of spin-curvature force 
alone.  Numerical residuals (relative errors) from expected strict 
conservation in energy and angular momentum are shown (dotted curves) for an 
orbit in which spin-curvature force is applied but the self-force is not.  
For contrast, the strictly geodesic quantities $\mathcal{E}^G$ and 
$\mathcal{L}_z^G$ (solid curves) can be seen to vary with an amplitude on 
the order of the mass ratio $\e$.  The orbital parameters in this example 
are $e_0=0.3$, $p_0=10$, $\iota_0=\pi/4$, 
$S^r_0 = S^\th_0 M = S^\vp_0 M = 0.3\mu^2$, $\e=10^{-3}$.}
\label{fig:energy} 
\end{figure}

To test whether our numerical code is functioning correctly we performed a
number of consistency checks. As discussed in Sec.~\ref{sec:spin_curvature},
if the self-force is not applied the spinning secondary's worldline admits 
a constant of motion for each of the spacetime's Killing vectors 
\cite{RuanVigeHugh15}.  The time-like killing vector of Schwarzschild 
spacetime results in a conserved specific energy $\mathcal{E}^S$.  The 
presence of spin perturbs the geodesic specific energy $\mathcal{E}^G$ by 
$\Delta \mathcal{E}$ 
\begin{align}
\mathcal{E}^S &= \mathcal{E}^G + \Delta \mathcal{E} ,
\\ 
\D \mathcal{E} &= \frac{M}{\mu}\sin{\th_p}(u^\th S^\vp-u^\vp S^\th) .
\label{eqn:delta_E} 
\end{align} 
Additionally, the three rotational killing vectors imply conservation of the 
$x$, $y$, and $z$ components of angular momentum.  The $z$-component is given 
by 
\begin{align} 
\mathcal{L}_z^S &=
\mathcal{L}^G_z + \Delta \mathcal{L}_z ,
\\ 
\mathcal{L}_z^G &= r_p^2\, u^\vp \sin^2{\th_p} ,
\\ 
\Delta \mathcal{L}_z &= \frac{1}{\mu} \Big( f_p r_p( S^t u^\th-S^\th u^t)
\sin{\th_p} 
\notag 		
\\ 
&\q\q\q\q\q + ( S^r u^t-S^t u^r) \cos{\th_p}
\Big) .  
\end{align} 
Similar results are straightforwardly obtained for the $x$
and $y$ components of the angular momentum. Fig.~\ref{fig:energy} demonstrates
that the perturbed energy $\mathcal{E}^S$ and $z$-component of angular momentum
$\mathcal{L}^S_z$ are conserved along the worldline if application of the 
self-force is withheld.  As a further consistency check we verified that the 
spin magnitude is invariant under parallel transport.

\subsection{Sample results}

\begin{figure*}
\includegraphics[width=\textwidth]{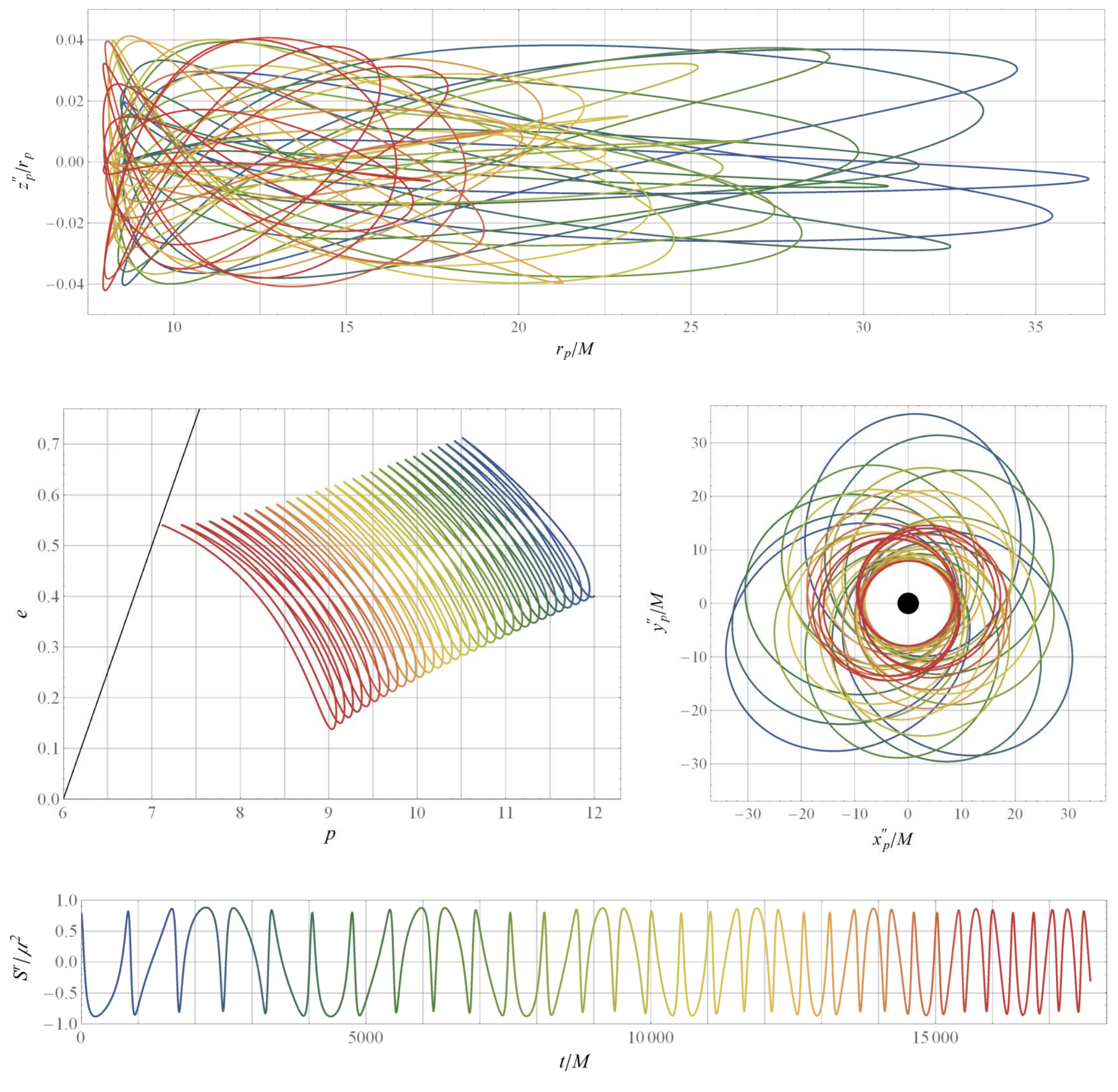} 
\caption{Sample inspiral
with mass ratio $\e=0.08$ and $|s| = 1$ and initial parameters $p_0 = 12$, $e_0
= 0.4$, $\iota_0 = \pi/4$, $S^r_0 = 0.7955930287154575 \mu^2$, $S^\th_0 M =
S^\vp_0 M = 0.03977965143577288 \mu^2$. Changes in color represent the passage
of time (blue is $t=0$ and red is immediately before plunge). Top:
$z^{\prime\prime}_p/r_p$ is plotted vs.~$r_p$ to demonstrate precession of the
orbital plane. With this large mass-ratio the $z$-coordinate of the position
vector reaches 4\% of $r_p$ at its maximum. Middle-left: Evolution of the
binary through $(p,e)$ space. The large oscillations in the best-fit 
(osculating) geodesics are a result of the high
mass ratio. The black line is the separatrix. Middle-right: A top-down view of
the trajectory in Schwarzschild coordinates. From this viewpoint the trajectory
resembles that of an equatorial inspiral because the precession is modest.
Bottom: The $r$-component of the spin-vector is plotted vs.~$t$. On short
timescales $S^r$ exhibits bi-periodicity where the fundamental frequencies are
the radial frequency and the frequency of spin precession.
\label{fig:generic}} 
\end{figure*}

The larger the mass ratio, the more prominent the precession of the orbital
plane will be. In order to make the precession due to the spin-curvature
force visible we computed an inspiral with mass-ratio $\e=0.08$ and spin 
magnitude $s=1$.  Computing an inspiral at this mass ratio is a slight abuse of
perturbation theory, as we are not sure of its validity for these values of
$\epsilon$ (though see e.g., \cite{LetiETC11} which suggests the range of
validity of black hole perturbation theory might be larger than once thought).
Our results for this $\epsilon=0.08$ inspiral are presented in
Fig.~\ref{fig:generic}.  
In displaying our results we rotate the
coordinates such that the new inclination parameter oscillates around zero
(recall that $\iota_0 = \pi/4$), or equivalently, the new $z$-coordinate of the
position vector, $z^{\prime\prime}_p$, oscillates with a minimized amplitude.
The transformation to the double primed coordinate system is given by
\begin{align} 
&\left[ \begin{array}{c} x_p \\ y_p \\ z_p \end{array} \right] =
\left[ \begin{array}{ccc} 1 & 0 & 0 \\ 0 & \cos{\iota_0} & -\sin{\iota_0} \\ 0 
& \sin{\iota_0} & \cos{\iota_0} \end{array} \right] \left[ \begin{array}{c}
x^{\prime\prime}_p \\ y^{\prime\prime}_p \\ z^{\prime\prime}_p \end{array}
\right] .  
\end{align} 
For the orbital parameters chosen in Fig.~\ref{fig:generic}, 
$z^{\prime\prime}_p$ reaches 4\% of $r_p$ at its maximum.  At this level, 
precession of the orbital plane is noticeable but modest. 

\section{Conclusions}\label{sec:conclusions}

In this work we have computed the effect of spin-curvature coupling on the
inspiral of a spinning body into a non-rotating black hole including all
first-order in the mass-ratio self-force effects. For binaries where the spin
and orbital angular momentum are aligned or anti-aligned we compute the
waveforms by evolving through a sequence of snapshot waveforms. We also computed
the dephasing of these waveforms with respect to non-spinning binaries, finding that the sign of the accumulated phase difference depends on the initial binary separation. For binaries with initial eccentricity $e_0 = 0$ the phase accumulation for a spin-aligned binary verses a non-spinning binary is negative for $p\gtrsim 30$ and positive for $p\lesssim30$ (spin anti-aligned binaries show the opposite behavior). For binaries with initial eccentricity $e_0=0.75$ the change in sign of the accumulated phase difference occurs for $p\simeq20$. The magnitude of the dephasing due to the spin-curvature force is similar in magnitude to the
conservative self-force corrections to the inspiral phase
\cite{OsbuWarbEvan16}. We leave it to future work to examine whether inspirals
computed using, e.g., a radiative approximation with a spinning secondary can
mimic inspirals computed using the full dissipative and conservative self-force
for a non-spinning binary. For orbits where the spin and orbital
angular-momentum are not aligned the secondary will precess out of the
equatorial plane. We extend the osculating element prescription of motion to
accommodate this precession and present an example inspiral in
Fig.~\ref{fig:generic}.

This work is naturally extended in a number of ways. First, our primary is not
rotating. For a rotating primary there are self-force calculations (in the
radiation gauge) \cite{vandeMeent:2016pee}, efficient calculations of the
spin-curvature force \cite{RuanVigeHugh15}, and osculating element schemes
\cite{GairETC11}. These three pieces could be combined to extend the results in
this work. Second, we do not include second-order in the mass-ratio fluxes or
fluxes related the spinning secondary in this work. Work progresses calculating
the former \cite{Poun12a,Gral12,Detw12, Pound:2014koa, WarbWard14, PounMill14,
WardWarb15, Pound:2015wva, Miller:2016hjv, Pound:2017psq}. The latter has been
calculated for circular orbits \cite{Harms:2016ctx,Harms:2015ixa,Han:2010tp}
but has not, to the best of our knowledge, been calculated for generic orbits.
As per the discussion in Sec.~\ref{sec:effects_of_SF}, inclusion of these pieces
is necessary to get a waveform accurate to post-1-adiabatic order. Third,
because of the geodesic self-force approximation we use in this work, it is
important to compare our inspirals against those from computed using
self-consistent evolutions \cite{DienETC12,Warb13,Warb14a,Dien15}.

Finally we note that, although our evolution scheme is primarily based upon
numerical self-force results, there is much scope for synergy with analytic
results. Recently, combining black hole perturbation and post-Newtonian theory
has allowed for the calculation of gauge invariant results to very high
post-Newtonian order (at first-order in the mass-ratio) \cite{Nolan:2015vpa,
Kavanagh:2015lva,Kavanagh:2016idg,ShahPoun15,Forseth:2015oua,Hopper:2015icj,
Bini:2013rfa,Bini:2014ica,Bini:2014zxa}.  These results are often to such 
high order that they reproduce strong-field
results to better than a fraction of a percent. As we argued in our previous
paper \cite{OsbuWarbEvan16} (and is also noted in \cite{Huerta:2011kt}), whilst
the leading-order flux needs to be calculated to better than one part in the
inverse mass-ratio, the terms that contribute to the post-1-adiabatic evolution
need only be calculated to one part in a thousand. Even in the strong-field,
the accuracy requirement for the latter is within the reach of high-order
post-Newtonian black hole perturbation theory results. We thus envisage these
high-order post-Newtonian results complementing high accuracy numerical
calculations like those presented in
this work.

\medskip
\acknowledgments

We thank Zach Nasipak for computational assistance.  This work was 
supported in part by NSF grant PHY-1506182.  N.W.~gratefully acknowledges 
support from a Marie Curie International Outgoing Fellowship 
(PIOF-GA-2012-627781). C.R.E.~acknowledges support from the Bahnson Fund at the University of North
Carolina-Chapel Hill.

\appendix

\begin{widetext}

\section{Osculating element equations}\label{apdx:osculating}

The coefficients that appear in Eq.~\eqref{eq:d_dchi} are given by
\begin{align} c^{(t)}_e &= \frac{ \left(2 e^2-p+6\right) \left[\left(3 e^2-4
p+12\right) \cos{v}+e \big(e \cos{(3v)}+(6-p) \cos{(2v)}\big)\right]+e \left(p
(3 p-32)+60-2 e^2 (p-10)\right)}{2 \mu M^{-1}p^{-1}(p-3-e^2)^{-1}[(p-2)^2-4
e^2]^{-1/2}[(p-6)^2-4 e^2] (1+e \cos{v})^4 \sqrt{p-6-2 e \cos{v}}} , \\
c^{(r)}_e &= \frac{M p^2(1-e^2) (p-3-e^2) \left((p-6)(p-2) + 4 e^2 \right)
\sin{v}}{\mu \left((p-6)^2-4 e^2\right) (p-2-2 e \cos{v}) (1+e \cos{v})^4},
\label{eqn:cre} \\ c^{(t)}_p &= \frac{M p^2 \left(p-3-e^2\right) [e^2 \cos{(2
v)}+e^2-2 p+6] \sqrt{(p-6-2 e \cos{v})[(p-2)^2-4 e^2]}}{\mu [(p-6)^2-4 e^2]
(1+e \cos{v})^4}, \\ c^{(r)}_p &= \frac{2 e M (p-4)^2 p^3 \left(e^2-p+3\right)
\sin{v}} {\mu (2 e \cos{v}-p+2) (1+e \cos{v})^4 [4 e^2-(p-6)^2]},
\label{eqn:crp} \\ c^{(t)}_{\chi_0} &= \frac{M p  \sin{v} \left(e^2-p+3\right)
\left[e \left(4 e^2-(p-6)^2\right) \cos{v}+(p-6) \left(e^2 \cos (2 v)+e^2-2
p+6\right)\right] \sqrt{(p-2)^2-4 e^2}}{\mu \, e \left((p-6)^2-4 e^2\right)
(1+e \cos{v})^4 \sqrt{p-6-2 e \cos{v}}}, \\ c^{(r)}_{\chi_0} &= \frac{M p^2
\left(p-3-e^2\right) \big[\left(4 e^4-e^2 ((p-8) p+24)-(p-6) (p-2)\right)
\cos{v}-2 e (p-4)^2\big]}{\mu \, e \left(4 e^2-(p-6)^2\right) (1+e \cos{v})^4
(2 e \cos{v}-p+2)} .  \end{align}

The coefficients that appear in Eq.~\eqref{eq:osc_els_i} are given by
\begin{align} c^{(\th)}_\iota &= \frac{M^2 p^{5/2} \cos{\vp'_p}\,
\left(e^2-p+3\right)}{\mu\cos{\iota}\,(1+e \cos{v})^4}
\sqrt{\frac{1-\sin^2{\iota}\, \sin^2{\vp'_p}}{p-6-2 e \cos{v}}} , \\
c^{(e)}_\iota &= \frac{\sin{\iota}\, \cos^2{\vp'_p} \left(e-(e^2-2 p+6) \cos{v}
\right)}{\cos{\iota}\,\left(e^2-p+3\right) (1+e \cos{v})} , \\ c^{(p)}_\iota &=
\frac{(6 + 2 e^2 - 3 p) \cos^2{\vp'_p} \sin{\iota}}{2 p\cos{\iota}\, (3 + e^2 -
p)} , \\ c^{(\chi_0)}_\iota &= -\frac{2 e \sin{\iota}\, \sin{v}\,
\cos^2{\vp'_p}}{\cos{\iota}\,(1+e \cos{v})} , \\ \label{eqn:cthPh}
c^{(\th)}_\Phi &= \frac{M^2 p^{5/2}  \sin{\vp'_p}\,
\left(p-3-e^2\right)}{\mu\sin{\iota}\, (1+e \cos{v})^4 }
\sqrt{\frac{1-\sin^2{\iota}\,\sin^2{\vp'_p}}{p-6- 2 e\cos{v}}} , \\
c^{(e)}_\Phi &= \frac{\sin{(2\vp'_p)} \left((e^2-2 p+6) \cos{v}-e\right)}{2
\left(e^2-p+3\right) (1+e \cos{v})} +\frac{ (2 e+p-6)\bar{F}-(p-6)\bar{E}}{e (2
e+p-6)} \sqrt{\frac{p}{p-6-2 e}} \notag \\&\hspace{58ex} -\frac{2 (p-6)
\sin{v}}{\left((p-6)^2-4 e^2\right)} \sqrt{\frac{p}{ (p-6-2 e \cos{v})}} , \\
c^{(p)}_\Phi &= -\frac{4 p^{1/2} \left(p-3-e^2\right) \sqrt{p-6-2 e} \left((2
e+p-6)\bar{F}-p \bar{E}\right)+ (2 e^2-3 p+6) (2 e-p+6) (2 e+p-6)
\sin{(2\vp'_p)}}{4 p (p-6-2 e) (p-6+2 e)(p-3-e^2)}  \notag \\&\hspace{60ex}
+\frac{2 e \sin{v}}{(p-6)^2-4 e^2 }\sqrt{\frac{p}{p-6-2 e\cos{v}}} , \\
c^{(\chi_0)}_\Phi &= \frac{e \sin{v}\,\sin{(2\vp'_p)}}{1+e
\cos{v}}-\sqrt{\frac{p}{p-6-2 e \cos{v}}}, \end{align} where $\bar{F}$ and
$\bar{E}$ are the incomplete elliptic integrals of the first and second kind,
respectively, each with the same arguments as the elliptic integral in
Eq.~\eqref{eqn:phiEllip}. The coefficients that appear in
Eq.~\eqref{eq:osc_els_O} are given by \begin{align} c^{(e)}_\O &= \frac{4
p^{1/2} \cos{\iota} \left(\sqrt{p-6-2 e} ( (p-6)\bar{E}- (2 e+p-6)\bar{F})
\sqrt{p-6-2 e \cos{v}}+2 e (p-6) \sin{v}\right)}{e (2 e-p+6) (2 e+p-6)
\sqrt{p-6-2 e \cos{v}} \left(2 \sin^2{\iota} \cos{(2 \vp'})+\cos{(2
\iota)}+3\right)}, \\ c^{(p)}_\O &= \frac{4 \cos{\iota} \left((2 e-p+6) ( (2
e+p-6)\bar{F}- p\bar{E}) \sqrt{p-6-2 e \cos{v}}+2 e p \sqrt{p-6-2 e}
\sin{v}\right)}{(p-6-2 e)^{3/2} (2 e+p-6) \sqrt{p (p-6-2 e \cos{v})} \left(2
\sin^2{\iota} \cos{(2 \vp)}+\cos{(2 \iota)}+3\right)}, \\ c^{(\chi_0)}_\O &=
\frac{4 \cos{\iota}}{2 \sin ^2{\iota}\, \cos (2 \vp'_p)+\cos (2 \iota)+3}
\sqrt{\frac{p}{p-6-2e\cos{v}}}, \\ c^{(\iota)}_\O &= \frac{2 \sin{\iota}\, \sin
(2\vp'_p)}{2 \sin ^2{\iota}\, \cos(2\vp'_p)+\cos (2 \iota )+3}, \\
c^{(\Phi)}_\O &= -\frac{4 \cos {\iota}}{2 \sin^2{\iota}\, \cos (2\vp'_p)+\cos
(2 \iota )+3}.  \end{align} \end{widetext}

\bibliography{spinspiral}

\begin{thebibliography}{124}%
\makeatletter
\providecommand \@ifxundefined [1]{%
 \@ifx{#1\undefined}
}%
\providecommand \@ifnum [1]{%
 \ifnum #1\expandafter \@firstoftwo
 \else \expandafter \@secondoftwo
 \fi
}%
\providecommand \@ifx [1]{%
 \ifx #1\expandafter \@firstoftwo
 \else \expandafter \@secondoftwo
 \fi
}%
\providecommand \natexlab [1]{#1}%
\providecommand \enquote  [1]{``#1''}%
\providecommand \bibnamefont  [1]{#1}%
\providecommand \bibfnamefont [1]{#1}%
\providecommand \citenamefont [1]{#1}%
\providecommand \href@noop [0]{\@secondoftwo}%
\providecommand \href [0]{\begingroup \@sanitize@url \@href}%
\providecommand \@href[1]{\@@startlink{#1}\@@href}%
\providecommand \@@href[1]{\endgroup#1\@@endlink}%
\providecommand \@sanitize@url [0]{\catcode `\\12\catcode `\$12\catcode
  `\&12\catcode `\#12\catcode `\^12\catcode `\_12\catcode `\%12\relax}%
\providecommand \@@startlink[1]{}%
\providecommand \@@endlink[0]{}%
\providecommand \url  [0]{\begingroup\@sanitize@url \@url }%
\providecommand \@url [1]{\endgroup\@href {#1}{\urlprefix }}%
\providecommand \urlprefix  [0]{URL }%
\providecommand \Eprint [0]{\href }%
\providecommand \doibase [0]{http://dx.doi.org/}%
\providecommand \selectlanguage [0]{\@gobble}%
\providecommand \bibinfo  [0]{\@secondoftwo}%
\providecommand \bibfield  [0]{\@secondoftwo}%
\providecommand \translation [1]{[#1]}%
\providecommand \BibitemOpen [0]{}%
\providecommand \bibitemStop [0]{}%
\providecommand \bibitemNoStop [0]{.\EOS\space}%
\providecommand \EOS [0]{\spacefactor3000\relax}%
\providecommand \BibitemShut  [1]{\csname bibitem#1\endcsname}%
\let\auto@bib@innerbib\@empty
\bibitem [{\citenamefont {Abbott}\ \emph
  {et~al.}(2016{\natexlab{a}})\citenamefont {Abbott} \emph {et~al.}}]{LIGO16a}%
  \BibitemOpen
  \bibfield  {author} {\bibinfo {author} {\bibfnamefont {B.~P.}\ \bibnamefont
  {Abbott}} \emph {et~al.} (\bibinfo {collaboration} {LIGO Scientific
  Collaboration and Virgo Collaboration}),\ }\href {\doibase
  10.1103/PhysRevLett.116.061102} {\bibfield  {journal} {\bibinfo  {journal}
  {Phys. Rev. Lett.}\ }\textbf {\bibinfo {volume} {116}},\ \bibinfo {pages}
  {061102} (\bibinfo {year} {2016}{\natexlab{a}})}\BibitemShut {NoStop}%
\bibitem [{\citenamefont {Abbott}\ \emph
  {et~al.}(2016{\natexlab{b}})\citenamefont {Abbott} \emph {et~al.}}]{LIGO16b}%
  \BibitemOpen
  \bibfield  {author} {\bibinfo {author} {\bibfnamefont {B.~P.}\ \bibnamefont
  {Abbott}} \emph {et~al.} (\bibinfo {collaboration} {LIGO Scientific
  Collaboration and Virgo Collaboration}),\ }\href {\doibase
  10.1103/PhysRevLett.116.241103} {\bibfield  {journal} {\bibinfo  {journal}
  {Phys. Rev. Lett.}\ }\textbf {\bibinfo {volume} {116}},\ \bibinfo {pages}
  {241103} (\bibinfo {year} {2016}{\natexlab{b}})},\ \Eprint
  {http://arxiv.org/abs/1606.04855} {arXiv:1606.04855} \BibitemShut {NoStop}%
\bibitem [{\citenamefont {Abbott}\ \emph {et~al.}(2017)\citenamefont {Abbott}
  \emph {et~al.}}]{LIGO17a}%
  \BibitemOpen
  \bibfield  {author} {\bibinfo {author} {\bibfnamefont {B.~P.}\ \bibnamefont
  {Abbott}} \emph {et~al.} (\bibinfo {collaboration} {LIGO Scientific
  Collaboration and Virgo Collaboration}),\ }\href {\doibase
  10.1103/PhysRevLett.118.221101} {\bibfield  {journal} {\bibinfo  {journal}
  {Phys. Rev. Lett.}\ }\textbf {\bibinfo {volume} {118}},\ \bibinfo {pages}
  {221101} (\bibinfo {year} {2017})},\ \Eprint
  {http://arxiv.org/abs/1706.01812} {arXiv:1706.01812} \BibitemShut {NoStop}%
\bibitem [{\citenamefont {{Amaro-Seoane}}\ \emph {et~al.}(2017)\citenamefont
  {{Amaro-Seoane}} \emph {et~al.}}]{LISA}%
  \BibitemOpen
  \bibfield  {author} {\bibinfo {author} {\bibfnamefont {P.}~\bibnamefont
  {{Amaro-Seoane}}} \emph {et~al.},\ }\href@noop {} {\bibfield  {journal}
  {\bibinfo  {journal} {ArXiv e-prints}\ } (\bibinfo {year} {2017})},\ \Eprint
  {http://arxiv.org/abs/1702.00786} {arXiv:1702.00786} \BibitemShut {NoStop}%
\bibitem [{\citenamefont {Babak}\ \emph {et~al.}(2017)\citenamefont {Babak},
  \citenamefont {Gair}, \citenamefont {Sesana}, \citenamefont {Barausse},
  \citenamefont {Sopuerta}, \citenamefont {Berry}, \citenamefont {Berti},
  \citenamefont {Amaro-Seoane}, \citenamefont {Petiteau},\ and\ \citenamefont
  {Klein}}]{Babak:2017tow}%
  \BibitemOpen
  \bibfield  {author} {\bibinfo {author} {\bibfnamefont {S.}~\bibnamefont
  {Babak}}, \bibinfo {author} {\bibfnamefont {J.}~\bibnamefont {Gair}},
  \bibinfo {author} {\bibfnamefont {A.}~\bibnamefont {Sesana}}, \bibinfo
  {author} {\bibfnamefont {E.}~\bibnamefont {Barausse}}, \bibinfo {author}
  {\bibfnamefont {C.~F.}\ \bibnamefont {Sopuerta}}, \bibinfo {author}
  {\bibfnamefont {C.~P.~L.}\ \bibnamefont {Berry}}, \bibinfo {author}
  {\bibfnamefont {E.}~\bibnamefont {Berti}}, \bibinfo {author} {\bibfnamefont
  {P.}~\bibnamefont {Amaro-Seoane}}, \bibinfo {author} {\bibfnamefont
  {A.}~\bibnamefont {Petiteau}}, \ and\ \bibinfo {author} {\bibfnamefont
  {A.}~\bibnamefont {Klein}},\ }\href@noop {} {\  (\bibinfo {year} {2017})},\
  \Eprint {http://arxiv.org/abs/1703.09722} {arXiv:1703.09722} \BibitemShut
  {NoStop}%
\bibitem [{\citenamefont {{Vigeland}}\ and\ \citenamefont
  {{Hughes}}(2010)}]{VigeHugh10}%
  \BibitemOpen
  \bibfield  {author} {\bibinfo {author} {\bibfnamefont {S.~J.}\ \bibnamefont
  {{Vigeland}}}\ and\ \bibinfo {author} {\bibfnamefont {S.~A.}\ \bibnamefont
  {{Hughes}}},\ }\href {\doibase 10.1103/PhysRevD.81.024030} {\bibfield
  {journal} {\bibinfo  {journal} {Phys. Rev. D}\ }\textbf {\bibinfo {volume}
  {81}},\ \bibinfo {eid} {024030} (\bibinfo {year} {2010})},\ \Eprint
  {http://arxiv.org/abs/0911.1756} {arXiv:0911.1756} \BibitemShut {NoStop}%
\bibitem [{\citenamefont {Barack}\ and\ \citenamefont
  {Cutler}(2007)}]{BaraCutl07}%
  \BibitemOpen
  \bibfield  {author} {\bibinfo {author} {\bibfnamefont {L.}~\bibnamefont
  {Barack}}\ and\ \bibinfo {author} {\bibfnamefont {C.}~\bibnamefont
  {Cutler}},\ }\href {\doibase 10.1103/PhysRevD.75.042003} {\bibfield
  {journal} {\bibinfo  {journal} {Phys. Rev. D}\ }\textbf {\bibinfo {volume}
  {75}},\ \bibinfo {pages} {042003} (\bibinfo {year} {2007})},\ \Eprint
  {http://arxiv.org/abs/gr-qc/0612029} {arXiv:gr-qc/0612029} \BibitemShut
  {NoStop}%
\bibitem [{\citenamefont {{Brown}}\ \emph {et~al.}(2007)\citenamefont
  {{Brown}}, \citenamefont {{Brink}}, \citenamefont {{Fang}}, \citenamefont
  {{Gair}}, \citenamefont {{Li}}, \citenamefont {{Lovelace}}, \citenamefont
  {{Mandel}},\ and\ \citenamefont {{Thorne}}}]{BrowETC07}%
  \BibitemOpen
  \bibfield  {author} {\bibinfo {author} {\bibfnamefont {D.~A.}\ \bibnamefont
  {{Brown}}}, \bibinfo {author} {\bibfnamefont {J.}~\bibnamefont {{Brink}}},
  \bibinfo {author} {\bibfnamefont {H.}~\bibnamefont {{Fang}}}, \bibinfo
  {author} {\bibfnamefont {J.~R.}\ \bibnamefont {{Gair}}}, \bibinfo {author}
  {\bibfnamefont {C.}~\bibnamefont {{Li}}}, \bibinfo {author} {\bibfnamefont
  {G.}~\bibnamefont {{Lovelace}}}, \bibinfo {author} {\bibfnamefont
  {I.}~\bibnamefont {{Mandel}}}, \ and\ \bibinfo {author} {\bibfnamefont
  {K.~S.}\ \bibnamefont {{Thorne}}},\ }\href {\doibase
  10.1103/PhysRevLett.99.201102} {\bibfield  {journal} {\bibinfo  {journal}
  {Physical Review Letters}\ }\textbf {\bibinfo {volume} {99}},\ \bibinfo {eid}
  {201102} (\bibinfo {year} {2007})},\ \Eprint
  {http://arxiv.org/abs/gr-qc/0612060} {gr-qc/0612060} \BibitemShut {NoStop}%
\bibitem [{\citenamefont {Gair}\ \emph {et~al.}(2013)\citenamefont {Gair},
  \citenamefont {Vallisneri}, \citenamefont {Larson},\ and\ \citenamefont
  {Baker}}]{Gair:2012nm}%
  \BibitemOpen
  \bibfield  {author} {\bibinfo {author} {\bibfnamefont {J.~R.}\ \bibnamefont
  {Gair}}, \bibinfo {author} {\bibfnamefont {M.}~\bibnamefont {Vallisneri}},
  \bibinfo {author} {\bibfnamefont {S.~L.}\ \bibnamefont {Larson}}, \ and\
  \bibinfo {author} {\bibfnamefont {J.~G.}\ \bibnamefont {Baker}},\ }\href
  {\doibase 10.12942/lrr-2013-7} {\bibfield  {journal} {\bibinfo  {journal}
  {Living Rev. Rel.}\ }\textbf {\bibinfo {volume} {16}},\ \bibinfo {pages} {7}
  (\bibinfo {year} {2013})},\ \Eprint {http://arxiv.org/abs/1212.5575}
  {arXiv:1212.5575} \BibitemShut {NoStop}%
\bibitem [{\citenamefont {Barausse}\ \emph {et~al.}(2014)\citenamefont
  {Barausse}, \citenamefont {Cardoso},\ and\ \citenamefont
  {Pani}}]{Barausse:2014tra}%
  \BibitemOpen
  \bibfield  {author} {\bibinfo {author} {\bibfnamefont {E.}~\bibnamefont
  {Barausse}}, \bibinfo {author} {\bibfnamefont {V.}~\bibnamefont {Cardoso}}, \
  and\ \bibinfo {author} {\bibfnamefont {P.}~\bibnamefont {Pani}},\ }\href
  {\doibase 10.1103/PhysRevD.89.104059} {\bibfield  {journal} {\bibinfo
  {journal} {Phys. Rev.}\ }\textbf {\bibinfo {volume} {D89}},\ \bibinfo {pages}
  {104059} (\bibinfo {year} {2014})},\ \Eprint {http://arxiv.org/abs/1404.7149}
  {arXiv:1404.7149} \BibitemShut {NoStop}%
\bibitem [{\citenamefont {Brown}\ \emph {et~al.}(2007)\citenamefont {Brown},
  \citenamefont {Fang}, \citenamefont {Gair}, \citenamefont {Li}, \citenamefont
  {Lovelace}, \citenamefont {Mandel},\ and\ \citenamefont
  {Thorne}}]{Brown:2006pj}%
  \BibitemOpen
  \bibfield  {author} {\bibinfo {author} {\bibfnamefont {D.~A.}\ \bibnamefont
  {Brown}}, \bibinfo {author} {\bibfnamefont {H.}~\bibnamefont {Fang}},
  \bibinfo {author} {\bibfnamefont {J.~R.}\ \bibnamefont {Gair}}, \bibinfo
  {author} {\bibfnamefont {C.}~\bibnamefont {Li}}, \bibinfo {author}
  {\bibfnamefont {G.}~\bibnamefont {Lovelace}}, \bibinfo {author}
  {\bibfnamefont {I.}~\bibnamefont {Mandel}}, \ and\ \bibinfo {author}
  {\bibfnamefont {K.~S.}\ \bibnamefont {Thorne}},\ }\href {\doibase
  10.1103/PhysRevLett.99.201102} {\bibfield  {journal} {\bibinfo  {journal}
  {Phys. Rev. Lett.}\ }\textbf {\bibinfo {volume} {99}},\ \bibinfo {pages}
  {201102} (\bibinfo {year} {2007})},\ \Eprint
  {http://arxiv.org/abs/gr-qc/0612060} {arXiv:gr-qc/0612060} \BibitemShut
  {NoStop}%
\bibitem [{\citenamefont {Huerta}\ and\ \citenamefont
  {Gair}(2011{\natexlab{a}})}]{Huerta:2010un}%
  \BibitemOpen
  \bibfield  {author} {\bibinfo {author} {\bibfnamefont {E.~A.}\ \bibnamefont
  {Huerta}}\ and\ \bibinfo {author} {\bibfnamefont {J.~R.}\ \bibnamefont
  {Gair}},\ }\href {\doibase 10.1103/PhysRevD.83.044020} {\bibfield  {journal}
  {\bibinfo  {journal} {Phys. Rev.}\ }\textbf {\bibinfo {volume} {D83}},\
  \bibinfo {pages} {044020} (\bibinfo {year} {2011}{\natexlab{a}})},\ \Eprint
  {http://arxiv.org/abs/1009.1985} {arXiv:1009.1985} \BibitemShut {NoStop}%
\bibitem [{\citenamefont {Huerta}\ and\ \citenamefont
  {Gair}(2011{\natexlab{b}})}]{Huerta:2010tp}%
  \BibitemOpen
  \bibfield  {author} {\bibinfo {author} {\bibfnamefont {E.~A.}\ \bibnamefont
  {Huerta}}\ and\ \bibinfo {author} {\bibfnamefont {J.~R.}\ \bibnamefont
  {Gair}},\ }\href {\doibase 10.1103/PhysRevD.83.044021} {\bibfield  {journal}
  {\bibinfo  {journal} {Phys. Rev.}\ }\textbf {\bibinfo {volume} {D83}},\
  \bibinfo {pages} {044021} (\bibinfo {year} {2011}{\natexlab{b}})},\ \Eprint
  {http://arxiv.org/abs/1011.0421} {arXiv:1011.0421} \BibitemShut {NoStop}%
\bibitem [{\citenamefont {Miller}(2002)}]{Miller:2002vg}%
  \BibitemOpen
  \bibfield  {author} {\bibinfo {author} {\bibfnamefont {M.~C.}\ \bibnamefont
  {Miller}},\ }\href {\doibase 10.1086/344156} {\bibfield  {journal} {\bibinfo
  {journal} {Astrophys. J.}\ }\textbf {\bibinfo {volume} {581}},\ \bibinfo
  {pages} {438} (\bibinfo {year} {2002})},\ \Eprint
  {http://arxiv.org/abs/astro-ph/0206404} {arXiv:astro-ph/0206404} \BibitemShut
  {NoStop}%
\bibitem [{\citenamefont {Sesana}(2016)}]{Sesana:2016ljz}%
  \BibitemOpen
  \bibfield  {author} {\bibinfo {author} {\bibfnamefont {A.}~\bibnamefont
  {Sesana}},\ }\href {\doibase 10.1103/PhysRevLett.116.231102} {\bibfield
  {journal} {\bibinfo  {journal} {Phys. Rev. Lett.}\ }\textbf {\bibinfo
  {volume} {116}},\ \bibinfo {pages} {231102} (\bibinfo {year} {2016})},\
  \Eprint {http://arxiv.org/abs/1602.06951} {arXiv:1602.06951} \BibitemShut
  {NoStop}%
\bibitem [{\citenamefont {Miller}(2004)}]{Miller:2004va}%
  \BibitemOpen
  \bibfield  {author} {\bibinfo {author} {\bibfnamefont {M.~C.}\ \bibnamefont
  {Miller}},\ }\href {\doibase 10.1086/425910} {\bibfield  {journal} {\bibinfo
  {journal} {Astrophys. J.}\ }\textbf {\bibinfo {volume} {618}},\ \bibinfo
  {pages} {426} (\bibinfo {year} {2004})},\ \Eprint
  {http://arxiv.org/abs/astro-ph/0409331} {arXiv:astro-ph/0409331 [astro-ph]}
  \BibitemShut {NoStop}%
\bibitem [{\citenamefont {Mandel}\ \emph {et~al.}(2008)\citenamefont {Mandel},
  \citenamefont {Brown}, \citenamefont {Gair},\ and\ \citenamefont
  {Miller}}]{Mandel:2007hi}%
  \BibitemOpen
  \bibfield  {author} {\bibinfo {author} {\bibfnamefont {I.}~\bibnamefont
  {Mandel}}, \bibinfo {author} {\bibfnamefont {D.~A.}\ \bibnamefont {Brown}},
  \bibinfo {author} {\bibfnamefont {J.~R.}\ \bibnamefont {Gair}}, \ and\
  \bibinfo {author} {\bibfnamefont {M.~C.}\ \bibnamefont {Miller}},\ }\href
  {\doibase 10.1086/588246} {\bibfield  {journal} {\bibinfo  {journal}
  {Astrophys. J.}\ }\textbf {\bibinfo {volume} {681}},\ \bibinfo {pages} {1431}
  (\bibinfo {year} {2008})},\ \Eprint {http://arxiv.org/abs/0705.0285}
  {arXiv:0705.0285} \BibitemShut {NoStop}%
\bibitem [{\citenamefont {Drasco}\ and\ \citenamefont
  {Hughes}(2004)}]{DrasHugh05}%
  \BibitemOpen
  \bibfield  {author} {\bibinfo {author} {\bibfnamefont {S.}~\bibnamefont
  {Drasco}}\ and\ \bibinfo {author} {\bibfnamefont {S.~A.}\ \bibnamefont
  {Hughes}},\ }\href {\doibase 10.1103/PhysRevD.69.044015} {\bibfield
  {journal} {\bibinfo  {journal} {Phys. Rev. D}\ }\textbf {\bibinfo {volume}
  {69}},\ \bibinfo {pages} {044015} (\bibinfo {year} {2004})}\BibitemShut
  {NoStop}%
\bibitem [{\citenamefont {Fujita}\ \emph {et~al.}(2009)\citenamefont {Fujita},
  \citenamefont {Hikida},\ and\ \citenamefont {Tagoshi}}]{FujiHikiTago09}%
  \BibitemOpen
  \bibfield  {author} {\bibinfo {author} {\bibfnamefont {R.}~\bibnamefont
  {Fujita}}, \bibinfo {author} {\bibfnamefont {W.}~\bibnamefont {Hikida}}, \
  and\ \bibinfo {author} {\bibfnamefont {H.}~\bibnamefont {Tagoshi}},\ }\href
  {\doibase 10.1143/PTP.121.843} {\bibfield  {journal} {\bibinfo  {journal}
  {Prog. Theor. Phys.}\ }\textbf {\bibinfo {volume} {121}},\ \bibinfo {pages}
  {843} (\bibinfo {year} {2009})},\ \Eprint {http://arxiv.org/abs/0904.3810}
  {arXiv:0904.3810} \BibitemShut {NoStop}%
\bibitem [{\citenamefont {Babak}\ \emph {et~al.}(2007)\citenamefont {Babak},
  \citenamefont {Fang}, \citenamefont {Gair}, \citenamefont {Glampedakis},\
  and\ \citenamefont {Hughes}}]{Babak:2006uv}%
  \BibitemOpen
  \bibfield  {author} {\bibinfo {author} {\bibfnamefont {S.}~\bibnamefont
  {Babak}}, \bibinfo {author} {\bibfnamefont {H.}~\bibnamefont {Fang}},
  \bibinfo {author} {\bibfnamefont {J.~R.}\ \bibnamefont {Gair}}, \bibinfo
  {author} {\bibfnamefont {K.}~\bibnamefont {Glampedakis}}, \ and\ \bibinfo
  {author} {\bibfnamefont {S.~A.}\ \bibnamefont {Hughes}},\ }\href {\doibase
  10.1103/PhysRevD.75.024005, 10.1103/PhysRevD.77.04990} {\bibfield  {journal}
  {\bibinfo  {journal} {Phys. Rev.}\ }\textbf {\bibinfo {volume} {D75}},\
  \bibinfo {pages} {024005} (\bibinfo {year} {2007})},\ \bibinfo {note}
  {[Erratum: Phys. Rev.D77,04990(2008)]},\ \Eprint
  {http://arxiv.org/abs/gr-qc/0607007} {arXiv:gr-qc/0607007} \BibitemShut
  {NoStop}%
\bibitem [{\citenamefont {Hinderer}\ and\ \citenamefont
  {Flanagan}(2008)}]{HindFlan08}%
  \BibitemOpen
  \bibfield  {author} {\bibinfo {author} {\bibfnamefont {T.}~\bibnamefont
  {Hinderer}}\ and\ \bibinfo {author} {\bibfnamefont {E.~E.}\ \bibnamefont
  {Flanagan}},\ }\href {\doibase 10.1103/PhysRevD.78.064028} {\bibfield
  {journal} {\bibinfo  {journal} {Phys. Rev. D}\ }\textbf {\bibinfo {volume}
  {78}},\ \bibinfo {pages} {064028} (\bibinfo {year} {2008})},\ \Eprint
  {http://arxiv.org/abs/0805.3337} {arXiv:0805.3337} \BibitemShut {NoStop}%
\bibitem [{\citenamefont {Poisson}\ \emph {et~al.}(2011)\citenamefont
  {Poisson}, \citenamefont {Pound},\ and\ \citenamefont
  {Vega}}]{PoisPounVega11}%
  \BibitemOpen
  \bibfield  {author} {\bibinfo {author} {\bibfnamefont {E.}~\bibnamefont
  {Poisson}}, \bibinfo {author} {\bibfnamefont {A.}~\bibnamefont {Pound}}, \
  and\ \bibinfo {author} {\bibfnamefont {I.}~\bibnamefont {Vega}},\ }\href@noop
  {} {\bibfield  {journal} {\bibinfo  {journal} {Living Rev. Rel.}\ }\textbf
  {\bibinfo {volume} {14}},\ \bibinfo {pages} {7} (\bibinfo {year} {2011})},\
  \Eprint {http://arxiv.org/abs/gr-qc/1102.0529} {arXiv:gr-qc/1102.0529}
  \BibitemShut {NoStop}%
\bibitem [{\citenamefont {Detweiler}(2008)}]{Detw08}%
  \BibitemOpen
  \bibfield  {author} {\bibinfo {author} {\bibfnamefont {S.}~\bibnamefont
  {Detweiler}},\ }\href {\doibase 10.1103/PhysRevD.77.124026} {\bibfield
  {journal} {\bibinfo  {journal} {Phys. Rev. D}\ }\textbf {\bibinfo {volume}
  {77}},\ \bibinfo {pages} {124026} (\bibinfo {year} {2008})},\ \Eprint
  {http://arxiv.org/abs/0804.3529} {arXiv:0804.3529} \BibitemShut {NoStop}%
\bibitem [{\citenamefont {Barack}\ and\ \citenamefont
  {Sago}(2009)}]{BaraSago09}%
  \BibitemOpen
  \bibfield  {author} {\bibinfo {author} {\bibfnamefont {L.}~\bibnamefont
  {Barack}}\ and\ \bibinfo {author} {\bibfnamefont {N.}~\bibnamefont {Sago}},\
  }\href {\doibase 10.1103/PhysRevLett.102.191101} {\bibfield  {journal}
  {\bibinfo  {journal} {Phys. Rev. Lett.}\ }\textbf {\bibinfo {volume} {102}},\
  \bibinfo {pages} {191101} (\bibinfo {year} {2009})},\ \Eprint
  {http://arxiv.org/abs/0902.0573} {arXiv:0902.0573} \BibitemShut {NoStop}%
\bibitem [{\citenamefont {{Barack}}\ and\ \citenamefont
  {{Sago}}(2011)}]{BaraSago11}%
  \BibitemOpen
  \bibfield  {author} {\bibinfo {author} {\bibfnamefont {L.}~\bibnamefont
  {{Barack}}}\ and\ \bibinfo {author} {\bibfnamefont {N.}~\bibnamefont
  {{Sago}}},\ }\href {\doibase 10.1103/PhysRevD.83.084023} {\bibfield
  {journal} {\bibinfo  {journal} {Phys. Rev. D}\ }\textbf {\bibinfo {volume}
  {83}},\ \bibinfo {pages} {084023} (\bibinfo {year} {2011})},\ \Eprint
  {http://arxiv.org/abs/1101.3331} {arXiv:1101.3331} \BibitemShut {NoStop}%
\bibitem [{\citenamefont {{Dolan}}\ \emph {et~al.}(2014)\citenamefont
  {{Dolan}}, \citenamefont {{Warburton}}, \citenamefont {{Harte}},
  \citenamefont {{Le Tiec}}, \citenamefont {{Wardell}},\ and\ \citenamefont
  {{Barack}}}]{DolaETC14a}%
  \BibitemOpen
  \bibfield  {author} {\bibinfo {author} {\bibfnamefont {S.~R.}\ \bibnamefont
  {{Dolan}}}, \bibinfo {author} {\bibfnamefont {N.}~\bibnamefont
  {{Warburton}}}, \bibinfo {author} {\bibfnamefont {A.~I.}\ \bibnamefont
  {{Harte}}}, \bibinfo {author} {\bibfnamefont {A.}~\bibnamefont {{Le Tiec}}},
  \bibinfo {author} {\bibfnamefont {B.}~\bibnamefont {{Wardell}}}, \ and\
  \bibinfo {author} {\bibfnamefont {L.}~\bibnamefont {{Barack}}},\ }\href
  {\doibase 10.1103/PhysRevD.89.064011} {\bibfield  {journal} {\bibinfo
  {journal} {Phys. Rev. D}\ }\textbf {\bibinfo {volume} {89}},\ \bibinfo {eid}
  {064011} (\bibinfo {year} {2014})},\ \Eprint {http://arxiv.org/abs/1312.0775}
  {arXiv:1312.0775} \BibitemShut {NoStop}%
\bibitem [{\citenamefont {Dolan}\ \emph {et~al.}(2015)\citenamefont {Dolan},
  \citenamefont {Nolan}, \citenamefont {Ottewill}, \citenamefont {Warburton},\
  and\ \citenamefont {Wardell}}]{DolaETC14b}%
  \BibitemOpen
  \bibfield  {author} {\bibinfo {author} {\bibfnamefont {S.~R.}\ \bibnamefont
  {Dolan}}, \bibinfo {author} {\bibfnamefont {P.}~\bibnamefont {Nolan}},
  \bibinfo {author} {\bibfnamefont {A.~C.}\ \bibnamefont {Ottewill}}, \bibinfo
  {author} {\bibfnamefont {N.}~\bibnamefont {Warburton}}, \ and\ \bibinfo
  {author} {\bibfnamefont {B.}~\bibnamefont {Wardell}},\ }\href {\doibase
  10.1103/PhysRevD.91.023009} {\bibfield  {journal} {\bibinfo  {journal} {Phys.
  Rev. D}\ }\textbf {\bibinfo {volume} {91}},\ \bibinfo {pages} {023009}
  (\bibinfo {year} {2015})},\ \Eprint {http://arxiv.org/abs/1406.4890}
  {arXiv:1406.4890} \BibitemShut {NoStop}%
\bibitem [{\citenamefont {Nolan}\ \emph {et~al.}(2015)\citenamefont {Nolan},
  \citenamefont {Kavanagh}, \citenamefont {Dolan}, \citenamefont {Ottewill},
  \citenamefont {Warburton},\ and\ \citenamefont {Wardell}}]{Nolan:2015vpa}%
  \BibitemOpen
  \bibfield  {author} {\bibinfo {author} {\bibfnamefont {P.}~\bibnamefont
  {Nolan}}, \bibinfo {author} {\bibfnamefont {C.}~\bibnamefont {Kavanagh}},
  \bibinfo {author} {\bibfnamefont {S.~R.}\ \bibnamefont {Dolan}}, \bibinfo
  {author} {\bibfnamefont {A.~C.}\ \bibnamefont {Ottewill}}, \bibinfo {author}
  {\bibfnamefont {N.}~\bibnamefont {Warburton}}, \ and\ \bibinfo {author}
  {\bibfnamefont {B.}~\bibnamefont {Wardell}},\ }\href {\doibase
  10.1103/PhysRevD.92.123008} {\bibfield  {journal} {\bibinfo  {journal} {Phys.
  Rev. D}\ }\textbf {\bibinfo {volume} {92}},\ \bibinfo {pages} {123008}
  (\bibinfo {year} {2015})},\ \Eprint {http://arxiv.org/abs/1505.04447}
  {arXiv:1505.04447} \BibitemShut {NoStop}%
\bibitem [{\citenamefont {Akcay}\ \emph {et~al.}(2015)\citenamefont {Akcay},
  \citenamefont {Le~Tiec}, \citenamefont {Barack}, \citenamefont {Sago},\ and\
  \citenamefont {Warburton}}]{Akcay:2015pza}%
  \BibitemOpen
  \bibfield  {author} {\bibinfo {author} {\bibfnamefont {S.}~\bibnamefont
  {Akcay}}, \bibinfo {author} {\bibfnamefont {A.}~\bibnamefont {Le~Tiec}},
  \bibinfo {author} {\bibfnamefont {L.}~\bibnamefont {Barack}}, \bibinfo
  {author} {\bibfnamefont {N.}~\bibnamefont {Sago}}, \ and\ \bibinfo {author}
  {\bibfnamefont {N.}~\bibnamefont {Warburton}},\ }\href {\doibase
  10.1103/PhysRevD.91.124014} {\bibfield  {journal} {\bibinfo  {journal} {Phys.
  Rev. D}\ }\textbf {\bibinfo {volume} {91}},\ \bibinfo {pages} {124014}
  (\bibinfo {year} {2015})},\ \Eprint {http://arxiv.org/abs/1503.01374}
  {arXiv:1503.01374} \BibitemShut {NoStop}%
\bibitem [{\citenamefont {Shah}\ and\ \citenamefont
  {Pound}(2015)}]{ShahPoun15}%
  \BibitemOpen
  \bibfield  {author} {\bibinfo {author} {\bibfnamefont {A.~G.}\ \bibnamefont
  {Shah}}\ and\ \bibinfo {author} {\bibfnamefont {A.}~\bibnamefont {Pound}},\
  }\href {\doibase 10.1103/PhysRevD.91.124022} {\bibfield  {journal} {\bibinfo
  {journal} {Phys. Rev. D}\ }\textbf {\bibinfo {volume} {91}},\ \bibinfo
  {pages} {124022} (\bibinfo {year} {2015})},\ \Eprint
  {http://arxiv.org/abs/1503.02414} {arXiv:1503.02414} \BibitemShut {NoStop}%
\bibitem [{\citenamefont {{Shah}}\ \emph {et~al.}(2014)\citenamefont {{Shah}},
  \citenamefont {{Friedman}},\ and\ \citenamefont
  {{Whiting}}}]{ShahFrieWhit14}%
  \BibitemOpen
  \bibfield  {author} {\bibinfo {author} {\bibfnamefont {A.~G.}\ \bibnamefont
  {{Shah}}}, \bibinfo {author} {\bibfnamefont {J.~L.}\ \bibnamefont
  {{Friedman}}}, \ and\ \bibinfo {author} {\bibfnamefont {B.~F.}\ \bibnamefont
  {{Whiting}}},\ }\href {\doibase 10.1103/PhysRevD.89.064042} {\bibfield
  {journal} {\bibinfo  {journal} {Phys. Rev. D}\ }\textbf {\bibinfo {volume}
  {89}},\ \bibinfo {eid} {064042} (\bibinfo {year} {2014})},\ \Eprint
  {http://arxiv.org/abs/1312.1952} {arXiv:1312.1952} \BibitemShut {NoStop}%
\bibitem [{\citenamefont {Isoyama}\ \emph {et~al.}(2014)\citenamefont
  {Isoyama}, \citenamefont {Barack}, \citenamefont {Dolan}, \citenamefont
  {Le~Tiec}, \citenamefont {Nakano}, \citenamefont {Shah}, \citenamefont
  {Tanaka},\ and\ \citenamefont {Warburton}}]{Isoyama:2014mja}%
  \BibitemOpen
  \bibfield  {author} {\bibinfo {author} {\bibfnamefont {S.}~\bibnamefont
  {Isoyama}}, \bibinfo {author} {\bibfnamefont {L.}~\bibnamefont {Barack}},
  \bibinfo {author} {\bibfnamefont {S.~R.}\ \bibnamefont {Dolan}}, \bibinfo
  {author} {\bibfnamefont {A.}~\bibnamefont {Le~Tiec}}, \bibinfo {author}
  {\bibfnamefont {H.}~\bibnamefont {Nakano}}, \bibinfo {author} {\bibfnamefont
  {A.~G.}\ \bibnamefont {Shah}}, \bibinfo {author} {\bibfnamefont
  {T.}~\bibnamefont {Tanaka}}, \ and\ \bibinfo {author} {\bibfnamefont
  {N.}~\bibnamefont {Warburton}},\ }\href {\doibase
  10.1103/PhysRevLett.113.161101} {\bibfield  {journal} {\bibinfo  {journal}
  {Phys. Rev. Lett.}\ }\textbf {\bibinfo {volume} {113}},\ \bibinfo {pages}
  {161101} (\bibinfo {year} {2014})},\ \Eprint {http://arxiv.org/abs/1404.6133}
  {arXiv:1404.6133} \BibitemShut {NoStop}%
\bibitem [{\citenamefont {van~de Meent}\ and\ \citenamefont
  {Shah}(2015)}]{VandShah15}%
  \BibitemOpen
  \bibfield  {author} {\bibinfo {author} {\bibfnamefont {M.}~\bibnamefont
  {van~de Meent}}\ and\ \bibinfo {author} {\bibfnamefont {A.~G.}\ \bibnamefont
  {Shah}},\ }\href {\doibase 10.1103/PhysRevD.92.064025} {\bibfield  {journal}
  {\bibinfo  {journal} {Phys. Rev. D}\ }\textbf {\bibinfo {volume} {92}},\
  \bibinfo {pages} {064025} (\bibinfo {year} {2015})},\ \Eprint
  {http://arxiv.org/abs/1506.04755} {arXiv:1506.04755} \BibitemShut {NoStop}%
\bibitem [{\citenamefont {van~de Meent}(2017)}]{vandeMeent:2016hel}%
  \BibitemOpen
  \bibfield  {author} {\bibinfo {author} {\bibfnamefont {M.}~\bibnamefont
  {van~de Meent}},\ }\href {\doibase 10.1103/PhysRevLett.118.011101} {\bibfield
   {journal} {\bibinfo  {journal} {Phys. Rev. Lett.}\ }\textbf {\bibinfo
  {volume} {118}},\ \bibinfo {pages} {011101} (\bibinfo {year} {2017})},\
  \Eprint {http://arxiv.org/abs/1610.03497} {arXiv:1610.03497} \BibitemShut
  {NoStop}%
\bibitem [{\citenamefont {Kavanagh}\ \emph {et~al.}(2015)\citenamefont
  {Kavanagh}, \citenamefont {Ottewill},\ and\ \citenamefont
  {Wardell}}]{Kavanagh:2015lva}%
  \BibitemOpen
  \bibfield  {author} {\bibinfo {author} {\bibfnamefont {C.}~\bibnamefont
  {Kavanagh}}, \bibinfo {author} {\bibfnamefont {A.~C.}\ \bibnamefont
  {Ottewill}}, \ and\ \bibinfo {author} {\bibfnamefont {B.}~\bibnamefont
  {Wardell}},\ }\href {\doibase 10.1103/PhysRevD.92.084025} {\bibfield
  {journal} {\bibinfo  {journal} {Phys. Rev. D}\ }\textbf {\bibinfo {volume}
  {92}},\ \bibinfo {pages} {084025} (\bibinfo {year} {2015})},\ \Eprint
  {http://arxiv.org/abs/1503.02334} {arXiv:1503.02334} \BibitemShut {NoStop}%
\bibitem [{\citenamefont {{Blanchet}}\ \emph {et~al.}(2010)\citenamefont
  {{Blanchet}}, \citenamefont {{Detweiler}}, \citenamefont {{Le Tiec}},\ and\
  \citenamefont {{Whiting}}}]{BlanETC09}%
  \BibitemOpen
  \bibfield  {author} {\bibinfo {author} {\bibfnamefont {L.}~\bibnamefont
  {{Blanchet}}}, \bibinfo {author} {\bibfnamefont {S.}~\bibnamefont
  {{Detweiler}}}, \bibinfo {author} {\bibfnamefont {A.}~\bibnamefont {{Le
  Tiec}}}, \ and\ \bibinfo {author} {\bibfnamefont {B.~F.}\ \bibnamefont
  {{Whiting}}},\ }\href {\doibase 10.1103/PhysRevD.81.064004} {\bibfield
  {journal} {\bibinfo  {journal} {Phys. Rev. D}\ }\textbf {\bibinfo {volume}
  {81}},\ \bibinfo {eid} {064004} (\bibinfo {year} {2010})},\ \Eprint
  {http://arxiv.org/abs/0910.0207} {arXiv:0910.0207} \BibitemShut {NoStop}%
\bibitem [{\citenamefont {{Le Tiec}}\ \emph {et~al.}(2011)\citenamefont {{Le
  Tiec}}, \citenamefont {{Mrou{\'e}}}, \citenamefont {{Barack}}, \citenamefont
  {{Buonanno}}, \citenamefont {{Pfeiffer}}, \citenamefont {{Sago}},\ and\
  \citenamefont {{Taracchini}}}]{LetiETC11}%
  \BibitemOpen
  \bibfield  {author} {\bibinfo {author} {\bibfnamefont {A.}~\bibnamefont {{Le
  Tiec}}}, \bibinfo {author} {\bibfnamefont {A.~H.}\ \bibnamefont
  {{Mrou{\'e}}}}, \bibinfo {author} {\bibfnamefont {L.}~\bibnamefont
  {{Barack}}}, \bibinfo {author} {\bibfnamefont {A.}~\bibnamefont
  {{Buonanno}}}, \bibinfo {author} {\bibfnamefont {H.~P.}\ \bibnamefont
  {{Pfeiffer}}}, \bibinfo {author} {\bibfnamefont {N.}~\bibnamefont {{Sago}}},
  \ and\ \bibinfo {author} {\bibfnamefont {A.}~\bibnamefont {{Taracchini}}},\
  }\href {\doibase 10.1103/PhysRevLett.107.141101} {\bibfield  {journal}
  {\bibinfo  {journal} {Phys. Rev. Lett.}\ }\textbf {\bibinfo {volume} {107}},\
  \bibinfo {eid} {141101} (\bibinfo {year} {2011})},\ \Eprint
  {http://arxiv.org/abs/1106.3278} {arXiv:1106.3278} \BibitemShut {NoStop}%
\bibitem [{\citenamefont {Akcay}\ \emph {et~al.}(2012)\citenamefont {Akcay},
  \citenamefont {Barack}, \citenamefont {Damour},\ and\ \citenamefont
  {Sago}}]{Akcay:2012ea}%
  \BibitemOpen
  \bibfield  {author} {\bibinfo {author} {\bibfnamefont {S.}~\bibnamefont
  {Akcay}}, \bibinfo {author} {\bibfnamefont {L.}~\bibnamefont {Barack}},
  \bibinfo {author} {\bibfnamefont {T.}~\bibnamefont {Damour}}, \ and\ \bibinfo
  {author} {\bibfnamefont {N.}~\bibnamefont {Sago}},\ }\href {\doibase
  10.1103/PhysRevD.86.104041} {\bibfield  {journal} {\bibinfo  {journal} {Phys.
  Rev. D}\ }\textbf {\bibinfo {volume} {86}},\ \bibinfo {pages} {104041}
  (\bibinfo {year} {2012})},\ \Eprint {http://arxiv.org/abs/1209.0964}
  {arXiv:1209.0964} \BibitemShut {NoStop}%
\bibitem [{\citenamefont {Bini}\ and\ \citenamefont
  {Damour}(2014{\natexlab{a}})}]{Bini:2013rfa}%
  \BibitemOpen
  \bibfield  {author} {\bibinfo {author} {\bibfnamefont {D.}~\bibnamefont
  {Bini}}\ and\ \bibinfo {author} {\bibfnamefont {T.}~\bibnamefont {Damour}},\
  }\href {\doibase 10.1103/PhysRevD.89.064063} {\bibfield  {journal} {\bibinfo
  {journal} {Phys. Rev. D}\ }\textbf {\bibinfo {volume} {89}},\ \bibinfo
  {pages} {064063} (\bibinfo {year} {2014}{\natexlab{a}})},\ \Eprint
  {http://arxiv.org/abs/1312.2503} {arXiv:1312.2503} \BibitemShut {NoStop}%
\bibitem [{\citenamefont {Bini}\ and\ \citenamefont
  {Damour}(2014{\natexlab{b}})}]{Bini:2014ica}%
  \BibitemOpen
  \bibfield  {author} {\bibinfo {author} {\bibfnamefont {D.}~\bibnamefont
  {Bini}}\ and\ \bibinfo {author} {\bibfnamefont {T.}~\bibnamefont {Damour}},\
  }\href {\doibase 10.1103/PhysRevD.90.024039} {\bibfield  {journal} {\bibinfo
  {journal} {Phys. Rev. D}\ }\textbf {\bibinfo {volume} {90}},\ \bibinfo
  {pages} {024039} (\bibinfo {year} {2014}{\natexlab{b}})},\ \Eprint
  {http://arxiv.org/abs/1404.2747} {arXiv:1404.2747} \BibitemShut {NoStop}%
\bibitem [{\citenamefont {Bini}\ and\ \citenamefont
  {Damour}(2014{\natexlab{c}})}]{Bini:2014zxa}%
  \BibitemOpen
  \bibfield  {author} {\bibinfo {author} {\bibfnamefont {D.}~\bibnamefont
  {Bini}}\ and\ \bibinfo {author} {\bibfnamefont {T.}~\bibnamefont {Damour}},\
  }\href {\doibase 10.1103/PhysRevD.90.124037} {\bibfield  {journal} {\bibinfo
  {journal} {Phys. Rev. D}\ }\textbf {\bibinfo {volume} {90}},\ \bibinfo
  {pages} {124037} (\bibinfo {year} {2014}{\natexlab{c}})},\ \Eprint
  {http://arxiv.org/abs/1409.6933} {arXiv:1409.6933} \BibitemShut {NoStop}%
\bibitem [{\citenamefont {Zimmerman}\ \emph {et~al.}(2016)\citenamefont
  {Zimmerman}, \citenamefont {Lewis},\ and\ \citenamefont
  {Pfeiffer}}]{Zimmerman:2016ajr}%
  \BibitemOpen
  \bibfield  {author} {\bibinfo {author} {\bibfnamefont {A.}~\bibnamefont
  {Zimmerman}}, \bibinfo {author} {\bibfnamefont {A.~G.~M.}\ \bibnamefont
  {Lewis}}, \ and\ \bibinfo {author} {\bibfnamefont {H.~P.}\ \bibnamefont
  {Pfeiffer}},\ }\href {\doibase 10.1103/PhysRevLett.117.191101} {\bibfield
  {journal} {\bibinfo  {journal} {Phys. Rev. Lett.}\ }\textbf {\bibinfo
  {volume} {117}},\ \bibinfo {pages} {191101} (\bibinfo {year} {2016})},\
  \Eprint {http://arxiv.org/abs/1606.08056} {arXiv:1606.08056} \BibitemShut
  {NoStop}%
\bibitem [{\citenamefont {Akcay}\ \emph {et~al.}(2017)\citenamefont {Akcay},
  \citenamefont {Dempsey},\ and\ \citenamefont {Dolan}}]{Akcay:2016dku}%
  \BibitemOpen
  \bibfield  {author} {\bibinfo {author} {\bibfnamefont {S.}~\bibnamefont
  {Akcay}}, \bibinfo {author} {\bibfnamefont {D.}~\bibnamefont {Dempsey}}, \
  and\ \bibinfo {author} {\bibfnamefont {S.}~\bibnamefont {Dolan}},\ }\href
  {\doibase 10.1088/1361-6382/aa61d6} {\bibfield  {journal} {\bibinfo
  {journal} {Class. Quant. Grav.}\ }\textbf {\bibinfo {volume} {34}},\ \bibinfo
  {pages} {084001} (\bibinfo {year} {2017})},\ \Eprint
  {http://arxiv.org/abs/1608.04811} {arXiv:1608.04811} \BibitemShut {NoStop}%
\bibitem [{\citenamefont {Flanagan}\ and\ \citenamefont
  {Hinderer}(2012)}]{FlanHind10}%
  \BibitemOpen
  \bibfield  {author} {\bibinfo {author} {\bibfnamefont {E.~E.}\ \bibnamefont
  {Flanagan}}\ and\ \bibinfo {author} {\bibfnamefont {T.}~\bibnamefont
  {Hinderer}},\ }\href {\doibase 10.1103/PhysRevLett.109.071102} {\bibfield
  {journal} {\bibinfo  {journal} {Phys. Rev. Lett.}\ }\textbf {\bibinfo
  {volume} {109}},\ \bibinfo {pages} {071102} (\bibinfo {year} {2012})},\
  \Eprint {http://arxiv.org/abs/1009.4923} {arXiv:1009.4923} \BibitemShut
  {NoStop}%
\bibitem [{\citenamefont {Brink}\ \emph {et~al.}(2015)\citenamefont {Brink},
  \citenamefont {Geyer},\ and\ \citenamefont {Hinderer}}]{Brink:2013nna}%
  \BibitemOpen
  \bibfield  {author} {\bibinfo {author} {\bibfnamefont {J.}~\bibnamefont
  {Brink}}, \bibinfo {author} {\bibfnamefont {M.}~\bibnamefont {Geyer}}, \ and\
  \bibinfo {author} {\bibfnamefont {T.}~\bibnamefont {Hinderer}},\ }\href
  {\doibase 10.1103/PhysRevLett.114.081102} {\bibfield  {journal} {\bibinfo
  {journal} {Phys. Rev. Lett.}\ }\textbf {\bibinfo {volume} {114}},\ \bibinfo
  {pages} {081102} (\bibinfo {year} {2015})},\ \Eprint
  {http://arxiv.org/abs/1304.0330} {arXiv:1304.0330} \BibitemShut {NoStop}%
\bibitem [{\citenamefont {Ruangsri}\ and\ \citenamefont
  {Hughes}(2014)}]{Ruangsri:2013hra}%
  \BibitemOpen
  \bibfield  {author} {\bibinfo {author} {\bibfnamefont {U.}~\bibnamefont
  {Ruangsri}}\ and\ \bibinfo {author} {\bibfnamefont {S.~A.}\ \bibnamefont
  {Hughes}},\ }\href {\doibase 10.1103/PhysRevD.89.084036} {\bibfield
  {journal} {\bibinfo  {journal} {Phys. Rev. D}\ }\textbf {\bibinfo {volume}
  {89}},\ \bibinfo {pages} {084036} (\bibinfo {year} {2014})},\ \Eprint
  {http://arxiv.org/abs/1307.6483} {arXiv:1307.6483} \BibitemShut {NoStop}%
\bibitem [{\citenamefont {Pound}(2012)}]{Poun12a}%
  \BibitemOpen
  \bibfield  {author} {\bibinfo {author} {\bibfnamefont {A.}~\bibnamefont
  {Pound}},\ }\href {\doibase 10.1103/PhysRevLett.109.051101} {\bibfield
  {journal} {\bibinfo  {journal} {Phys. Rev. Lett.}\ }\textbf {\bibinfo
  {volume} {109}},\ \bibinfo {pages} {051101} (\bibinfo {year} {2012})},\
  \Eprint {http://arxiv.org/abs/1201.5089} {arXiv:1201.5089} \BibitemShut
  {NoStop}%
\bibitem [{\citenamefont {Gralla}(2012)}]{Gral12}%
  \BibitemOpen
  \bibfield  {author} {\bibinfo {author} {\bibfnamefont {S.~E.}\ \bibnamefont
  {Gralla}},\ }\href {\doibase 10.1103/PhysRevD.85.124011} {\bibfield
  {journal} {\bibinfo  {journal} {Phys. Rev. D}\ }\textbf {\bibinfo {volume}
  {85}},\ \bibinfo {pages} {124011} (\bibinfo {year} {2012})},\ \Eprint
  {http://arxiv.org/abs/1203.3189} {arXiv:1203.3189} \BibitemShut {NoStop}%
\bibitem [{\citenamefont {Detweiler}(2012)}]{Detw12}%
  \BibitemOpen
  \bibfield  {author} {\bibinfo {author} {\bibfnamefont {S.}~\bibnamefont
  {Detweiler}},\ }\href {\doibase 10.1103/PhysRevD.85.044048} {\bibfield
  {journal} {\bibinfo  {journal} {Phys. Rev. D}\ }\textbf {\bibinfo {volume}
  {85}},\ \bibinfo {pages} {044048} (\bibinfo {year} {2012})},\ \Eprint
  {http://arxiv.org/abs/1107.2098} {arXiv:1107.2098} \BibitemShut {NoStop}%
\bibitem [{\citenamefont {Pound}(2014)}]{Pound:2014koa}%
  \BibitemOpen
  \bibfield  {author} {\bibinfo {author} {\bibfnamefont {A.}~\bibnamefont
  {Pound}},\ }\href {\doibase 10.1103/PhysRevD.90.084039} {\bibfield  {journal}
  {\bibinfo  {journal} {Phys. Rev.}\ }\textbf {\bibinfo {volume} {D90}},\
  \bibinfo {pages} {084039} (\bibinfo {year} {2014})},\ \Eprint
  {http://arxiv.org/abs/1404.1543} {arXiv:1404.1543} \BibitemShut {NoStop}%
\bibitem [{\citenamefont {{Warburton}}\ and\ \citenamefont
  {{Wardell}}(2014)}]{WarbWard14}%
  \BibitemOpen
  \bibfield  {author} {\bibinfo {author} {\bibfnamefont {N.}~\bibnamefont
  {{Warburton}}}\ and\ \bibinfo {author} {\bibfnamefont {B.}~\bibnamefont
  {{Wardell}}},\ }\href {\doibase 10.1103/PhysRevD.89.044046} {\bibfield
  {journal} {\bibinfo  {journal} {Phys. Rev. D}\ }\textbf {\bibinfo {volume}
  {89}},\ \bibinfo {eid} {044046} (\bibinfo {year} {2014})},\ \Eprint
  {http://arxiv.org/abs/1311.3104} {arXiv:1311.3104} \BibitemShut {NoStop}%
\bibitem [{\citenamefont {{Pound}}\ and\ \citenamefont
  {{Miller}}(2014)}]{PounMill14}%
  \BibitemOpen
  \bibfield  {author} {\bibinfo {author} {\bibfnamefont {A.}~\bibnamefont
  {{Pound}}}\ and\ \bibinfo {author} {\bibfnamefont {J.}~\bibnamefont
  {{Miller}}},\ }\href {\doibase 10.1103/PhysRevD.89.104020} {\bibfield
  {journal} {\bibinfo  {journal} {Phys. Rev. D}\ }\textbf {\bibinfo {volume}
  {89}},\ \bibinfo {eid} {104020} (\bibinfo {year} {2014})},\ \Eprint
  {http://arxiv.org/abs/1403.1843} {arXiv:1403.1843} \BibitemShut {NoStop}%
\bibitem [{\citenamefont {Wardell}\ and\ \citenamefont
  {Warburton}(2015)}]{WardWarb15}%
  \BibitemOpen
  \bibfield  {author} {\bibinfo {author} {\bibfnamefont {B.}~\bibnamefont
  {Wardell}}\ and\ \bibinfo {author} {\bibfnamefont {N.}~\bibnamefont
  {Warburton}},\ }\href {\doibase 10.1103/PhysRevD.92.084019} {\bibfield
  {journal} {\bibinfo  {journal} {Phys. Rev. D}\ }\textbf {\bibinfo {volume}
  {92}},\ \bibinfo {pages} {084019} (\bibinfo {year} {2015})},\ \Eprint
  {http://arxiv.org/abs/1505.07841} {arXiv:1505.07841} \BibitemShut {NoStop}%
\bibitem [{\citenamefont {Pound}(2015{\natexlab{a}})}]{Pound:2015wva}%
  \BibitemOpen
  \bibfield  {author} {\bibinfo {author} {\bibfnamefont {A.}~\bibnamefont
  {Pound}},\ }\href {\doibase 10.1103/PhysRevD.92.104047} {\bibfield  {journal}
  {\bibinfo  {journal} {Phys. Rev.}\ }\textbf {\bibinfo {volume} {D92}},\
  \bibinfo {pages} {104047} (\bibinfo {year} {2015}{\natexlab{a}})},\ \Eprint
  {http://arxiv.org/abs/1510.05172} {arXiv:1510.05172} \BibitemShut {NoStop}%
\bibitem [{\citenamefont {Miller}\ \emph {et~al.}(2016)\citenamefont {Miller},
  \citenamefont {Wardell},\ and\ \citenamefont {Pound}}]{Miller:2016hjv}%
  \BibitemOpen
  \bibfield  {author} {\bibinfo {author} {\bibfnamefont {J.}~\bibnamefont
  {Miller}}, \bibinfo {author} {\bibfnamefont {B.}~\bibnamefont {Wardell}}, \
  and\ \bibinfo {author} {\bibfnamefont {A.}~\bibnamefont {Pound}},\ }\href
  {\doibase 10.1103/PhysRevD.94.104018} {\bibfield  {journal} {\bibinfo
  {journal} {Phys. Rev.}\ }\textbf {\bibinfo {volume} {D94}},\ \bibinfo {pages}
  {104018} (\bibinfo {year} {2016})},\ \Eprint
  {http://arxiv.org/abs/1608.06783} {arXiv:1608.06783} \BibitemShut {NoStop}%
\bibitem [{\citenamefont {Pound}(2017)}]{Pound:2017psq}%
  \BibitemOpen
  \bibfield  {author} {\bibinfo {author} {\bibfnamefont {A.}~\bibnamefont
  {Pound}},\ }\href {\doibase 10.1103/PhysRevD.95.104056} {\bibfield  {journal}
  {\bibinfo  {journal} {Phys. Rev.}\ }\textbf {\bibinfo {volume} {D95}},\
  \bibinfo {pages} {104056} (\bibinfo {year} {2017})},\ \Eprint
  {http://arxiv.org/abs/1703.02836} {arXiv:1703.02836} \BibitemShut {NoStop}%
\bibitem [{\citenamefont {Warburton}\ \emph {et~al.}(2012)\citenamefont
  {Warburton}, \citenamefont {Akcay}, \citenamefont {Barack}, \citenamefont
  {Gair},\ and\ \citenamefont {Sago}}]{WarbETC12}%
  \BibitemOpen
  \bibfield  {author} {\bibinfo {author} {\bibfnamefont {N.}~\bibnamefont
  {Warburton}}, \bibinfo {author} {\bibfnamefont {S.}~\bibnamefont {Akcay}},
  \bibinfo {author} {\bibfnamefont {L.}~\bibnamefont {Barack}}, \bibinfo
  {author} {\bibfnamefont {J.~R.}\ \bibnamefont {Gair}}, \ and\ \bibinfo
  {author} {\bibfnamefont {N.}~\bibnamefont {Sago}},\ }\href {\doibase
  10.1103/PhysRevD.85.061501} {\bibfield  {journal} {\bibinfo  {journal} {Phys.
  Rev. D}\ }\textbf {\bibinfo {volume} {85}},\ \bibinfo {pages} {061501}
  (\bibinfo {year} {2012})},\ \Eprint {http://arxiv.org/abs/1111.6908}
  {arXiv:1111.6908} \BibitemShut {NoStop}%
\bibitem [{\citenamefont {Osburn}\ \emph {et~al.}(2016)\citenamefont {Osburn},
  \citenamefont {Warburton},\ and\ \citenamefont {Evans}}]{OsbuWarbEvan16}%
  \BibitemOpen
  \bibfield  {author} {\bibinfo {author} {\bibfnamefont {T.}~\bibnamefont
  {Osburn}}, \bibinfo {author} {\bibfnamefont {N.}~\bibnamefont {Warburton}}, \
  and\ \bibinfo {author} {\bibfnamefont {C.~R.}\ \bibnamefont {Evans}},\ }\href
  {\doibase 10.1103/PhysRevD.93.064024} {\bibfield  {journal} {\bibinfo
  {journal} {Phys. Rev. D}\ }\textbf {\bibinfo {volume} {93}},\ \bibinfo
  {pages} {064024} (\bibinfo {year} {2016})},\ \Eprint
  {http://arxiv.org/abs/1511.01498} {arXiv:1511.01498} \BibitemShut {NoStop}%
\bibitem [{\citenamefont {{Diener}}\ \emph {et~al.}(2012)\citenamefont
  {{Diener}}, \citenamefont {{Vega}}, \citenamefont {{Wardell}},\ and\
  \citenamefont {{Detweiler}}}]{DienETC12}%
  \BibitemOpen
  \bibfield  {author} {\bibinfo {author} {\bibfnamefont {P.}~\bibnamefont
  {{Diener}}}, \bibinfo {author} {\bibfnamefont {I.}~\bibnamefont {{Vega}}},
  \bibinfo {author} {\bibfnamefont {B.}~\bibnamefont {{Wardell}}}, \ and\
  \bibinfo {author} {\bibfnamefont {S.}~\bibnamefont {{Detweiler}}},\ }\href
  {\doibase 10.1103/PhysRevLett.108.191102} {\bibfield  {journal} {\bibinfo
  {journal} {Physical Review Letters}\ }\textbf {\bibinfo {volume} {108}},\
  \bibinfo {eid} {191102} (\bibinfo {year} {2012})},\ \Eprint
  {http://arxiv.org/abs/1112.4821} {arXiv:1112.4821} \BibitemShut {NoStop}%
\bibitem [{\citenamefont {Huerta}\ and\ \citenamefont
  {Gair}(2011{\natexlab{c}})}]{Huerta:2011kt}%
  \BibitemOpen
  \bibfield  {author} {\bibinfo {author} {\bibfnamefont {E.~A.}\ \bibnamefont
  {Huerta}}\ and\ \bibinfo {author} {\bibfnamefont {J.~R.}\ \bibnamefont
  {Gair}},\ }\href {\doibase 10.1103/PhysRevD.84.064023} {\bibfield  {journal}
  {\bibinfo  {journal} {Phys. Rev.}\ }\textbf {\bibinfo {volume} {D84}},\
  \bibinfo {pages} {064023} (\bibinfo {year} {2011}{\natexlab{c}})},\ \Eprint
  {http://arxiv.org/abs/1105.3567} {arXiv:1105.3567} \BibitemShut {NoStop}%
\bibitem [{\citenamefont {Huerta}\ \emph {et~al.}(2012)\citenamefont {Huerta},
  \citenamefont {Gair},\ and\ \citenamefont {Brown}}]{Huerta:2011zi}%
  \BibitemOpen
  \bibfield  {author} {\bibinfo {author} {\bibfnamefont {E.~A.}\ \bibnamefont
  {Huerta}}, \bibinfo {author} {\bibfnamefont {J.~R.}\ \bibnamefont {Gair}}, \
  and\ \bibinfo {author} {\bibfnamefont {D.~A.}\ \bibnamefont {Brown}},\ }\href
  {\doibase 10.1103/PhysRevD.85.064023} {\bibfield  {journal} {\bibinfo
  {journal} {Phys. Rev.}\ }\textbf {\bibinfo {volume} {D85}},\ \bibinfo {pages}
  {064023} (\bibinfo {year} {2012})},\ \Eprint {http://arxiv.org/abs/1111.3243}
  {arXiv:1111.3243} \BibitemShut {NoStop}%
\bibitem [{\citenamefont {Burko}\ and\ \citenamefont
  {Khanna}(2015)}]{Burko:2015sqa}%
  \BibitemOpen
  \bibfield  {author} {\bibinfo {author} {\bibfnamefont {L.~M.}\ \bibnamefont
  {Burko}}\ and\ \bibinfo {author} {\bibfnamefont {G.}~\bibnamefont {Khanna}},\
  }\href {\doibase 10.1103/PhysRevD.91.104017} {\bibfield  {journal} {\bibinfo
  {journal} {Phys. Rev.}\ }\textbf {\bibinfo {volume} {D91}},\ \bibinfo {pages}
  {104017} (\bibinfo {year} {2015})},\ \Eprint
  {http://arxiv.org/abs/1503.05097} {arXiv:1503.05097} \BibitemShut {NoStop}%
\bibitem [{\citenamefont {Papapetrou}(1951)}]{Papa51}%
  \BibitemOpen
  \bibfield  {author} {\bibinfo {author} {\bibfnamefont {A.}~\bibnamefont
  {Papapetrou}},\ }\href@noop {} {\bibfield  {journal} {\bibinfo  {journal}
  {Proc. R. Soc. Lond.}\ }\textbf {\bibinfo {volume} {A209}},\ \bibinfo {pages}
  {248} (\bibinfo {year} {1951})}\BibitemShut {NoStop}%
\bibitem [{\citenamefont {Dixon}(1970)}]{Dixo70}%
  \BibitemOpen
  \bibfield  {author} {\bibinfo {author} {\bibfnamefont {W.~G.}\ \bibnamefont
  {Dixon}},\ }\href {\doibase 10.1098/rspa.1970.0020} {\bibfield  {journal}
  {\bibinfo  {journal} {Royal Society of London Proceedings Series A}\ }\textbf
  {\bibinfo {volume} {314}},\ \bibinfo {pages} {499} (\bibinfo {year}
  {1970})}\BibitemShut {NoStop}%
\bibitem [{\citenamefont {Hughes}\ \emph {et~al.}(2005)\citenamefont {Hughes},
  \citenamefont {Drasco}, \citenamefont {Flanagan},\ and\ \citenamefont
  {Franklin}}]{HughETC05}%
  \BibitemOpen
  \bibfield  {author} {\bibinfo {author} {\bibfnamefont {S.~A.}\ \bibnamefont
  {Hughes}}, \bibinfo {author} {\bibfnamefont {S.}~\bibnamefont {Drasco}},
  \bibinfo {author} {\bibfnamefont {E.~E.}\ \bibnamefont {Flanagan}}, \ and\
  \bibinfo {author} {\bibfnamefont {J.}~\bibnamefont {Franklin}},\ }\href
  {\doibase 10.1103/PhysRevLett.94.221101} {\bibfield  {journal} {\bibinfo
  {journal} {Phys. Rev. Lett.}\ }\textbf {\bibinfo {volume} {94}},\ \bibinfo
  {pages} {221101} (\bibinfo {year} {2005})},\ \Eprint
  {http://arxiv.org/abs/gr-qc/0504015} {arXiv:gr-qc/0504015} \BibitemShut
  {NoStop}%
\bibitem [{\citenamefont {Drasco}\ \emph {et~al.}(2005)\citenamefont {Drasco},
  \citenamefont {Flanagan},\ and\ \citenamefont {Hughes}}]{DrasFlanHugh05}%
  \BibitemOpen
  \bibfield  {author} {\bibinfo {author} {\bibfnamefont {S.}~\bibnamefont
  {Drasco}}, \bibinfo {author} {\bibfnamefont {E.~E.}\ \bibnamefont
  {Flanagan}}, \ and\ \bibinfo {author} {\bibfnamefont {S.~A.}\ \bibnamefont
  {Hughes}},\ }\href@noop {} {\bibfield  {journal} {\bibinfo  {journal}
  {Classical and Quantum Gravity}\ }\textbf {\bibinfo {volume} {22}},\ \bibinfo
  {pages} {S801} (\bibinfo {year} {2005})}\BibitemShut {NoStop}%
\bibitem [{\citenamefont {{Tanaka}}(2006)}]{Tanaka:2006}%
  \BibitemOpen
  \bibfield  {author} {\bibinfo {author} {\bibfnamefont {T.}~\bibnamefont
  {{Tanaka}}},\ }\href {\doibase 10.1143/PTPS.163.120} {\bibfield  {journal}
  {\bibinfo  {journal} {Progress of Theoretical Physics Supplement}\ }\textbf
  {\bibinfo {volume} {163}},\ \bibinfo {pages} {120} (\bibinfo {year}
  {2006})},\ \Eprint {http://arxiv.org/abs/gr-qc/0508114} {gr-qc/0508114}
  \BibitemShut {NoStop}%
\bibitem [{\citenamefont {{Drasco}}\ and\ \citenamefont
  {{Hughes}}(2006)}]{DrasHugh06}%
  \BibitemOpen
  \bibfield  {author} {\bibinfo {author} {\bibfnamefont {S.}~\bibnamefont
  {{Drasco}}}\ and\ \bibinfo {author} {\bibfnamefont {S.~A.}\ \bibnamefont
  {{Hughes}}},\ }\href {\doibase 10.1103/PhysRevD.73.024027} {\bibfield
  {journal} {\bibinfo  {journal} {Phys. Rev. D}\ }\textbf {\bibinfo {volume}
  {73}},\ \bibinfo {eid} {024027} (\bibinfo {year} {2006})},\ \Eprint
  {http://arxiv.org/abs/gr-qc/0509101} {gr-qc/0509101} \BibitemShut {NoStop}%
\bibitem [{\citenamefont {Hartl}(2003)}]{Hartl:2002ig}%
  \BibitemOpen
  \bibfield  {author} {\bibinfo {author} {\bibfnamefont {M.~D.}\ \bibnamefont
  {Hartl}},\ }\href {\doibase 10.1103/PhysRevD.67.024005} {\bibfield  {journal}
  {\bibinfo  {journal} {Phys. Rev.}\ }\textbf {\bibinfo {volume} {D67}},\
  \bibinfo {pages} {024005} (\bibinfo {year} {2003})},\ \Eprint
  {http://arxiv.org/abs/gr-qc/0210042} {arXiv:gr-qc/0210042} \BibitemShut
  {NoStop}%
\bibitem [{\citenamefont {Akcay}\ \emph {et~al.}(2013)\citenamefont {Akcay},
  \citenamefont {Warburton},\ and\ \citenamefont {Barack}}]{AkcaWarbBara13}%
  \BibitemOpen
  \bibfield  {author} {\bibinfo {author} {\bibfnamefont {S.}~\bibnamefont
  {Akcay}}, \bibinfo {author} {\bibfnamefont {N.}~\bibnamefont {Warburton}}, \
  and\ \bibinfo {author} {\bibfnamefont {L.}~\bibnamefont {Barack}},\ }\href
  {\doibase 10.1103/PhysRevD.88.104009} {\bibfield  {journal} {\bibinfo
  {journal} {Phys. Rev. D}\ }\textbf {\bibinfo {volume} {88}},\ \bibinfo
  {pages} {104009} (\bibinfo {year} {2013})},\ \Eprint
  {http://arxiv.org/abs/1308.5223} {arXiv:1308.5223} \BibitemShut {NoStop}%
\bibitem [{\citenamefont {Osburn}\ \emph {et~al.}(2014)\citenamefont {Osburn},
  \citenamefont {Forseth}, \citenamefont {Evans},\ and\ \citenamefont
  {Hopper}}]{OsbuETC14}%
  \BibitemOpen
  \bibfield  {author} {\bibinfo {author} {\bibfnamefont {T.}~\bibnamefont
  {Osburn}}, \bibinfo {author} {\bibfnamefont {E.}~\bibnamefont {Forseth}},
  \bibinfo {author} {\bibfnamefont {C.~R.}\ \bibnamefont {Evans}}, \ and\
  \bibinfo {author} {\bibfnamefont {S.}~\bibnamefont {Hopper}},\ }\href
  {\doibase 10.1103/PhysRevD.90.104031} {\bibfield  {journal} {\bibinfo
  {journal} {Phys. Rev. D}\ }\textbf {\bibinfo {volume} {90}},\ \bibinfo
  {pages} {104031} (\bibinfo {year} {2014})},\ \Eprint
  {http://arxiv.org/abs/1409.4419} {arXiv:1409.4419} \BibitemShut {NoStop}%
\bibitem [{\citenamefont {Pound}(2015{\natexlab{b}})}]{Pound:2015tma}%
  \BibitemOpen
  \bibfield  {author} {\bibinfo {author} {\bibfnamefont {A.}~\bibnamefont
  {Pound}},\ }\bibfield  {booktitle} {\emph {\bibinfo {booktitle}
  {{Proceedings, 524th WE-Heraeus-Seminar: Equations of Motion in Relativistic
  Gravity (EOM 2013)}}},\ }\href {\doibase 10.1007/978-3-319-18335-0_13}
  {\bibfield  {journal} {\bibinfo  {journal} {Fund. Theor. Phys.}\ }\textbf
  {\bibinfo {volume} {179}},\ \bibinfo {pages} {399} (\bibinfo {year}
  {2015}{\natexlab{b}})},\ \Eprint {http://arxiv.org/abs/1506.06245}
  {arXiv:1506.06245} \BibitemShut {NoStop}%
\bibitem [{\citenamefont {Warburton}(2013)}]{Warb13}%
  \BibitemOpen
  \bibfield  {author} {\bibinfo {author} {\bibfnamefont {N.}~\bibnamefont
  {Warburton}}\ }(\bibinfo {year} {2013})\ \bibinfo {note} {{, Talk presented
  at the 16th Capra meeting held at University College Dublin,
  \url{http://maths.ucd.ie/capra16/schedule/}}}\BibitemShut {NoStop}%
\bibitem [{\citenamefont {Warburton}(2014)}]{Warb14a}%
  \BibitemOpen
  \bibfield  {author} {\bibinfo {author} {\bibfnamefont {N.}~\bibnamefont
  {Warburton}}\ }(\bibinfo {year} {2014})\ \bibinfo {note} {{, Talk presented
  at the 17th Capra meeting held at Caltech,
  \url{http://www.tapir.caltech.edu/~capra17/Schedule.shtml}}}\BibitemShut
  {NoStop}%
\bibitem [{\citenamefont {Diener}(2015)}]{Dien15}%
  \BibitemOpen
  \bibfield  {author} {\bibinfo {author} {\bibfnamefont {P.}~\bibnamefont
  {Diener}}\ }(\bibinfo {year} {2015})\ \bibinfo {note} {{, Talk presented at
  the 18th Capra meeting held at Kyoto University,
  \url{http://www2.yukawa.kyoto-u.ac.jp/~capra18/}}}\BibitemShut {NoStop}%
\bibitem [{\citenamefont {Pound}\ and\ \citenamefont
  {Poisson}(2008)}]{PounPois08b}%
  \BibitemOpen
  \bibfield  {author} {\bibinfo {author} {\bibfnamefont {A.}~\bibnamefont
  {Pound}}\ and\ \bibinfo {author} {\bibfnamefont {E.}~\bibnamefont
  {Poisson}},\ }\href {\doibase 10.1103/PhysRevD.77.044013} {\bibfield
  {journal} {\bibinfo  {journal} {Phys. Rev. D}\ }\textbf {\bibinfo {volume}
  {77}},\ \bibinfo {pages} {044013} (\bibinfo {year} {2008})},\ \Eprint
  {http://arxiv.org/abs/0708.3033} {arXiv:0708.3033} \BibitemShut {NoStop}%
\bibitem [{\citenamefont {{Gair}}\ \emph {et~al.}(2011)\citenamefont {{Gair}},
  \citenamefont {{Flanagan}}, \citenamefont {{Drasco}}, \citenamefont
  {{Hinderer}},\ and\ \citenamefont {{Babak}}}]{GairETC11}%
  \BibitemOpen
  \bibfield  {author} {\bibinfo {author} {\bibfnamefont {J.~R.}\ \bibnamefont
  {{Gair}}}, \bibinfo {author} {\bibfnamefont {{\'E}.~{\'E}.}\ \bibnamefont
  {{Flanagan}}}, \bibinfo {author} {\bibfnamefont {S.}~\bibnamefont
  {{Drasco}}}, \bibinfo {author} {\bibfnamefont {T.}~\bibnamefont
  {{Hinderer}}}, \ and\ \bibinfo {author} {\bibfnamefont {S.}~\bibnamefont
  {{Babak}}},\ }\href {\doibase 10.1103/PhysRevD.83.044037} {\bibfield
  {journal} {\bibinfo  {journal} {Phys. Rev. D}\ }\textbf {\bibinfo {volume}
  {83}},\ \bibinfo {eid} {044037} (\bibinfo {year} {2011})},\ \Eprint
  {http://arxiv.org/abs/1012.5111} {arXiv:1012.5111} \BibitemShut {NoStop}%
\bibitem [{\citenamefont {Harms}\ \emph
  {et~al.}(2016{\natexlab{a}})\citenamefont {Harms}, \citenamefont
  {Lukes-Gerakopoulos}, \citenamefont {Bernuzzi},\ and\ \citenamefont
  {Nagar}}]{Harms:2016ctx}%
  \BibitemOpen
  \bibfield  {author} {\bibinfo {author} {\bibfnamefont {E.}~\bibnamefont
  {Harms}}, \bibinfo {author} {\bibfnamefont {G.}~\bibnamefont
  {Lukes-Gerakopoulos}}, \bibinfo {author} {\bibfnamefont {S.}~\bibnamefont
  {Bernuzzi}}, \ and\ \bibinfo {author} {\bibfnamefont {A.}~\bibnamefont
  {Nagar}},\ }\href {\doibase 10.1103/PhysRevD.94.104010} {\bibfield  {journal}
  {\bibinfo  {journal} {Phys. Rev.}\ }\textbf {\bibinfo {volume} {D94}},\
  \bibinfo {pages} {104010} (\bibinfo {year} {2016}{\natexlab{a}})},\ \Eprint
  {http://arxiv.org/abs/1609.00356} {arXiv:1609.00356} \BibitemShut {NoStop}%
\bibitem [{\citenamefont {Harms}\ \emph
  {et~al.}(2016{\natexlab{b}})\citenamefont {Harms}, \citenamefont
  {Lukes-Gerakopoulos}, \citenamefont {Bernuzzi},\ and\ \citenamefont
  {Nagar}}]{Harms:2015ixa}%
  \BibitemOpen
  \bibfield  {author} {\bibinfo {author} {\bibfnamefont {E.}~\bibnamefont
  {Harms}}, \bibinfo {author} {\bibfnamefont {G.}~\bibnamefont
  {Lukes-Gerakopoulos}}, \bibinfo {author} {\bibfnamefont {S.}~\bibnamefont
  {Bernuzzi}}, \ and\ \bibinfo {author} {\bibfnamefont {A.}~\bibnamefont
  {Nagar}},\ }\href {\doibase 10.1103/PhysRevD.93.044015} {\bibfield  {journal}
  {\bibinfo  {journal} {Phys. Rev.}\ }\textbf {\bibinfo {volume} {D93}},\
  \bibinfo {pages} {044015} (\bibinfo {year} {2016}{\natexlab{b}})},\ \Eprint
  {http://arxiv.org/abs/1510.05548} {arXiv:1510.05548} \BibitemShut {NoStop}%
\bibitem [{\citenamefont {Han}(2010)}]{Han:2010tp}%
  \BibitemOpen
  \bibfield  {author} {\bibinfo {author} {\bibfnamefont {W.-B.}\ \bibnamefont
  {Han}},\ }\href {\doibase 10.1103/PhysRevD.82.084013} {\bibfield  {journal}
  {\bibinfo  {journal} {Phys. Rev.}\ }\textbf {\bibinfo {volume} {D82}},\
  \bibinfo {pages} {084013} (\bibinfo {year} {2010})},\ \Eprint
  {http://arxiv.org/abs/1008.3324} {arXiv:1008.3324 [gr-qc]} \BibitemShut
  {NoStop}%
\bibitem [{\citenamefont {Cutler}\ \emph {et~al.}(1994)\citenamefont {Cutler},
  \citenamefont {Kennefick},\ and\ \citenamefont {Poisson}}]{CutlKennPois94}%
  \BibitemOpen
  \bibfield  {author} {\bibinfo {author} {\bibfnamefont {C.}~\bibnamefont
  {Cutler}}, \bibinfo {author} {\bibfnamefont {D.}~\bibnamefont {Kennefick}}, \
  and\ \bibinfo {author} {\bibfnamefont {E.}~\bibnamefont {Poisson}},\ }\href
  {\doibase 10.1103/PhysRevD.50.3816} {\bibfield  {journal} {\bibinfo
  {journal} {Phys. Rev. D}\ }\textbf {\bibinfo {volume} {50}},\ \bibinfo
  {pages} {3816} (\bibinfo {year} {1994})}\BibitemShut {NoStop}%
\bibitem [{\citenamefont {{Darwin}}(1959)}]{Darw59}%
  \BibitemOpen
  \bibfield  {author} {\bibinfo {author} {\bibfnamefont {C.}~\bibnamefont
  {{Darwin}}},\ }\href {\doibase 10.1098/rspa.1959.0015} {\bibfield  {journal}
  {\bibinfo  {journal} {Proc. R. Soc. Lond. A}\ }\textbf {\bibinfo {volume}
  {249}},\ \bibinfo {pages} {180} (\bibinfo {year} {1959})}\BibitemShut
  {NoStop}%
\bibitem [{\citenamefont {{Fujita}}\ and\ \citenamefont
  {{Hikida}}(2009)}]{FujiHiki09}%
  \BibitemOpen
  \bibfield  {author} {\bibinfo {author} {\bibfnamefont {R.}~\bibnamefont
  {{Fujita}}}\ and\ \bibinfo {author} {\bibfnamefont {W.}~\bibnamefont
  {{Hikida}}},\ }\href {\doibase 10.1088/0264-9381/26/13/135002} {\bibfield
  {journal} {\bibinfo  {journal} {Classical and Quantum Gravity}\ }\textbf
  {\bibinfo {volume} {26}},\ \bibinfo {eid} {135002} (\bibinfo {year}
  {2009})},\ \Eprint {http://arxiv.org/abs/0906.1420} {arXiv:0906.1420}
  \BibitemShut {NoStop}%
\bibitem [{\citenamefont {Detweiler}\ and\ \citenamefont
  {Poisson}(2004)}]{DetwPois04}%
  \BibitemOpen
  \bibfield  {author} {\bibinfo {author} {\bibfnamefont {S.~L.}\ \bibnamefont
  {Detweiler}}\ and\ \bibinfo {author} {\bibfnamefont {E.}~\bibnamefont
  {Poisson}},\ }\href {\doibase 10.1103/PhysRevD.69.084019} {\bibfield
  {journal} {\bibinfo  {journal} {Phys. Rev. D}\ }\textbf {\bibinfo {volume}
  {69}},\ \bibinfo {pages} {084019} (\bibinfo {year} {2004})},\ \Eprint
  {http://arxiv.org/abs/gr-qc/0312010} {arXiv:gr-qc/0312010} \BibitemShut
  {NoStop}%
\bibitem [{\citenamefont {Barack}\ and\ \citenamefont
  {Lousto}(2005)}]{BaraLous05}%
  \BibitemOpen
  \bibfield  {author} {\bibinfo {author} {\bibfnamefont {L.}~\bibnamefont
  {Barack}}\ and\ \bibinfo {author} {\bibfnamefont {C.~O.}\ \bibnamefont
  {Lousto}},\ }\href {\doibase 10.1103/PhysRevD.72.104026} {\bibfield
  {journal} {\bibinfo  {journal} {Phys. Rev. D}\ }\textbf {\bibinfo {volume}
  {72}},\ \bibinfo {pages} {104026} (\bibinfo {year} {2005})},\ \Eprint
  {http://arxiv.org/abs/gr-qc/0510019} {arXiv:gr-qc/0510019} \BibitemShut
  {NoStop}%
\bibitem [{\citenamefont {Berndston}(2007)}]{Bern07}%
  \BibitemOpen
  \bibfield  {author} {\bibinfo {author} {\bibfnamefont {M.}~\bibnamefont
  {Berndston}},\ }\emph {\bibinfo {title} {Harmonic Gauge Perturbations of the
  Schwarzschild Metric}},\ \href@noop {} {Ph.D. thesis},\ \bibinfo  {school}
  {University of Colorado} (\bibinfo {year} {2007}),\ \Eprint
  {http://arxiv.org/abs/0904.0033v1} {arXiv:0904.0033v1} \BibitemShut {NoStop}%
\bibitem [{\citenamefont {Barack}\ and\ \citenamefont
  {Sago}(2010)}]{BaraSago10}%
  \BibitemOpen
  \bibfield  {author} {\bibinfo {author} {\bibfnamefont {L.}~\bibnamefont
  {Barack}}\ and\ \bibinfo {author} {\bibfnamefont {N.}~\bibnamefont {Sago}},\
  }\href@noop {} {\bibfield  {journal} {\bibinfo  {journal} {Phys. Rev. D}\
  }\textbf {\bibinfo {volume} {81}},\ \bibinfo {pages} {084021} (\bibinfo
  {year} {2010})},\ \Eprint {http://arxiv.org/abs/1002.2386} {arXiv:1002.2386}
  \BibitemShut {NoStop}%
\bibitem [{\citenamefont {Hopper}\ and\ \citenamefont
  {Evans}(2010)}]{HoppEvan10}%
  \BibitemOpen
  \bibfield  {author} {\bibinfo {author} {\bibfnamefont {S.}~\bibnamefont
  {Hopper}}\ and\ \bibinfo {author} {\bibfnamefont {C.~R.}\ \bibnamefont
  {Evans}},\ }\href {\doibase 10.1103/PhysRevD.82.084010} {\bibfield  {journal}
  {\bibinfo  {journal} {Phys. Rev. D}\ }\textbf {\bibinfo {volume} {82}},\
  \bibinfo {pages} {084010} (\bibinfo {year} {2010})},\ \Eprint
  {http://arxiv.org/abs/1006.4907} {arXiv:1006.4907} \BibitemShut {NoStop}%
\bibitem [{\citenamefont {Akcay}(2011)}]{Akca11}%
  \BibitemOpen
  \bibfield  {author} {\bibinfo {author} {\bibfnamefont {S.}~\bibnamefont
  {Akcay}},\ }\href {\doibase 10.1103/PhysRevD.83.124026} {\bibfield  {journal}
  {\bibinfo  {journal} {Phys. Rev. D}\ }\textbf {\bibinfo {volume} {83}},\
  \bibinfo {pages} {124026} (\bibinfo {year} {2011})}\BibitemShut {NoStop}%
\bibitem [{\citenamefont {{Hopper}}\ and\ \citenamefont
  {{Evans}}(2013)}]{HoppEvan13}%
  \BibitemOpen
  \bibfield  {author} {\bibinfo {author} {\bibfnamefont {S.}~\bibnamefont
  {{Hopper}}}\ and\ \bibinfo {author} {\bibfnamefont {C.~R.}\ \bibnamefont
  {{Evans}}},\ }\href {\doibase 10.1103/PhysRevD.87.064008} {\bibfield
  {journal} {\bibinfo  {journal} {Phys. Rev. D}\ }\textbf {\bibinfo {volume}
  {87}},\ \bibinfo {eid} {064008} (\bibinfo {year} {2013})},\ \Eprint
  {http://arxiv.org/abs/1210.7969} {arXiv:1210.7969} \BibitemShut {NoStop}%
\bibitem [{\citenamefont {{Hopper}}\ \emph {et~al.}(2015)\citenamefont
  {{Hopper}}, \citenamefont {{Forseth}}, \citenamefont {{Osburn}},\ and\
  \citenamefont {{Evans}}}]{HoppETC15}%
  \BibitemOpen
  \bibfield  {author} {\bibinfo {author} {\bibfnamefont {S.}~\bibnamefont
  {{Hopper}}}, \bibinfo {author} {\bibfnamefont {E.}~\bibnamefont {{Forseth}}},
  \bibinfo {author} {\bibfnamefont {T.}~\bibnamefont {{Osburn}}}, \ and\
  \bibinfo {author} {\bibfnamefont {C.~R.}\ \bibnamefont {{Evans}}},\ }\href
  {\doibase 10.1103/PhysRevD.92.044048} {\bibfield  {journal} {\bibinfo
  {journal} {Phys. Rev. D}\ }\textbf {\bibinfo {volume} {92}},\ \bibinfo {eid}
  {044048} (\bibinfo {year} {2015})},\ \Eprint
  {http://arxiv.org/abs/1506.04742} {arXiv:1506.04742} \BibitemShut {NoStop}%
\bibitem [{\citenamefont {Mino}\ \emph {et~al.}(1997)\citenamefont {Mino},
  \citenamefont {Sasaki},\ and\ \citenamefont {Tanaka}}]{MinoSasaTana97}%
  \BibitemOpen
  \bibfield  {author} {\bibinfo {author} {\bibfnamefont {Y.}~\bibnamefont
  {Mino}}, \bibinfo {author} {\bibfnamefont {M.}~\bibnamefont {Sasaki}}, \ and\
  \bibinfo {author} {\bibfnamefont {T.}~\bibnamefont {Tanaka}},\ }\href@noop {}
  {\bibfield  {journal} {\bibinfo  {journal} {Phys. Rev. D}\ }\textbf {\bibinfo
  {volume} {55}},\ \bibinfo {pages} {3457} (\bibinfo {year}
  {1997})}\BibitemShut {NoStop}%
\bibitem [{\citenamefont {Quinn}\ and\ \citenamefont
  {Wald}(1997)}]{QuinWald97}%
  \BibitemOpen
  \bibfield  {author} {\bibinfo {author} {\bibfnamefont {T.~C.}\ \bibnamefont
  {Quinn}}\ and\ \bibinfo {author} {\bibfnamefont {R.~M.}\ \bibnamefont
  {Wald}},\ }\href {\doibase 10.1103/PhysRevD.56.3381} {\bibfield  {journal}
  {\bibinfo  {journal} {Phys. Rev. D}\ }\textbf {\bibinfo {volume} {56}},\
  \bibinfo {pages} {3381} (\bibinfo {year} {1997})}\BibitemShut {NoStop}%
\bibitem [{\citenamefont {Barack}\ and\ \citenamefont {Ori}(2000)}]{BaraOri00}%
  \BibitemOpen
  \bibfield  {author} {\bibinfo {author} {\bibfnamefont {L.}~\bibnamefont
  {Barack}}\ and\ \bibinfo {author} {\bibfnamefont {A.}~\bibnamefont {Ori}},\
  }\href {\doibase 10.1103/PhysRevD.61.061502} {\bibfield  {journal} {\bibinfo
  {journal} {Phys. Rev. D}\ }\textbf {\bibinfo {volume} {61}},\ \bibinfo
  {pages} {061502} (\bibinfo {year} {2000})},\ \Eprint
  {http://arxiv.org/abs/gr-qc/9912010} {arXiv:gr-qc/9912010} \BibitemShut
  {NoStop}%
\bibitem [{\citenamefont {Barack}(2001)}]{Bara01}%
  \BibitemOpen
  \bibfield  {author} {\bibinfo {author} {\bibfnamefont {L.}~\bibnamefont
  {Barack}},\ }\href {\doibase 10.1103/PhysRevD.64.084021} {\bibfield
  {journal} {\bibinfo  {journal} {Phys. Rev.}\ }\textbf {\bibinfo {volume}
  {D64}},\ \bibinfo {pages} {084021} (\bibinfo {year} {2001})},\ \Eprint
  {http://arxiv.org/abs/gr-qc/0105040} {arXiv:gr-qc/0105040} \BibitemShut
  {NoStop}%
\bibitem [{\citenamefont {{Barack}}\ \emph {et~al.}(2002)\citenamefont
  {{Barack}}, \citenamefont {{Mino}}, \citenamefont {{Nakano}}, \citenamefont
  {{Ori}},\ and\ \citenamefont {{Sasaki}}}]{BaraETC02}%
  \BibitemOpen
  \bibfield  {author} {\bibinfo {author} {\bibfnamefont {L.}~\bibnamefont
  {{Barack}}}, \bibinfo {author} {\bibfnamefont {Y.}~\bibnamefont {{Mino}}},
  \bibinfo {author} {\bibfnamefont {H.}~\bibnamefont {{Nakano}}}, \bibinfo
  {author} {\bibfnamefont {A.}~\bibnamefont {{Ori}}}, \ and\ \bibinfo {author}
  {\bibfnamefont {M.}~\bibnamefont {{Sasaki}}},\ }\href {\doibase
  10.1103/PhysRevLett.88.091101} {\bibfield  {journal} {\bibinfo  {journal}
  {Physical Review Letters}\ }\textbf {\bibinfo {volume} {88}},\ \bibinfo {eid}
  {091101} (\bibinfo {year} {2002})},\ \Eprint
  {http://arxiv.org/abs/gr-qc/0111001} {gr-qc/0111001} \BibitemShut {NoStop}%
\bibitem [{\citenamefont {Barack}\ and\ \citenamefont {Ori}(2003)}]{BaraOri03}%
  \BibitemOpen
  \bibfield  {author} {\bibinfo {author} {\bibfnamefont {L.}~\bibnamefont
  {Barack}}\ and\ \bibinfo {author} {\bibfnamefont {A.}~\bibnamefont {Ori}},\
  }\href {\doibase 10.1103/PhysRevD.67.024029} {\bibfield  {journal} {\bibinfo
  {journal} {Phys. Rev.}\ }\textbf {\bibinfo {volume} {D67}},\ \bibinfo {pages}
  {024029} (\bibinfo {year} {2003})},\ \Eprint
  {http://arxiv.org/abs/gr-qc/0209072} {arXiv:gr-qc/0209072} \BibitemShut
  {NoStop}%
\bibitem [{\citenamefont {Detweiler}\ \emph {et~al.}(2003)\citenamefont
  {Detweiler}, \citenamefont {Messaritaki},\ and\ \citenamefont
  {Whiting}}]{DetwMessWhit03}%
  \BibitemOpen
  \bibfield  {author} {\bibinfo {author} {\bibfnamefont {S.}~\bibnamefont
  {Detweiler}}, \bibinfo {author} {\bibfnamefont {E.}~\bibnamefont
  {Messaritaki}}, \ and\ \bibinfo {author} {\bibfnamefont {B.~F.}\ \bibnamefont
  {Whiting}},\ }\href {\doibase 10.1103/PhysRevD.67.104016} {\bibfield
  {journal} {\bibinfo  {journal} {Phys. Rev. D}\ }\textbf {\bibinfo {volume}
  {67}},\ \bibinfo {pages} {104016} (\bibinfo {year} {2003})},\ \Eprint
  {http://arxiv.org/abs/gr-qc/0205079} {arXiv:gr-qc/0205079} \BibitemShut
  {NoStop}%
\bibitem [{\citenamefont {Heffernan}\ \emph {et~al.}(2012)\citenamefont
  {Heffernan}, \citenamefont {Ottewill},\ and\ \citenamefont
  {Wardell}}]{HeffOtteWard12a}%
  \BibitemOpen
  \bibfield  {author} {\bibinfo {author} {\bibfnamefont {A.}~\bibnamefont
  {Heffernan}}, \bibinfo {author} {\bibfnamefont {A.}~\bibnamefont {Ottewill}},
  \ and\ \bibinfo {author} {\bibfnamefont {B.}~\bibnamefont {Wardell}},\ }\href
  {\doibase 10.1103/PhysRevD.86.104023} {\bibfield  {journal} {\bibinfo
  {journal} {Phys. Rev. D}\ }\textbf {\bibinfo {volume} {86}},\ \bibinfo
  {pages} {104023} (\bibinfo {year} {2012})},\ \Eprint
  {http://arxiv.org/abs/1204.0794} {arXiv:1204.0794} \BibitemShut {NoStop}%
\bibitem [{\citenamefont {Heffernan}\ \emph {et~al.}(2014)\citenamefont
  {Heffernan}, \citenamefont {Ottewill},\ and\ \citenamefont
  {Wardell}}]{HeffOtteWard12b}%
  \BibitemOpen
  \bibfield  {author} {\bibinfo {author} {\bibfnamefont {A.}~\bibnamefont
  {Heffernan}}, \bibinfo {author} {\bibfnamefont {A.}~\bibnamefont {Ottewill}},
  \ and\ \bibinfo {author} {\bibfnamefont {B.}~\bibnamefont {Wardell}},\ }\href
  {\doibase 10.1103/PhysRevD.89.024030} {\bibfield  {journal} {\bibinfo
  {journal} {Phys. Rev. D}\ }\textbf {\bibinfo {volume} {89}},\ \bibinfo
  {pages} {024030} (\bibinfo {year} {2014})},\ \Eprint
  {http://arxiv.org/abs/1211.6446} {arXiv:1211.6446} \BibitemShut {NoStop}%
\bibitem [{\citenamefont {Pound}\ \emph {et~al.}(2014)\citenamefont {Pound},
  \citenamefont {Merlin},\ and\ \citenamefont {Barack}}]{PounMerlBara13}%
  \BibitemOpen
  \bibfield  {author} {\bibinfo {author} {\bibfnamefont {A.}~\bibnamefont
  {Pound}}, \bibinfo {author} {\bibfnamefont {C.}~\bibnamefont {Merlin}}, \
  and\ \bibinfo {author} {\bibfnamefont {L.}~\bibnamefont {Barack}},\ }\href
  {\doibase 10.1103/PhysRevD.89.024009} {\bibfield  {journal} {\bibinfo
  {journal} {Phys. Rev. D}\ }\textbf {\bibinfo {volume} {89}},\ \bibinfo
  {pages} {024009} (\bibinfo {year} {2014})},\ \Eprint
  {http://arxiv.org/abs/1310.1513} {arXiv:1310.1513} \BibitemShut {NoStop}%
\bibitem [{\citenamefont {Barack}\ and\ \citenamefont
  {Sago}(2007)}]{BaraSago07}%
  \BibitemOpen
  \bibfield  {author} {\bibinfo {author} {\bibfnamefont {L.}~\bibnamefont
  {Barack}}\ and\ \bibinfo {author} {\bibfnamefont {N.}~\bibnamefont {Sago}},\
  }\href {\doibase 10.1103/PhysRevD.75.064021} {\bibfield  {journal} {\bibinfo
  {journal} {Phys. Rev. D}\ }\textbf {\bibinfo {volume} {75}},\ \bibinfo
  {pages} {064021} (\bibinfo {year} {2007})},\ \Eprint
  {http://arxiv.org/abs/gr-qc/0701069} {arXiv:gr-qc/0701069} \BibitemShut
  {NoStop}%
\bibitem [{\citenamefont {Dolan}\ and\ \citenamefont
  {Barack}(2013)}]{DolaBara13}%
  \BibitemOpen
  \bibfield  {author} {\bibinfo {author} {\bibfnamefont {S.~R.}\ \bibnamefont
  {Dolan}}\ and\ \bibinfo {author} {\bibfnamefont {L.}~\bibnamefont {Barack}},\
  }\href {\doibase 10.1103/PhysRevD.87.084066} {\bibfield  {journal} {\bibinfo
  {journal} {Phys. Rev. D}\ }\textbf {\bibinfo {volume} {87}},\ \bibinfo
  {pages} {084066} (\bibinfo {year} {2013})},\ \Eprint
  {http://arxiv.org/abs/1211.4586} {arXiv:1211.4586} \BibitemShut {NoStop}%
\bibitem [{\citenamefont {Regge}\ and\ \citenamefont
  {Wheeler}(1957)}]{ReggWhee57}%
  \BibitemOpen
  \bibfield  {author} {\bibinfo {author} {\bibfnamefont {T.}~\bibnamefont
  {Regge}}\ and\ \bibinfo {author} {\bibfnamefont {J.}~\bibnamefont
  {Wheeler}},\ }\href@noop {} {\bibfield  {journal} {\bibinfo  {journal} {Phys.
  Rev.}\ }\textbf {\bibinfo {volume} {108}},\ \bibinfo {pages} {1063} (\bibinfo
  {year} {1957})}\BibitemShut {NoStop}%
\bibitem [{\citenamefont {Zerilli}(1970)}]{Zeri70}%
  \BibitemOpen
  \bibfield  {author} {\bibinfo {author} {\bibfnamefont {F.}~\bibnamefont
  {Zerilli}},\ }\href@noop {} {\bibfield  {journal} {\bibinfo  {journal} {Phys.
  Rev. D}\ }\textbf {\bibinfo {volume} {2}},\ \bibinfo {pages} {2141} (\bibinfo
  {year} {1970})}\BibitemShut {NoStop}%
\bibitem [{\citenamefont {Mathisson}(2010)}]{Mathisson2010}%
  \BibitemOpen
  \bibfield  {author} {\bibinfo {author} {\bibfnamefont {M.}~\bibnamefont
  {Mathisson}},\ }\href {\doibase 10.1007/s10714-010-0939-y} {\bibfield
  {journal} {\bibinfo  {journal} {General Relativity and Gravitation}\ }\textbf
  {\bibinfo {volume} {42}},\ \bibinfo {pages} {1011} (\bibinfo {year}
  {2010})}\BibitemShut {NoStop}%
\bibitem [{\citenamefont {Costa}\ and\ \citenamefont
  {Natário}(2015)}]{Costa:2014nta}%
  \BibitemOpen
  \bibfield  {author} {\bibinfo {author} {\bibfnamefont {L.~F.~O.}\
  \bibnamefont {Costa}}\ and\ \bibinfo {author} {\bibfnamefont
  {J.}~\bibnamefont {Natário}},\ }\bibfield  {booktitle} {\emph {\bibinfo
  {booktitle} {{Proceedings, 524th WE-Heraeus-Seminar: Equations of Motion in
  Relativistic Gravity (EOM 2013): Bad Honnef, Germany, February 17-23,
  2013}}},\ }\href {\doibase 10.1007/978-3-319-18335-0_6} {\bibfield  {journal}
  {\bibinfo  {journal} {Fund. Theor. Phys.}\ }\textbf {\bibinfo {volume}
  {179}},\ \bibinfo {pages} {215} (\bibinfo {year} {2015})},\ \Eprint
  {http://arxiv.org/abs/1410.6443} {arXiv:1410.6443} \BibitemShut {NoStop}%
\bibitem [{\citenamefont {{Kyrian}}\ and\ \citenamefont
  {{Semer{\'a}k}}(2007)}]{2007MNRAS.382.1922K}%
  \BibitemOpen
  \bibfield  {author} {\bibinfo {author} {\bibfnamefont {K.}~\bibnamefont
  {{Kyrian}}}\ and\ \bibinfo {author} {\bibfnamefont {O.}~\bibnamefont
  {{Semer{\'a}k}}},\ }\href {\doibase 10.1111/j.1365-2966.2007.12502.x}
  {\bibfield  {journal} {\bibinfo  {journal} {Mon. Not. R. Astron. Soc}\
  }\textbf {\bibinfo {volume} {382}},\ \bibinfo {pages} {1922} (\bibinfo {year}
  {2007})}\BibitemShut {NoStop}%
\bibitem [{\citenamefont {Lukes-Gerakopoulos}\ \emph
  {et~al.}(2014)\citenamefont {Lukes-Gerakopoulos}, \citenamefont {Seyrich},\
  and\ \citenamefont {Kunst}}]{Lukes-Gerakopoulos:2014dma}%
  \BibitemOpen
  \bibfield  {author} {\bibinfo {author} {\bibfnamefont {G.}~\bibnamefont
  {Lukes-Gerakopoulos}}, \bibinfo {author} {\bibfnamefont {J.}~\bibnamefont
  {Seyrich}}, \ and\ \bibinfo {author} {\bibfnamefont {D.}~\bibnamefont
  {Kunst}},\ }\href {\doibase 10.1103/PhysRevD.90.104019} {\bibfield  {journal}
  {\bibinfo  {journal} {Phys. Rev.}\ }\textbf {\bibinfo {volume} {D90}},\
  \bibinfo {pages} {104019} (\bibinfo {year} {2014})}\BibitemShut {NoStop}%
\bibitem [{\citenamefont {{Pirani}}(1956)}]{Pira56}%
  \BibitemOpen
  \bibfield  {author} {\bibinfo {author} {\bibfnamefont {F.~A.~E.}\
  \bibnamefont {{Pirani}}},\ }\href@noop {} {\bibfield  {journal} {\bibinfo
  {journal} {Acta Physica Polonica}\ }\textbf {\bibinfo {volume} {15}},\
  \bibinfo {pages} {389} (\bibinfo {year} {1956})}\BibitemShut {NoStop}%
\bibitem [{\citenamefont {Ruangsri}\ \emph {et~al.}(2016)\citenamefont
  {Ruangsri}, \citenamefont {Vigeland},\ and\ \citenamefont
  {Hughes}}]{RuanVigeHugh15}%
  \BibitemOpen
  \bibfield  {author} {\bibinfo {author} {\bibfnamefont {U.}~\bibnamefont
  {Ruangsri}}, \bibinfo {author} {\bibfnamefont {S.~J.}\ \bibnamefont
  {Vigeland}}, \ and\ \bibinfo {author} {\bibfnamefont {S.~A.}\ \bibnamefont
  {Hughes}},\ }\href {\doibase 10.1103/PhysRevD.94.044008} {\bibfield
  {journal} {\bibinfo  {journal} {Phys. Rev.}\ }\textbf {\bibinfo {volume}
  {D94}},\ \bibinfo {pages} {044008} (\bibinfo {year} {2016})},\ \Eprint
  {http://arxiv.org/abs/1512.00376} {arXiv:1512.00376} \BibitemShut {NoStop}%
\bibitem [{\citenamefont {Zenginoglu}\ and\ \citenamefont
  {Khanna}(2011)}]{Zenginoglu:2011zz}%
  \BibitemOpen
  \bibfield  {author} {\bibinfo {author} {\bibfnamefont {A.}~\bibnamefont
  {Zenginoglu}}\ and\ \bibinfo {author} {\bibfnamefont {G.}~\bibnamefont
  {Khanna}},\ }\href {\doibase 10.1103/PhysRevX.1.021017} {\bibfield  {journal}
  {\bibinfo  {journal} {Phys. Rev.}\ }\textbf {\bibinfo {volume} {X1}},\
  \bibinfo {pages} {021017} (\bibinfo {year} {2011})},\ \Eprint
  {http://arxiv.org/abs/1108.1816} {arXiv:1108.1816} \BibitemShut {NoStop}%
\bibitem [{\citenamefont {Sundararajan}\ \emph {et~al.}(2008)\citenamefont
  {Sundararajan}, \citenamefont {Khanna}, \citenamefont {Hughes},\ and\
  \citenamefont {Drasco}}]{Sundararajan:2008zm}%
  \BibitemOpen
  \bibfield  {author} {\bibinfo {author} {\bibfnamefont {P.~A.}\ \bibnamefont
  {Sundararajan}}, \bibinfo {author} {\bibfnamefont {G.}~\bibnamefont
  {Khanna}}, \bibinfo {author} {\bibfnamefont {S.~A.}\ \bibnamefont {Hughes}},
  \ and\ \bibinfo {author} {\bibfnamefont {S.}~\bibnamefont {Drasco}},\ }\href
  {\doibase 10.1103/PhysRevD.78.024022} {\bibfield  {journal} {\bibinfo
  {journal} {Phys. Rev.}\ }\textbf {\bibinfo {volume} {D78}},\ \bibinfo {pages}
  {024022} (\bibinfo {year} {2008})},\ \Eprint {http://arxiv.org/abs/0803.0317}
  {arXiv:0803.0317} \BibitemShut {NoStop}%
\bibitem [{\citenamefont {Barack}\ and\ \citenamefont
  {Cutler}(2004)}]{Barack:2003fp}%
  \BibitemOpen
  \bibfield  {author} {\bibinfo {author} {\bibfnamefont {L.}~\bibnamefont
  {Barack}}\ and\ \bibinfo {author} {\bibfnamefont {C.}~\bibnamefont
  {Cutler}},\ }\href {\doibase 10.1103/PhysRevD.69.082005} {\bibfield
  {journal} {\bibinfo  {journal} {Phys. Rev.}\ }\textbf {\bibinfo {volume}
  {D69}},\ \bibinfo {pages} {082005} (\bibinfo {year} {2004})},\ \Eprint
  {http://arxiv.org/abs/gr-qc/0310125} {arXiv:gr-qc/0310125} \BibitemShut
  {NoStop}%
\bibitem [{\citenamefont {Chua}\ and\ \citenamefont
  {Gair}(2015)}]{Chua:2015mua}%
  \BibitemOpen
  \bibfield  {author} {\bibinfo {author} {\bibfnamefont {A.~J.~K.}\
  \bibnamefont {Chua}}\ and\ \bibinfo {author} {\bibfnamefont {J.~R.}\
  \bibnamefont {Gair}},\ }\href {\doibase 10.1088/0264-9381/32/23/232002}
  {\bibfield  {journal} {\bibinfo  {journal} {Class. Quant. Grav.}\ }\textbf
  {\bibinfo {volume} {32}},\ \bibinfo {pages} {232002} (\bibinfo {year}
  {2015})},\ \Eprint {http://arxiv.org/abs/1510.06245} {arXiv:1510.06245}
  \BibitemShut {NoStop}%
\bibitem [{\citenamefont {Teukolsky}(1973)}]{Teuk73}%
  \BibitemOpen
  \bibfield  {author} {\bibinfo {author} {\bibfnamefont {S.}~\bibnamefont
  {Teukolsky}},\ }\href@noop {} {\bibfield  {journal} {\bibinfo  {journal}
  {Astrophys. J.}\ }\textbf {\bibinfo {volume} {185}},\ \bibinfo {pages} {635}
  (\bibinfo {year} {1973})}\BibitemShut {NoStop}%
\bibitem [{\citenamefont {Moncrief}(1974)}]{Monc74}%
  \BibitemOpen
  \bibfield  {author} {\bibinfo {author} {\bibfnamefont {V.}~\bibnamefont
  {Moncrief}},\ }\href {\doibase 10.1016/0003-4916(74)90173-0} {\bibfield
  {journal} {\bibinfo  {journal} {Ann. Phys.}\ }\textbf {\bibinfo {volume}
  {88}},\ \bibinfo {pages} {323} (\bibinfo {year} {1974})}\BibitemShut
  {NoStop}%
\bibitem [{\citenamefont {Cunningham}\ \emph {et~al.}(1978)\citenamefont
  {Cunningham}, \citenamefont {Price},\ and\ \citenamefont
  {Moncrief}}]{CunnPricMonc78}%
  \BibitemOpen
  \bibfield  {author} {\bibinfo {author} {\bibfnamefont {C.}~\bibnamefont
  {Cunningham}}, \bibinfo {author} {\bibfnamefont {R.}~\bibnamefont {Price}}, \
  and\ \bibinfo {author} {\bibfnamefont {V.}~\bibnamefont {Moncrief}},\ }\href
  {\doibase 10.1086/156413} {\bibfield  {journal} {\bibinfo  {journal}
  {Astrophys. J.}\ }\textbf {\bibinfo {volume} {224}},\ \bibinfo {pages} {643}
  (\bibinfo {year} {1978})}\BibitemShut {NoStop}%
\bibitem [{\citenamefont {Cunningham}\ \emph {et~al.}(1979)\citenamefont
  {Cunningham}, \citenamefont {Price},\ and\ \citenamefont
  {Moncrief}}]{CunnPricMonc79}%
  \BibitemOpen
  \bibfield  {author} {\bibinfo {author} {\bibfnamefont {C.}~\bibnamefont
  {Cunningham}}, \bibinfo {author} {\bibfnamefont {R.}~\bibnamefont {Price}}, \
  and\ \bibinfo {author} {\bibfnamefont {V.}~\bibnamefont {Moncrief}},\ }\href
  {\doibase 10.1086/157147} {\bibfield  {journal} {\bibinfo  {journal}
  {Astrophys. J.}\ }\textbf {\bibinfo {volume} {230}},\ \bibinfo {pages} {870}
  (\bibinfo {year} {1979})}\BibitemShut {NoStop}%
\bibitem [{\citenamefont {Campanelli}\ \emph {et~al.}(2006)\citenamefont
  {Campanelli}, \citenamefont {Lousto},\ and\ \citenamefont
  {Zlochower}}]{Campanelli:2006uy}%
  \BibitemOpen
  \bibfield  {author} {\bibinfo {author} {\bibfnamefont {M.}~\bibnamefont
  {Campanelli}}, \bibinfo {author} {\bibfnamefont {C.~O.}\ \bibnamefont
  {Lousto}}, \ and\ \bibinfo {author} {\bibfnamefont {Y.}~\bibnamefont
  {Zlochower}},\ }\href {\doibase 10.1103/PhysRevD.74.041501} {\bibfield
  {journal} {\bibinfo  {journal} {Phys. Rev.}\ }\textbf {\bibinfo {volume}
  {D74}},\ \bibinfo {pages} {041501} (\bibinfo {year} {2006})},\ \Eprint
  {http://arxiv.org/abs/gr-qc/0604012} {arXiv:gr-qc/0604012} \BibitemShut
  {NoStop}%
\bibitem [{\citenamefont {van~de Meent}(2016)}]{vandeMeent:2016pee}%
  \BibitemOpen
  \bibfield  {author} {\bibinfo {author} {\bibfnamefont {M.}~\bibnamefont
  {van~de Meent}},\ }\href {\doibase 10.1103/PhysRevD.94.044034} {\bibfield
  {journal} {\bibinfo  {journal} {Phys. Rev.}\ }\textbf {\bibinfo {volume}
  {D94}},\ \bibinfo {pages} {044034} (\bibinfo {year} {2016})},\ \Eprint
  {http://arxiv.org/abs/1606.06297} {arXiv:1606.06297} \BibitemShut {NoStop}%
\bibitem [{\citenamefont {Kavanagh}\ \emph {et~al.}(2016)\citenamefont
  {Kavanagh}, \citenamefont {Ottewill},\ and\ \citenamefont
  {Wardell}}]{Kavanagh:2016idg}%
  \BibitemOpen
  \bibfield  {author} {\bibinfo {author} {\bibfnamefont {C.}~\bibnamefont
  {Kavanagh}}, \bibinfo {author} {\bibfnamefont {A.~C.}\ \bibnamefont
  {Ottewill}}, \ and\ \bibinfo {author} {\bibfnamefont {B.}~\bibnamefont
  {Wardell}},\ }\href {\doibase 10.1103/PhysRevD.93.124038} {\bibfield
  {journal} {\bibinfo  {journal} {Phys. Rev.}\ }\textbf {\bibinfo {volume}
  {D93}},\ \bibinfo {pages} {124038} (\bibinfo {year} {2016})},\ \Eprint
  {http://arxiv.org/abs/1601.03394} {arXiv:1601.03394} \BibitemShut {NoStop}%
\bibitem [{\citenamefont {Forseth}\ \emph {et~al.}(2016)\citenamefont
  {Forseth}, \citenamefont {Evans},\ and\ \citenamefont
  {Hopper}}]{Forseth:2015oua}%
  \BibitemOpen
  \bibfield  {author} {\bibinfo {author} {\bibfnamefont {E.}~\bibnamefont
  {Forseth}}, \bibinfo {author} {\bibfnamefont {C.~R.}\ \bibnamefont {Evans}},
  \ and\ \bibinfo {author} {\bibfnamefont {S.}~\bibnamefont {Hopper}},\ }\href
  {\doibase 10.1103/PhysRevD.93.064058} {\bibfield  {journal} {\bibinfo
  {journal} {Phys. Rev.}\ }\textbf {\bibinfo {volume} {D93}},\ \bibinfo {pages}
  {064058} (\bibinfo {year} {2016})},\ \Eprint
  {http://arxiv.org/abs/1512.03051} {arXiv:1512.03051} \BibitemShut {NoStop}%
\bibitem [{\citenamefont {Hopper}\ \emph {et~al.}(2016)\citenamefont {Hopper},
  \citenamefont {Kavanagh},\ and\ \citenamefont {Ottewill}}]{Hopper:2015icj}%
  \BibitemOpen
  \bibfield  {author} {\bibinfo {author} {\bibfnamefont {S.}~\bibnamefont
  {Hopper}}, \bibinfo {author} {\bibfnamefont {C.}~\bibnamefont {Kavanagh}}, \
  and\ \bibinfo {author} {\bibfnamefont {A.~C.}\ \bibnamefont {Ottewill}},\
  }\href {\doibase 10.1103/PhysRevD.93.044010} {\bibfield  {journal} {\bibinfo
  {journal} {Phys. Rev.}\ }\textbf {\bibinfo {volume} {D93}},\ \bibinfo {pages}
  {044010} (\bibinfo {year} {2016})},\ \Eprint
  {http://arxiv.org/abs/1512.01556} {arXiv:1512.01556} \BibitemShut {NoStop}%
\end{thebibliography}%

\end{document}